\newcommand{\chem}[2]{\ensuremath{^{#2}\kern-0.8pt\mathrm{#1}}}
\newcommand{\reac}[6]{\ensuremath{\,^{#2}\kern-0.8pt\mathrm{#1}\,({#3}\,,{#4})\,{}^{#6}\kern-0.8pt\mathrm{#5}\,}}

\newcommand{\deltaTSub}{\ensuremath{\Delta T_\mathrm{sub}}}
\newcommand{\Ab}{\ensuremath{\mathcal{A}}}

\newcommand{\AbSmooth}{\ensuremath{\mathcal{A}_\mathrm{smooth}}}
\newcommand{\AbSub}{{\mathcal{A}_\mathrm{sub}}}
\newcommand{\meanAbSubs}{\ensuremath{\overline{\AbSub}}}
\newcommand{\excessAb}{\ensuremath{\mathcal{E}_\mathcal{A}}}
\newcommand{\diagram}{\ensuremath{\Ab \; \mathrm{versus} \; \excessAb}}
\newcommand{\EROSR}{\ensuremath{R_\mathrm{EROS}}}
\newcommand{\EROSB}{\ensuremath{B_\mathrm{EROS}}}
\newcommand{\field}[1]{\texttt{#1}}

\documentclass{aa}  
\usepackage{graphicx}
\usepackage{enumerate}
\usepackage{txfonts}
\usepackage{natbib}

\usepackage{ulem}
\usepackage{soul}
\usepackage{color}

\begin{document}
   \title{Searching transients in large-scale surveys}

   \subtitle{A method based on the Abbe value}

   \author{N. Mowlavi
          }

   \institute{Department of Astronomy, University of Geneva, 51 chemin des Maillettes, 1290 Versoix, Switzerland\\
                \email{Nami.Mowlavi@unige.ch
                       }
             }

   \date{Received ...; accepted ...}

 
  \abstract
   {}
   {A new method is presented to identify transient candidates in large-scale surveys based on the variability pattern in their light curves.
   }
   {The method is based on the Abbe value, $\Ab$, that estimates the smoothness of a light curve, and on a newly introduced value called the excess Abbe and denoted $\excessAb$, that estimates the regularity of the light curve variability pattern over the duration of the observations.
   }
   {Based on simulated light curves, transients are shown to occupy a specific region in the $\diagram$ diagram, distinct from sources presenting pulsating-like features in their light curves or having featureless light curves.
   
   The method is tested on real light curves taken from EROS-2 and OGLE-II surveys in a $0.50^{\circ} \times 0.17^{\circ}$ field of the sky in the LMC centered at RA(J2000)=5h25m56.5s and DEC(J2000)=-69d29m43.3s.
   The method identifies 43 EROS-2 transient candidates out of a total of $\sim$1300 variable stars, and 19 more OGLE-II candidates, 10 of which do not have any EROS-2 variable star matches and which would need further confirmation to assess their reliability.
    The efficiency of the method is further tested by comparing the list of transient candidates with known Be stars in the literature.
    It is shown that all Be stars known in the studied field of view with detectable bursts or outbursts are successfully extracted by the method.
    In addition, four new transient candidates displaying bursts and/or outbursts are found in the field, of which at least two are good new Be candidates.
   }
   {The new method proves to be a potentially powerful tool to extract transient candidates from large-scale multi-epoch surveys.
   The better the photometric measurement uncertainties are, the cleaner the list of detected transient candidates is.
   In addition, the $\diagram$ diagram is shown to be a good diagnostic tool to check the data quality of multi-epoch photometric surveys.
   A trend of instrumental and/or data reduction origin, for example, will manifest itself by an unexpected distribution of points in the diagram.
   }

   \keywords{Methods: data analysis -- Stars: variables: general -- Stars: emission-line, Be
            }

   \maketitle

%
\section{Introduction}
\label{Sect:introduction}

The analysis of transient phenomena has gained momentum since the advent of large-scale multi-epoch surveys in the 1990s.
These initial surveys, such as the Exp\'erience pour la Recherche d'Objets Sombres \cite[EROS-1, 1990-1995;][]{AubourgBareyreBrehin93,RenaultAubourgBareyre98}, the Massive Compact Halo Object experiment \citep[MACHO, 1992-1999;][]{AlcockAllsmanAlves97}, and the Optical Gravitational Lensing Experiment \citep[OGLE-I, 1992-1995;][]{UdalskiSzymanskiKaluzny92}, were initially devoted to the search of galactic microlensing events.
As a side product, they provided a wealth of data for stellar variability studies.
Based on their successes, two of those surveys continued their programs with upgraded instruments.
EROS-2 operated from 1996 to 2003 \citep{Palanque-DelabrouilleAfonsoAlfert_etal98,TisserandLeGuillouAfonso_etal07}, while the OGLE experiment had three major updates, OGLE-II \citep[1996-2000;][]{Szymanski05}, OGLE-III \citep[2001-2009;][]{UdalskiSzymanskiSoszynski_etal08}, and OGLE-IV \citep[since 2010; e.g.,][]{PoleskiUdalskiSkowron_etal11}.
In addition, multi-epoch surveys specifically dedicated to the search of transients have been developed.
Examples are the Robotic Optical Transient Search Experiment \citep[ROTSE-III;][]{AkerlofKehoeMcKay03}, the Palomar Transient Factory survey \citep[PTF;][]{LawKulkarniDekany_etal09}, and the Catalina Real-time Transient Survey \citep[CRTS;][]{DrakeDjorgovskiMahabal09}.
One of the challenges of those large-scale surveys in the field of transients resides in the automated detection of those objects.
Current techniques include image subtraction (e.g., used in PTF) and brightness increase detection above a given threshold (e.g., used in CRTS where a threshold of two magnitudes is adopted).
They target transient detection in real time.
New and future multi-epoch large-scale missions, such as the ESA Gaia space mission launched in December 2013 \citep{PerrymanDeBoerGilmore_etal01,EyerHollPourbaix_etal13} or the future Large Synoptic Survey Telescope (LSST) planned to enter science operations in 2022 \citep{IvezicTysonAcosta_etal08}, will also benefit the study of transients.

In this paper, I present a new method for identifying transients in a survey.
It is based on the Abbe value \citep{vonNeumann41,vonNeumann42}, denoted $\Ab$ in this paper, which quantifies the `smoothness' of a time series (see Sect.~\ref{Sect:method}). 
It has a value of about one for purely noisy time series of a constant function, and decreases to zero for time series displaying a smooth pattern of variability between successive measurements.
The method compares the Abbe value $\Ab$ of the whole time series to the mean of Abbe values computed on subtime intervals, denoted $\meanAbSubs$.
The amount by which this mean Abbe value exceeds the Abbe value of the whole time series, which I call the \textit{excess Abbe value} and denote $\excessAb$, quantifies the degree to which a time series contains transient features.
The larger $\excessAb$, the more likely the presence of transient features in the time series.
The transient nature of a time series is best summarized in the ($\Ab$, $\excessAb$) diagram, which I call the \textit{excess Abbe diagram}.
In this diagram, time series are spread over four distinct regions depending on their variability pattern (see Sect.~\ref{Sect:diagramSummary}, and Fig.~\ref{Fig:summaryDiagram} in particular), one of which corresponds to time series most likely containing transient events.

The paper consists of two parts.
In the first part, I describe the method (Sect.~\ref{Sect:method}) and test it on simulated light curves of different variability types (Sect.~\ref{Sect:simulations}).
In the second part, I apply the method to real light curves taken from EROS-2 (Sect.~\ref{Sect:EROS}) and OGLE-II (Sect.~\ref{Sect:OGLE-II}) surveys, and check in Sect.~\ref{Sect:literature} its efficiency in detecting Be stars, for which a census is available in the literature for the LMC field of view selected here.
The potential power of the method as a diagnostic tool to check the quality of multi-epoch surveys is also addressed in Sect.~\ref{Sect:EROS}.
Conclusions are drawn in Sect.~\ref{Sect:conclusions}.

\section{The method}
\label{Sect:method}

We consider a time series $\{t_i, y_i\}$ of values $y_i$ measured at times $t_i$, where $i$ is an index running over the number of measurements $n$.
The Abbe value $\Ab$ is defined as \citep{vonNeumann41,vonNeumann42}
\begin{equation}
  \Ab = \frac{n}{2(n-1)} \frac{\sum_{i=1}^{n-1}(y_{i+1}-y_i)^2}{\sum_{i=1}^{n}(y_i-\bar{y})^2} \,,
\label{Eq:Abbe}
\end{equation}
where $\bar{y}$ is the mean of $\{y_i\}$.
It quantifies the smoothness of a time series by comparing the sum of the squared differences between two successive measurements,  $(y_{i+1}-y_i)^2$, with the standard deviation of the time series.
The Abbe value decreases to zero for time series displaying a high degree of smoothness,
while the normalization factor in Eq.~\ref{Eq:Abbe} ensures that it tends to one for a purely noisy time series.

The Abbe value $\Ab$ is computed over the time span $\Delta T = t_n-t_1$ of the whole time series.
An Abbe value can also be computed on a subtime interval $[t_i-\frac{1}{2}\deltaTSub, t_i+\frac{1}{2}\deltaTSub]$ of duration $\deltaTSub < \Delta T$ centered at time $t_i$.
The resulting Abbe value, denoted by $\AbSub_{,i}$, will significantly differ from the Abbe value $\Ab$ of the whole time series only if the variability pattern in the subtime interval differs significantly from that of the whole time series.
I use this property to check whether a time series has a stable variability pattern over time, or whether specific features exist at specific times in the time series.
More specifically, I compute the mean $\meanAbSubs$ of all $\AbSub_{,i}$ values (i.e., computed at all times $t_i$ of the time series), and compare it to the Abbe value $\Ab$ of the whole time series through the \textit{excess Abbe} value defined by
\begin{equation}
  \excessAb \equiv \meanAbSubs-\Ab \;.
\label{Eq:excessAb}
\end{equation}
The excess Abbe value is close to zero for time series that display a constant variability pattern (with respect to $\deltaTSub$, see remark below),  and can increase up to about $1-\Ab$ for time series with non-uniform variability patterns (with respect to $\deltaTSub$).
The source is said to be a transient candidate if $\excessAb$ is greater than 0.2 (see Sect.~\ref{Sect:diagramSummary}).

The choice of $\deltaTSub$ determines the timescale above which a transient is detected, as well as the number of false positives.
Periodic stars with periods greater than $\deltaTSub$, for example, will pollute the ensemble of transient candidates extracted by the method.
If we consider a sine curve with period $P$, taking $\deltaTSub\gtrsim P$ ensures that the variability pattern is similar in any subtime interval of duration $\deltaTSub$.
In other words, $\AbSub_{,i}$ will have about the same value at all times $t_i$, assuming of course a regularly sampled time series.
This will no longer be the case if we take $\deltaTSub\lesssim P$.
The variability patterns observed around light curve maxima and minima (at phases $\pi/2$ or $3\pi/2$, respectively), for example, will differ from those observed at phases 0 or $\pi$ if $\deltaTSub\lesssim P$.
As a consequence, a sine curve will be considered as a transient candidate if $P \gtrsim \deltaTSub$.
This is an important point to keep in mind when assessing the reality of a transient candidate identified by the method.

The location of a time series in the $\diagram$ diagram is thus expected to depend on the type of variability present in the light curve.
This is explored in the next section using simulations and analytical models, a summary of which is presented in Sect.~\ref{Sect:diagramSummary}.
It will be shown, in particular, that transient sources populate a specific region of the diagram at high $\excessAb$ values.

\section{The $\diagram$ diagram}
\label{Sect:simulations}

In the following subsections, I explore the locations in the $\diagram$ diagram of several types of light curves using simulations and analytic developments.
All simulated time series cover 760~days, are regularly sampled with one simulated observation per night, and include various levels of noise.
A value of $\deltaTSub=100$~d is used throughout the section.
Based on the discussion in Sect.~\ref{Sect:method}, this value should lead to the identification of transients with timescales larger than $\sim$100~days, but the ensemble of transient candidates is expected to be polluted by periodic variables with periods larger than $\sim$100~days as well.

The location of periodic variables in the $\diagram$ diagram is first analyzed in Sect.~\ref{Sect:simuPeriodic}.
Light curves displaying trends are then addressed in Sect.~\ref{Sect:simuTrends}, and light curves with various transient patterns in Sect.~\ref{Sect:simuTransients}.
A summary of the $\diagram$ diagram is presented in Sect.~\ref{Sect:diagramSummary}, where the choice of $\deltaTSub$ is further discussed.

\subsection{Pulsating-like light curves}
\label{Sect:simuPeriodic}

\begin{figure}
  \centering
  \includegraphics[width=\columnwidth]{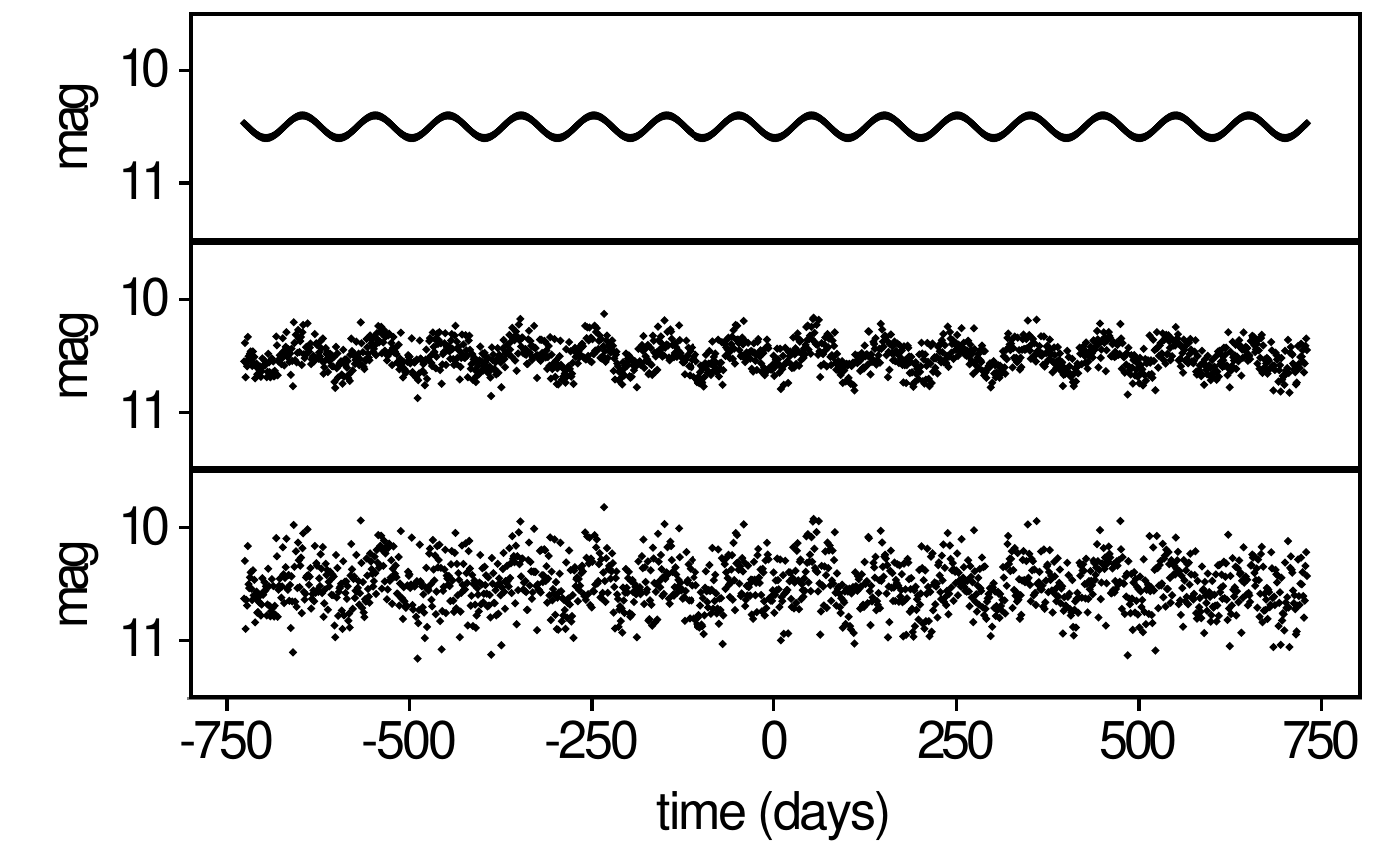}
  \caption{Simulated time series of a sinusoidal light curve with a sine amplitude of 0.1~mag and a period of 100~days.
  One point is simulated per night, at the same (arbitrary) time of the night each night.
  \textbf{Top panel:} Without noise.
  \textbf{Middle panel:} With Gaussian-distributed noise of 0.1~mag amplitude.
  \textbf{Bottom panel:} With Gaussian-distributed noise of 0.2~mag amplitude.
  }
\label{Fig:simuPerioricTs}
\end{figure}

Time series of sinusoidal light curves are simulated with periods $P_\mathrm{sine}=100$, 200, and 400~d, and with an (arbitrary) amplitude $A_\mathrm{sine}=0.1$~mag.
Noise is added to the time series with an amplitude randomly generated from a Gaussian distribution with standard deviation $\sigma_\mathrm{noise}$.
In our simulations, the noise level $\sigma_\mathrm{noise}/A_\mathrm{sine}$ ranges from 0 (noise-free time series) to 3.
Simulated time series for $\sigma_\mathrm{noise}/A_\mathrm{sine}=0$, 1, and 2 are shown in Fig.~\ref{Fig:simuPerioricTs} for $P_\mathrm{sine}=100$~d.

\begin{figure}
  \centering
  \includegraphics[width=\columnwidth]{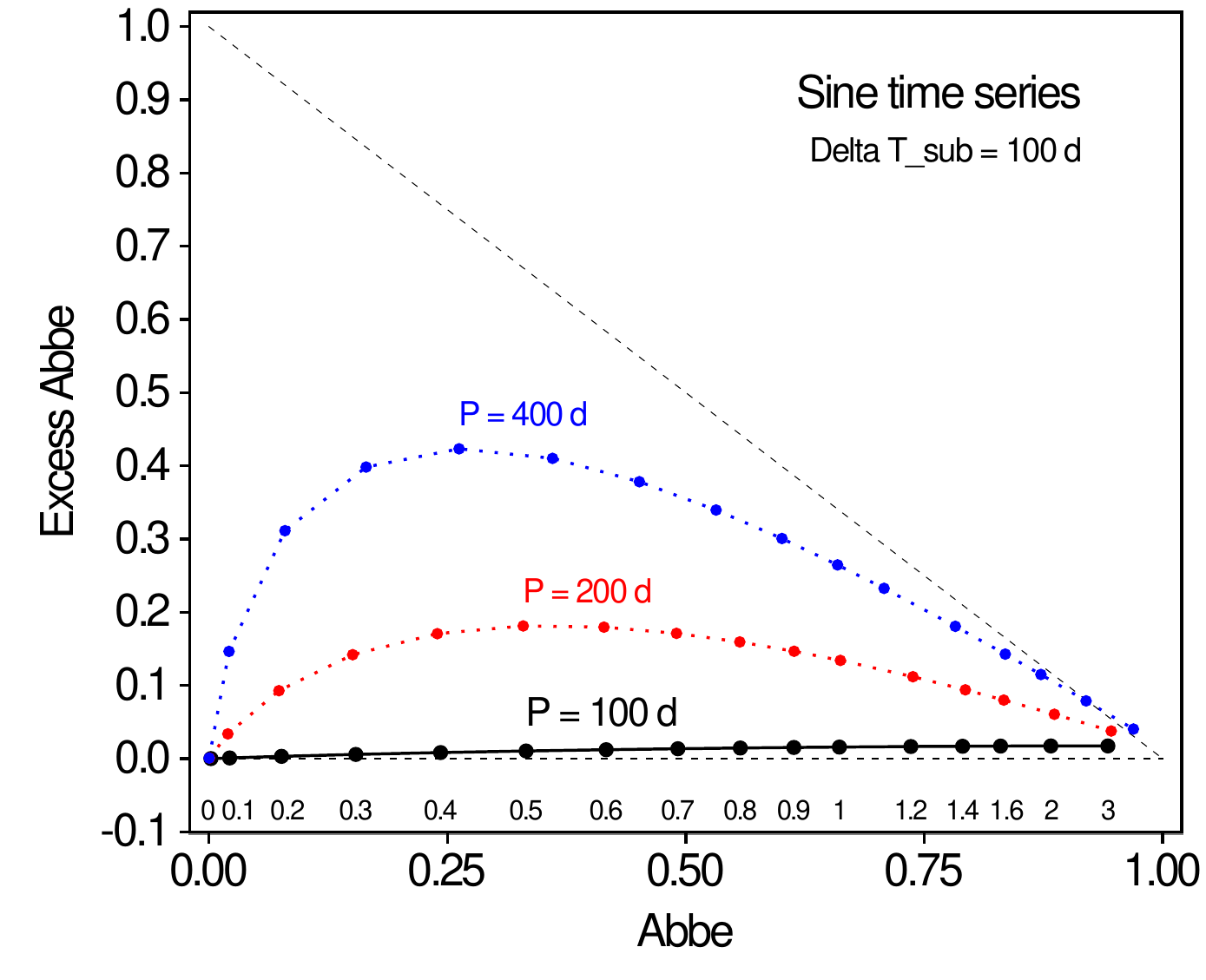}
  \caption{Locations in the $\diagram$ diagram of three sets of sixteen sinusoidal time series with different periods and noise levels.
  A value of $\deltaTSub=100$~d has been used in the calculation of $\excessAb$ for all time series.
  Black larger markers connected with a continuous line identify time series with a period of 100~d (i.e., equal to $\deltaTSub$), while red and blue smaller markers connected with dotted lines identify times series with periods of 200 and 400~d, respectively, as indicated above the lines.
  Each marker on a line locates a time series with a specific noise level, increasing from $\sigma_\mathrm{noise}/A_\mathrm{sine}=0$ to 3 from left to right, as indicated below the markers of the $P=100$~d time series.
  The horizontal dashed line is an guideline to $\excessAb=0$ and the diagonal dashed line an guideline to $\excessAb=1-\Ab$.
  }
\label{Fig:simuPeriodicDiagram}
\end{figure}

According to the discussion in Sect.~\ref{Sect:method}, light curves with periods shorter than $\deltaTSub$ (=100~d in the simulations) should lead to excess Abbe values $\excessAb$ close to zero.
Simulations with $P_\mathrm{sine}=100$~d and various noise levels confirm this prediction.
Their locations in the $\diagram$ diagram are shown in Fig.~\ref{Fig:simuPeriodicDiagram} by black markers connected with solid lines.
Their excess Abbe values are all less than 0.02.
We also note that the Abbe value increases with noise level, as expected.
For $\sigma_\mathrm{noise}/A_\mathrm{sine}=0$, 0.2, 0.5, 1, 2, and 3, for example, $\Ab$\,=\,0.02, 0.10, 0.35, 0.67, 0.88, and 0.94, respectively.

Periods longer than $\deltaTSub$ should lead to increasingly larger excess Abbe values.
The locations in the $\diagram$ diagram of sine time series simulated with $P_\mathrm{sine}=200$ and 400~d are shown in Fig.~\ref{Fig:simuPeriodicDiagram} by red and blue dots, respectively.
They confirm the expectation of larger excess Abbe values.
The amplitude of the excess, however, depends on the noise level.
In the noise-free time series ($\Ab \cong 0)$, the excess remains very small for the periods considered here.
In noisy time series, $\excessAb$ increases with increasing noise level, reaches a maximum at a given noise level limit, and then returns to its previous level at higher noise levels.
The maximum $\meanAbSubs$ value equals 0.18 for $P_\mathrm{sine} = 200$~d ($=2\,\deltaTSub$) and 0.42 for $P_\mathrm{sine} = 400$~d ($=4\,\deltaTSub$).
The decrease in $\excessAb$ at higher noise levels is mainly due to the increase in $\Ab$, the maximum possible value for $\excessAb$ being $1-\Ab$.

The simulations thus show that $\excessAb$ of a time series remains close to zero for all periods and at all noise levels, provided that the period of the signal is smaller or similar to the duration $\deltaTSub$ of the subintervals used to compute $\excessAb$.
Time series with larger periods will have higher excess Abbe values, and will be considered as transient candidates (see Sect.~\ref{Sect:diagramSummary}).
These conclusions are not limited to strictly sinusoidal signals, but to any light curve whose variability pattern resembles that of periodic signals.
I call them \textit{pulsating-like} light curves.

\subsection{Light curves with trends}
\label{Sect:simuTrends}

In this section, I identify the location in the $\diagram$ diagram of light curves with trends.
I first address the question analytically in Sect.~\ref{Sect:simuTrendAnalytical}, highlighting the main parameters affecting their location in the diagram.
I then confirm the results with numerical simulations in Sect.~\ref{Sect:simuTrendSimulations}.
The use of the $\diagram$ diagram as a diagnostic tool to check the data quality of a survey is highlighted in Sect.~\ref{Sect:simuTrendDiagnosticTool}.

\subsubsection{Light curves with trends: analytical study}
\label{Sect:simuTrendAnalytical}

We consider a time series $\{t_i, y_i\}$ regularly sampled at $n$ times over a duration $\Delta T$, that can be decomposed in a linear trend component $\{t_i, a \, t_i\}$ of slope $a$, and in a detrended component $\{t_i, f_i\}$, such that $y_i=f_i+a\,t_i$.
It is shown in Appendix \ref{SectAppendix:AbbeWithTrend} (see Eq.~\ref{Eq:AbbeWithTrend:3}) that
\begin{equation}
  \Ab \simeq \frac{\Ab_f + 6 \, \beta / n^2}{1 + \beta}  \,,
  \label{Eq:AbbeWithTrend}
\end{equation}
where $\Ab_f$ is the Abbe value of the detrended time series, and $\beta$ is a parameter quantifying the amplitude of the trend ($a\,\Delta T$) relative to the variability level of the detrended signal ($\sigma_f$) given by (see Eq.~\ref{Eq:beta})
\begin{equation}
  \beta = \frac{1}{12} \frac{(a \, \Delta T)^2}{\sigma_f^2} \;.
\label{Eq:betaDefinition}
\end{equation}

In the absence of any trend ($\beta=0$), Eq.~\ref{Eq:AbbeWithTrend} gives $\Ab_{\;\beta=0} = \Ab_f$, as expected.
When a linear trend is present ($\beta$\,$>$\,0), a term is added to both numerator and denominator of Eq.~\ref{Eq:AbbeWithTrend}, but the term added in the numerator is $n^2/6$ times smaller than the term added in the denominator.
As a consequence, $\Ab$ always decreases with increasing trend amplitudes, starting with $\Ab=\Ab_f$ for $\beta=0$.
The deviation of $\Ab$ from $\Ab_f$ becomes noticeable only when $\beta\gtrsim 1$, i.e., when the amplitude $a\,\Delta T$ of the trend becomes comparable to the standard deviation $\sigma_f$ of the detrended time series.
This corresponds to the condition for the trend to become visible in the light curve.
When this condition is reached, the value of $\Ab$ significantly decreases with increasing trend amplitudes.
At the limit of very high trend coefficients ($\beta \rightarrow \infty$), we have
$\Ab_{\,\beta \rightarrow \infty} \rightarrow 6/n^2$.

\begin{figure}
  \centering
  \includegraphics[width=\columnwidth]{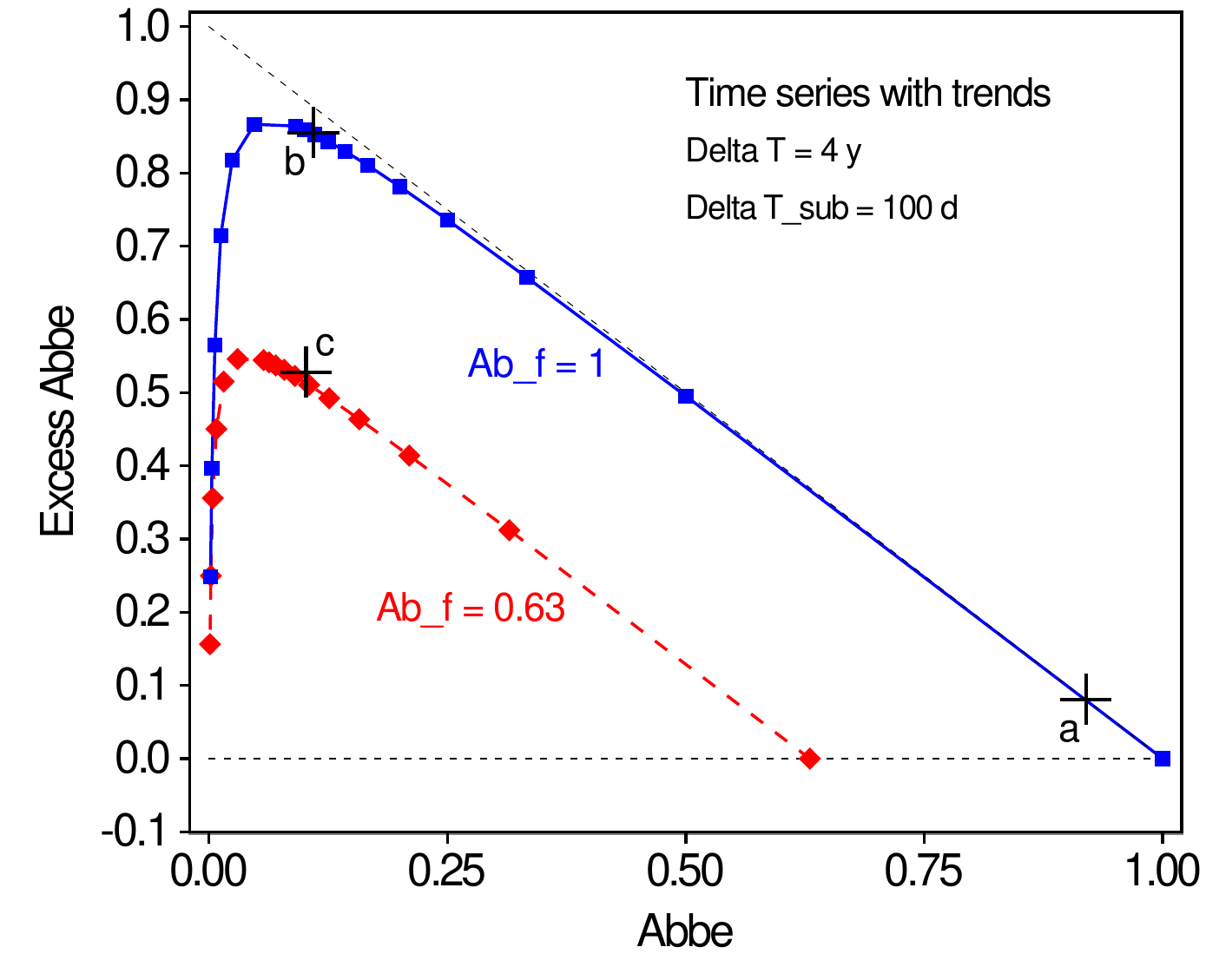}
  \caption{Locations in the $\diagram$ diagram of light curves with trends, determined analytically from Eqs.~\ref{Eq:excessAb}, \ref{Eq:AbbeWithTrend}, and \ref{Eq:approximateAbbeSubWithTrend}, using $\Delta T = 4$~y, $n=1000$, $\Delta T_\mathrm{sub} = 100$~d, and various trend parameters $\beta$.
  Blue markers connected with a continuous blue line (upper line) locate light curves with no variability pattern in the detrended component (i.e., $\Ab_f=1$).
  Red markers connected with a dashed red line (lower line) assume the presence of a variability pattern in the detrended component with $\Ab_f=0.63$.
  The markers on each line locate time series with increasing trend parameter $\beta$ of, from right to left, 0 to 10 by steps of 1, then from 10 to 640 by multiplication factors of 2.
  The black cross markers labeled a, b, and c locate the simulated time series shown in the top, middle, and bottom panels of Fig.~\ref{Fig:simuTrendTs}, respectively.
  The horizontal dashed line is an guideline to $\excessAb=0$ and the diagonal dashed line an guideline to $\excessAb=1-\Ab$.
  }
\label{Fig:simuTrendDiagram}
\end{figure}

We now need to estimate $\excessAb$ from Eq.~\ref{Eq:excessAb}.
To do this, we have to calculate the Abbe value $\AbSub$ of all subtime series of duration $\deltaTSub$, and take their mean value $\meanAbSubs$.
Replacing $\Delta T$ with $\deltaTSub$ in Eqs.~\ref{Eq:AbbeWithTrend} and \ref{Eq:betaDefinition},
we get
\begin{eqnarray}
  \AbSub & \simeq & \frac{\Ab_f +  \left(\frac{\deltaTSub}{\Delta T}\right)^2 \; 6\,\beta/n^2}
                    {1 +  \left(\frac{\deltaTSub}{\Delta T}\right)^2 \; \beta}  \;.
  \label{Eq:approximateAbbeSubWithTrend}
\end{eqnarray}
If the variability pattern of the detrended light curve is time independent, which is the case if we exclude transient-type light curves for the detrended component, then $\Ab_f$ has similar values in all subtime series.
We thus have $\meanAbSubs \simeq \AbSub$, and the value of $\meanAbSubs$ is given by Eq.~\ref{Eq:approximateAbbeSubWithTrend}.

We see that $\meanAbSubs$ given by Eq.~\ref{Eq:approximateAbbeSubWithTrend} is much less sensitive to the trend amplitude $\beta$ than $\Ab$ is, since the additive terms in both numerator and denominator of Eq.~\ref{Eq:approximateAbbeSubWithTrend} are multiplied by $(\deltaTSub / \Delta T)^2 \ll 1$.
Therefore, $\meanAbSubs$ keeps around $\Ab_f$ for a large range of values of $\beta$, and we can write $\excessAb \simeq \Ab_f - \Ab$.
In other words, time series with a trend will populate a line parallel to the upper diagonal limit in the $\diagram$ diagram.
In particular, time series with no variability pattern in their detrended component ($\Ab_f \simeq 1$) will basically populate the upper diagonal limit in the $\diagram$ diagram.
This property is used below (Sect.~\ref{Sect:simuTrendDiagnosticTool}) to provide a tool to test the quality of large sets of time series.

The analytical approximations given by Eqs.~\ref{Eq:AbbeWithTrend} and \ref{Eq:approximateAbbeSubWithTrend} are plotted in Fig.~\ref{Fig:simuTrendDiagram} for two values of $\Ab_f$.
They confirm the conclusions drawn above.
In particular, the representative points for $\Ab_f=1$ (connected by solid lines) are seen to remain close to the upper diagonal limit for all values of $\Ab$ down to 0.1, and to drop significantly for $\Ab$ values below 0.1.
The choice of the number of points $n$ in Eqs.~\ref{Eq:AbbeWithTrend} and \ref{Eq:approximateAbbeSubWithTrend} is not critical; a value of $n=100$ gives essentially the same locations in the $\diagram$ diagram as the value of $n=1000$ adopted in Fig.~\ref{Fig:simuTrendDiagram}.

\subsubsection{Light curves with trends: simulations}
\label{Sect:simuTrendSimulations}

\begin{figure}
  \centering
  \includegraphics[width=\columnwidth]{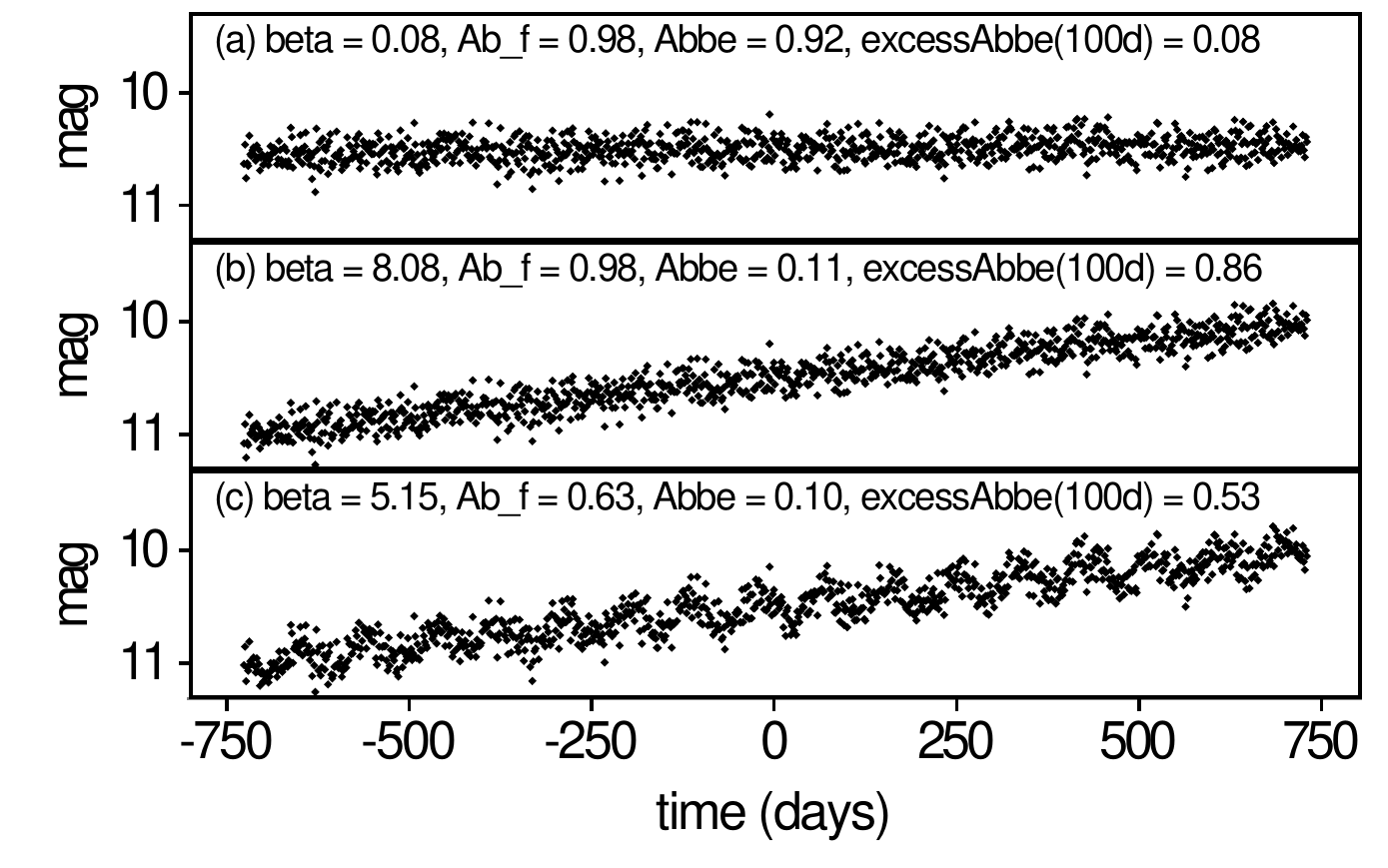}
  \caption{Simulated time series of light curves with a linear trend.
  The light curves are sampled with 1000 regularly spaced points over a total time interval $\Delta T$\,=\,4~years.
  \textbf{(a)} Time series with a trend coefficient $a$\,=\,$-0.1$~mag\,/\,$\Delta T$.
  \textbf{(b)} Same as top panel, but with $a$\,=\,$-1.0$~mag\,/\,$\Delta T$.
  \textbf{(c)} Same as middle panel, but with a sine of period 90~days and amplitude 0.1~mag added to the time series.
  Noise is added in all the time series with a Gaussian-distributed amplitude of 0.1~mag.
  The $\beta$ parameter of the trend (Eq.~\ref{Eq:betaDefinition}), the Abbe value $\Ab_f$ of the detrended component of the time series, and the resulting values of $\Ab$ and $\excessAb$ (computed with $\Delta T_\mathrm{sub}=100$~d) are indicated at the top of each panel.
  }
\label{Fig:simuTrendTs}
\end{figure}

Simulations of time series with trends support the conclusions of Sect.~\ref{Sect:simuTrendAnalytical} based on analytical considerations.
Three light curves are simulated, shown in Fig.~\ref{Fig:simuTrendTs}.

A first light curve is simulated with no variability pattern in the detrended time series and with a small trend coefficient of $a = -0.68\;10^{-4}$~mag/d (top panel of Fig.~\ref{Fig:simuTrendTs}).
It corresponds to a trend amplitude $\beta=0.08$.
Its representative point in the $\diagram$ diagram is shown by the marker labeled "a" in Fig.~\ref{Fig:simuTrendDiagram}.
It has a high Abbe value of 0.98 and lies on the upper diagonal limit, as expected since it resembles a featureless time series with no trend.

The second light curve is also simulated with no variability pattern in the detrended time series, but with a trend coefficient ten times larger than in the first simulation, i.e., $a = -6.85\;10^{-4}$~mag/d (middle panel of Fig.~\ref{Fig:simuTrendTs}).
This corresponds to a trend amplitude one hundred times larger than in the first simulation, i.e., $\beta=8.08$.
Its Abbe value is decreased to 0.11, and its representative point in the $\diagram$ diagram is shown by the marker labeled "b" in Fig.~\ref{Fig:simuTrendDiagram}.
Its location in the diagram agrees very well with the analytical predictions of the previous section.

Finally, the third simulation copies the second one, but with a periodic signal (a sine with an amplitude of 0.1~mag and a period of 90~d) added to the trend.
The resulting Abbe value of the detrended time series (i.e., of the sine) is $\Ab_f=0.63$.
We note that the trend amplitude is reduced to $\beta=5.15$ compared to 8.08 in the second simulation, even though the trend coefficient $a$ is the same in both cases.
This is due to the standard deviation $\sigma_f$ of the detrended light curve, which is larger (due to the sine) than in the second simulation.
The representative point in the $\diagram$ diagram, shown by the marker labeled "c" in Fig.~\ref{Fig:simuTrendDiagram}, also agrees remarkably well with the analytical predictions.

We thus conclude that light curves with trends will populate, in the $\diagram$ diagram, a diagonal below and parallel to the upper diagonal limit.
Time series with trends but with no specific variability pattern in their detrended component will populate the upper diagonal limit.

\subsubsection{The $\diagram$ diagram: a data-quality diagnostic tool}
\label{Sect:simuTrendDiagnosticTool}

A trend in a time series can originate from a long-term transient, a long-period variable sampled on a time interval shorter than one fourth of its period, an instrument drift, or a problem in data reduction, to cite only a few cases.

If trends due to instrument drift or data reduction issues are present in the data base of a survey, we expect the region around the upper diagonal limit in the $\diagram$ to be abnormally populated by a high number of points, according to the results of the previous sections.
This region should probably be limited to $\Ab \gtrsim 0.5$, unless the trend amplitudes are very large.
If we thus locate all sources of a given survey in the $\diagram$ diagram, an inspection of the density of points on the diagonal relative to the density of points in other parts of the diagram should provide a diagnostic tool of the quality of the data base.

On the other hand, if the data quality of a survey is good, only trends of astrophysical origin will populate the upper diagonal limit of the $\diagram$.
In that case, we expect the density of points in that region of the diagram to be small compared to the region populated by sources with pulsating-like light curves.  

The $\diagram$ diagram thus provides an interesting diagnostic tool to check the data quality of a survey.
An illustration of this potentiality using real case as an example is given in Sect.~\ref{Sect:EROS}.

\subsection{Transient-type light curves}
\label{Sect:simuTransients}

\begin{figure}
  \centering
  \includegraphics[width=\columnwidth]{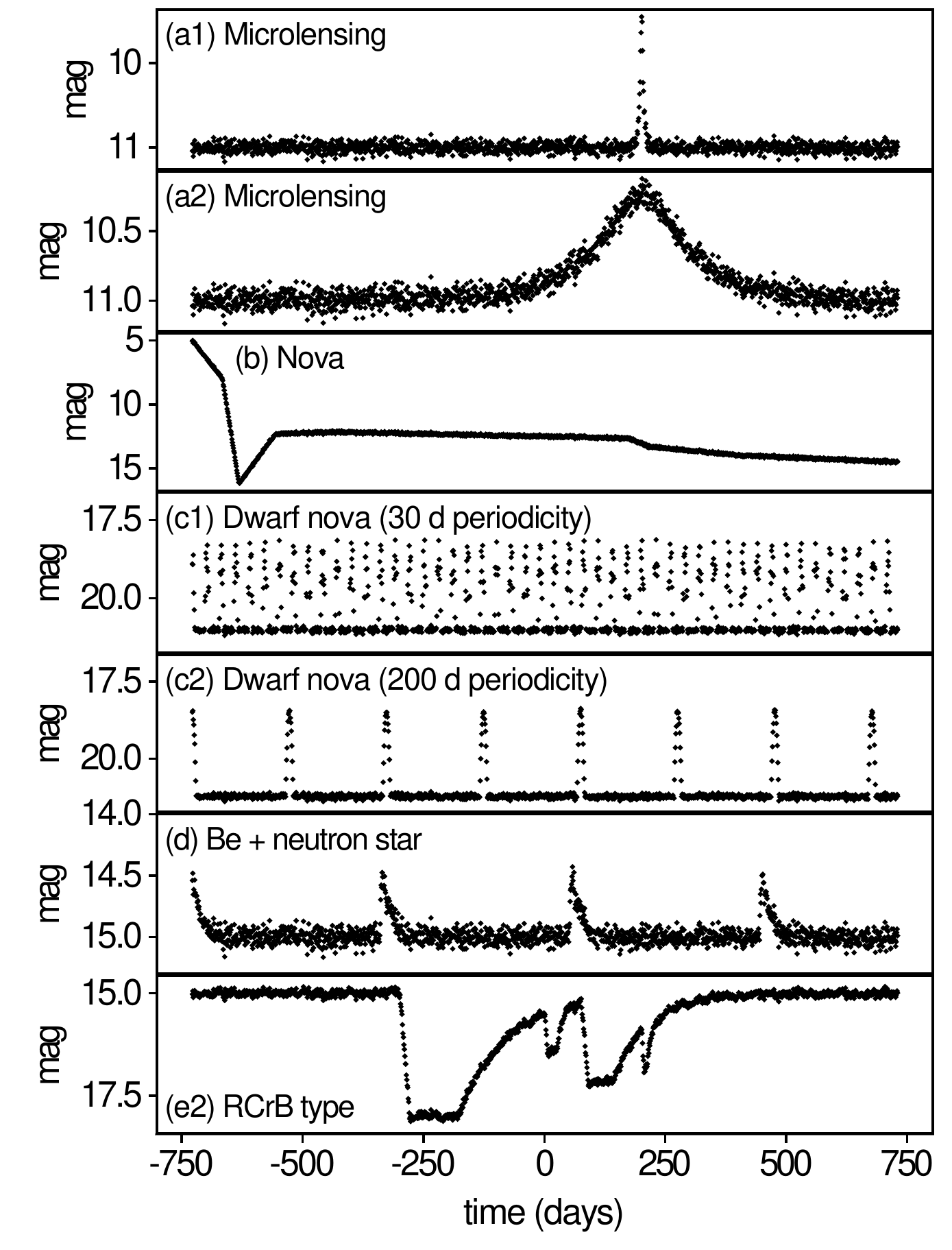}
  \caption{
  Simulations of various transient-type time series sampled every 1~day and with a Gaussian-distributed noise of 50~mmag (see text for more details).
  From top to bottom:
  \textbf{(a1)} microlensing with an Einstein radius crossing time of 7.5~d, based on OGLE source \texttt{SC4-267762};
  \textbf{(a2)} microlensing with an Einstein radius crossing time of 153.5~d, based on OGLE source \texttt{SC33-553617};
  \textbf{(b)} nova event, based on V705~Cas;
  \textbf{(c1)} and \textbf{(c2)} dwarf nova events, based on sources \texttt{OGLE-GD-DN-019} and \texttt{OGLE-GD-DN-022}, respectively;
  \textbf{(d)} Be+neutron star system displaying recurrent bursts, based on source \texttt{AX~J0049.4-7323};
  \textbf{(e2)} RCrB-type variability with intrinsic pulsations.
  }
\label{Fig:simuTransientTs}
\end{figure}

In this section, I investigate the locations of several types of transients in the $\diagram$ diagram.
Microlensing events are studied in Sect.~\ref{Sect:simuTransientsMicrolensing}, nova light curves in Sect.~\ref{Sect:simuTransientsNovae}, dwarf nova light curves and light curves with bursts in Sect.~\ref{Sect:simuTransientsDwarfNovae}, and light curves of R Coronae Borealis (RCrB) stars in Sect.~\ref{Sect:simuTransientsRCrB}.

\subsubsection{Microlensing events}
\label{Sect:simuTransientsMicrolensing}

\begin{figure}
  \centering
  \includegraphics[width=\columnwidth]{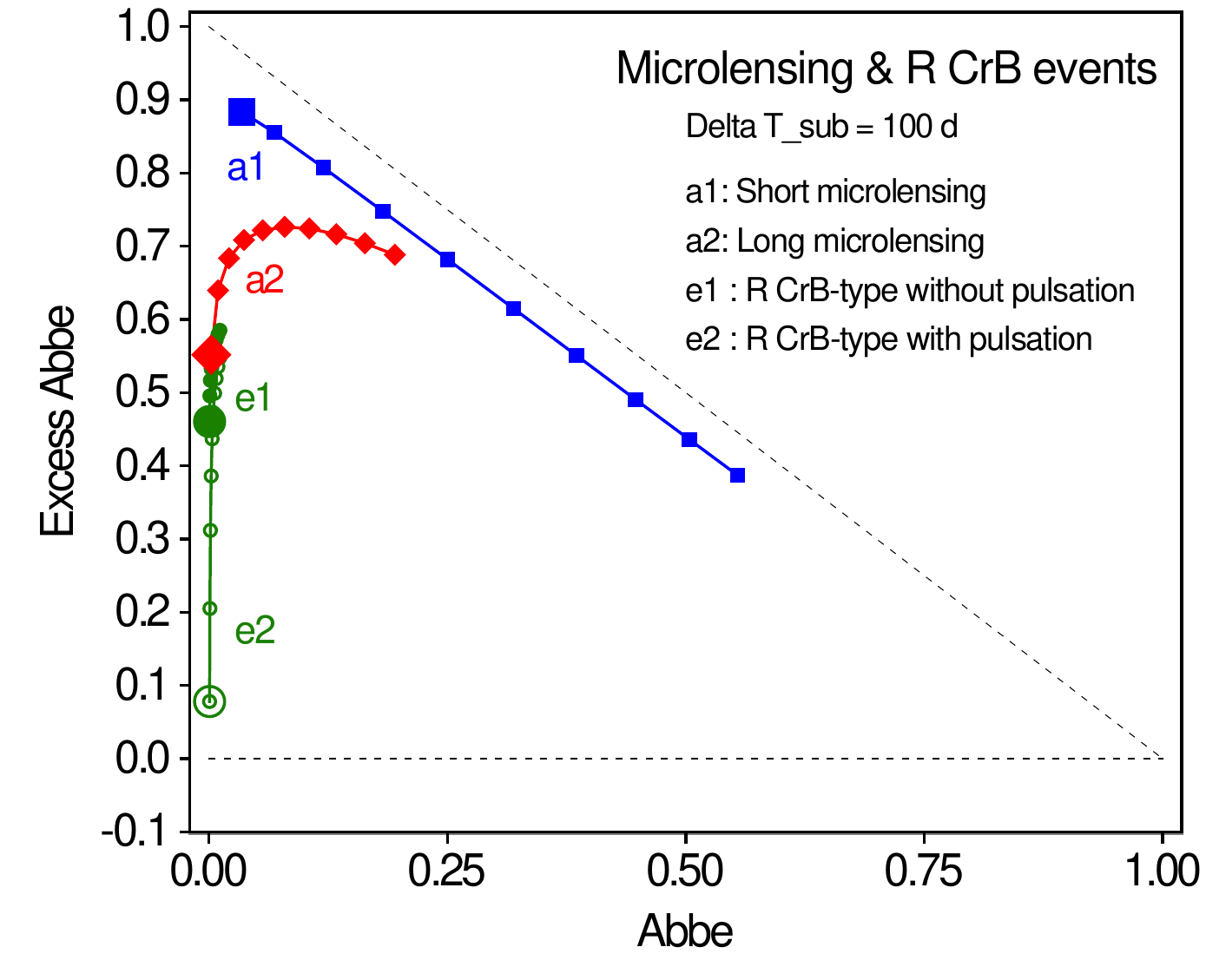}
  \caption{
  Locations in the $\diagram$ diagram of two simulated microlensing events (sequences a1 and a2) and two simulated RCrB events (sequences e1 and e2).
  \textbf{(a1)} Blue squares: OGLE source \texttt{SC4-267762} with $t_{\mathrm{E}}\!=\!7.5\mathrm{~d}$.
  \textbf{(a2)} Red triangles: OGLE source \texttt{SC33-553617} with $t_{\mathrm{E}}\!=\!153.5\mathrm{~d}$.
  \textbf{(e1)} Green filled circles: RCrB with four dip events.
  \textbf{(e2)} Green open circles: same as sequence e1, but with a two-frequency pulsation added to the light curves.
  See text for more details.
  The parameters of the time series are the same as in Fig.~\ref{Fig:simuTransientTs}, except for the noise level which ranges from $\sigma_\mathrm{noise}$\,=\,10 to 100~mmag by steps of 10~mmag (markers on each line, the bigger markers indicating the simulations with the smallest noise level of $\sigma_\mathrm{noise}$\,=\,10~mmag).
  All excess Abbe values are computed with $\Delta T_\mathrm{sub}=100$~d.
  The horizontal dashed line is an guideline to $\excessAb=0$ and the diagonal dashed line an guideline to $\excessAb=1-\Ab$.
  }
\label{Fig:simuMicrolensingDiagram}
\end{figure}

Two microlensing events are simulated with a Paczynski light curve of the form \citep[][\citeyear{Paczynski96}]{Paczynski86b} 
\begin{equation}
  mag(t) = -2.5 \log_{10} \left[ 1 + f_{\mathrm{S}} \times \left( \frac{u^2 +2}{u \sqrt{u^2+4}}-1 \right) \right] \,,
\label{Eq:microlensing}
\end{equation}
where
\begin{equation}
  u^2 = u_{\mathrm{min}}^2 + \left(\frac{t-t_{\mathrm{max}}}{t_{\mathrm{E}}}\right)^2 \;\;, \nonumber
\end{equation}
$f_{\mathrm{S}}$ is the ratio between the flux of the source (lensed object) and the total flux of the system (lensed + lensing objects), both fluxes being considered outside the microlensing event,
$u_{\mathrm{min}}$ the impact parameter (relative to the Einstein radius),
$t_{\mathrm{E}}$ the Einstein radius crossing time, and $t_\mathrm{max}$ the time at maximum amplification.
The parameters adopted for the simulated light curves are taken from two OGLE-II microlensing events identified in \citet[Table~4]{SumiWozniakUdalski_etal06}.
The first source is \texttt{SC4-267762} with a short $t_{\mathrm{E}}\!=\!7.5\mathrm{~d}$ (and $u_{\mathrm{min}}\!=\!0.214, f_{\mathrm{S}}\!=\!0.84$).
The second source is \texttt{SC33-553617} with a long $t_{\mathrm{E}}\!=\!153.5\mathrm{~d}$ (and $u_{\mathrm{min}}\!=\!0.452, f_{\mathrm{S}}\!=\!0.75$).
The time of maximum amplification is (arbitrarily) put at $t_\mathrm{max}$\,=\,200~d in the simulations.

Ten time series are constructed on this basis for each of the two simulated events, with Gaussian-distributed noise of $\sigma_\mathrm{noise}$\,=\, 10 to 100~mmag by steps of 10~mmag.
The time series simulated with 50~mmag noise level are shown in the top two panels of Fig.~\ref{Fig:simuTransientTs}.

The locations in the $\diagram$ diagram of the simulated time series are shown in Fig.~\ref{Fig:simuMicrolensingDiagram} by the markers connected with solid lines labeled "a1" and "a2" for the short and long Einstein radius crossing times, respectively.
They populate the upper-left region of the diagram, distinct from the region populated by pulsating-like light curves.

\subsubsection{Nova events}
\label{Sect:simuTransientsNovae}

A nova eruption leads to the brightening of a faint star by 8 to 20 magnitudes within a few days.
The source then fades by a few magnitudes on a timescale of weeks to months, then more slowly back to the quiescent level.
The light curve geometries are diverse, and can further be altered by dust, oscillations, and/or flares \citep[e.g.,][]{StropeSchaeferHenden10}.

The location of novae in the $\diagram$ diagram is analyzed with a light curve model of V705~Cas.
A simplified model is constructed based on the binned and averaged AAVSO light curve published by \cite{StropeSchaeferHenden10} in their Fig.~1.
The light curve is divided into eight time segments, between times 0.00, 62.50, 96.67, 171.19, 289.25, 904.57, 945.01, 1129.43, and 1400 days after outburst peak.
Straight line fits are adopted in each time segment with ($slope, intercept$) values equal to (0.048, 5.0), (0.24, -7.000), (-0.051892, 21.216), (-0.001561, 12.6), (0.000793, 11.919), (0.016134, -1.958), (0.003692, 9.8), and (0.001567, 12.2), respectively, in units of (mag/d, mag).
Ten time series are simulated on this basis, with Gaussian-distributed noise of $\sigma_\mathrm{noise}$\,=\, 10 to 100~mmag by steps of 10~mmag.
The time series simulated with 50~mmag noise level is shown in Fig.~\ref{Fig:simuTransientTs}, third panel from top.

The locations in the $\diagram$ diagram of the simulated time series are shown in Fig.~\ref{Fig:simuBurstsDiagram} by the blue markers connected with solid line labeled "b".
The Abbe value is close to $\Ab=0$ for all the simulated time series and due to the very large amplitude of variability compared to the noise levels.
The excess Abbe value, on the other hand, ranges from $\sim$0.1 for the time series with the smallest simulated noise level to $\sim$0.7 for the one with the largest simulated noise level.
The overall pattern of the light curve actually resembles that of a light curve with a strong trend and small noise (compare sequence "b" in Fig.~\ref{Fig:simuBurstsDiagram} with the low-$\Ab$ part of sequence $\Ab_f\!\!=\!\!1$ in Fig.~\ref{Fig:simuTrendDiagram}).
Nova light curves affected by dust, oscillations, and/or flares would also be located in the left part of the $\diagram$ diagram.

\subsubsection{Dwarf novae and multiple-burst events}
\label{Sect:simuTransientsDwarfNovae}

\begin{figure}
  \centering
  \includegraphics[width=\columnwidth]{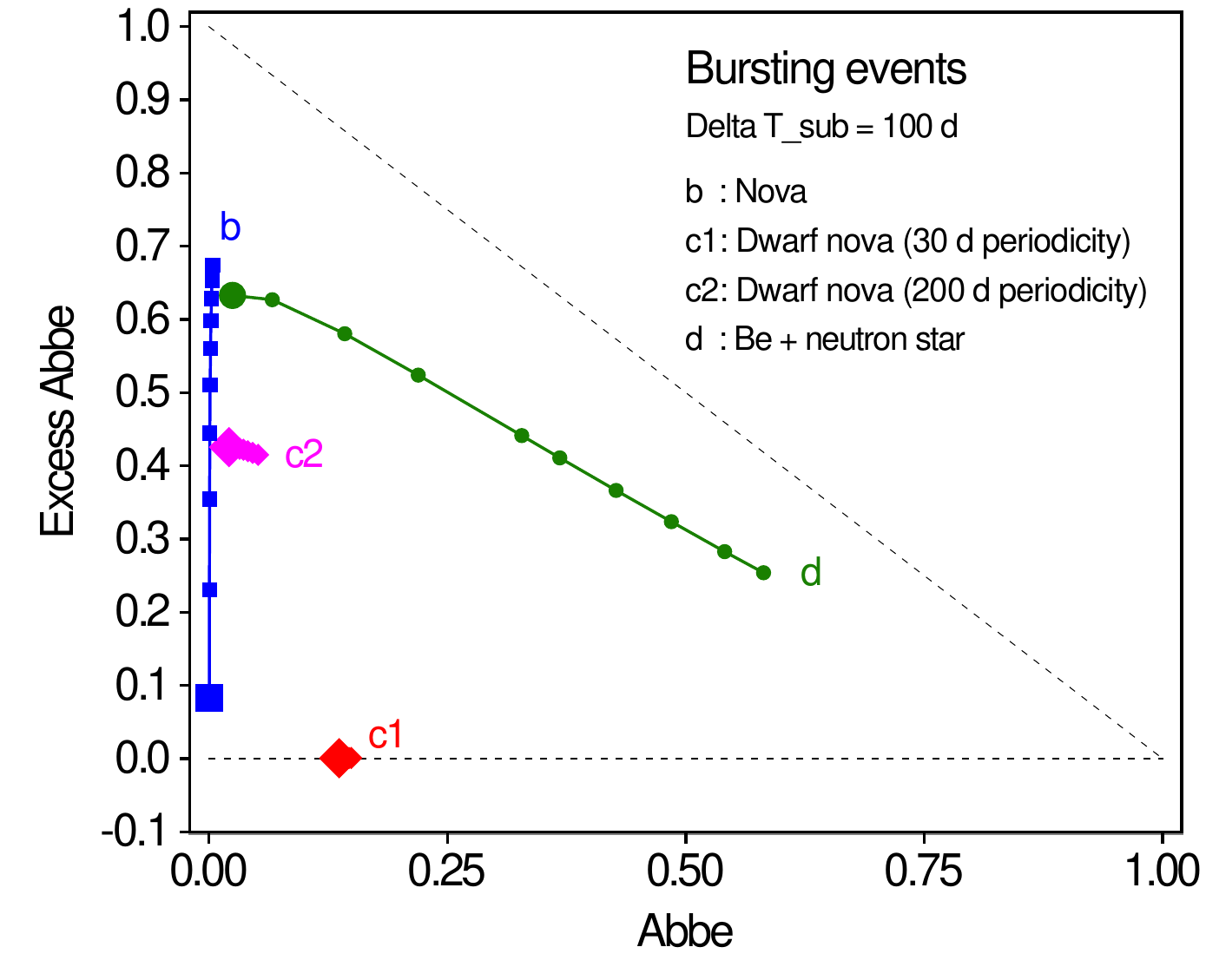}
  \caption{Same as Fig.~\ref{Fig:simuMicrolensingDiagram}, but for simulations of 
  \textbf{(b)} blue squares: a nova;
  \textbf{(c1)} red diamonds:  a dwarf nova with 30~d periodicity of the bursts;
  \textbf{(c2)} magenta diamonds:  same as sequence c1, but with bursts every 200 days;
  \textbf{(d)} green circles:  a Be+neutron star system.
  See text for more details.
  }
\label{Fig:simuBurstsDiagram}
\end{figure}

Dwarf novae are characterized by recurrent bursts, with typical amplitudes between 1.5~mag and 2.5~mag, and a variety of burst and quiescent durations \citep[e.g.,][]{MrozPietrukowiczPoleski13}.
Oscillations are detected in the light curves of a fraction of them as a result of the orbital motion of the binary, with periods of less than a day and typical peak-to-peak amplitudes of 0.5~mag.

The location of the light curves in the $\diagram$ diagram depends on the burst recurrence time $P_\mathrm{recur}$ relative to the $\Delta T_\mathrm{sub}$ parameter used in the computation of $\excessAb$.
In particular, $\excessAb$ should be close to zero if $P_\mathrm{recur}\lesssim\Delta T_\mathrm{sub}$, and positive otherwise.
To test this, I produce two sets of simulated light curves, one with $P_\mathrm{recur}<\Delta T_\mathrm{sub}$ (which is 100~d), and one with $P_\mathrm{recur}>\Delta T_\mathrm{sub}$.

The first set is based on the characteristics of the dwarf nova \texttt{OGLE-GD-DN-019} listed in \cite{MrozPietrukowiczPoleski13}.
From their Figs.~4 and 6 and Table~2, I adopt $P_\mathrm{recur}=30$~d, a burst duration $\Delta t_\mathrm{burst}=10$~d, a burst amplitude $A_\mathrm{burst}=2.39$~mag, an oscillation peak-to-peak amplitude $A_\mathrm{osc}=1$~mag and an oscillation period $P_\mathrm{osc}=0.427$~d.
The bursts are modeled with a parabola, and a series of ten simulated light curves are produced with various noise levels between 10 and 100~mmag.
The light curve with 50~mmag noise level is displayed in Fig.~\ref{Fig:simuTransientTs}, fourth panel from top.

The locations in the $\diagram$ diagram of the light curves of this first set, with $P_\mathrm{recur}=30$~d, are shown in Fig.~\ref{Fig:simuBurstsDiagram}, sequence labeled "c1".
They are actually all mingled at $\Ab\simeq 0.15$ and $\excessAb\simeq 0.001$.
The excess Abbe is almost null thanks to the pulsating-like character of the light curve in 100~d subintervals used for the computation of $\excessAb$.

The second set of simulated dwarf nova light curves is based on the characteristics of source \texttt{OGLE-GD-DN-022} listed in \cite{MrozPietrukowiczPoleski13}, adopting $P_\mathrm{recur}=200$~d (the real value is not known from observations).
The other parameters are $\Delta t_\mathrm{burst}=15.5$~d, $A_\mathrm{burst}=2.78$~mag, and no oscillation due to orbital motion.
The light curve simulated with 50~mmag noise level is displayed in Fig.~\ref{Fig:simuTransientTs}, fifth panel from top.

The locations in the $\diagram$ diagram of the second set of light curves, with $P_\mathrm{recur}=200$~d, are shown by the sequence labeled "c2" in Fig.~\ref{Fig:simuBurstsDiagram}.
They are mingled at $\excessAb$ values between 0.40 and 0.45, higher than those of the first set, because the inter-burst intervals are longer than $\Delta T_\mathrm{sub}$, and during those intervals the source behaves like a constant star.
The relatively small impact of photometric noise level on the location in the $\diagram$ diagram is due to the large dwarf novae burst amplitudes (2-3~mag) relative to the noise amplitudes considered in the simulations (0.01-0.1~mag).

Finally, in the category of recurrent bursting stars, I consider the case of a binary system composed of a Be and a neutron star, that undergoes periodic bursts as the neutron star crosses the disk of the companion every orbital revolution.
I take the example of \texttt{AX J0049.4-7323} in the Small Magellanic Cloud, the light curve of which presents bursts every 393.1 days with an amplitude of $\sim$0.5~mag \citep[][Fig.~29]{SchmidtkeCowleyUdalski13}.
I simulate the bursts with a simple model of linear flux increase over 6 days followed by an exponential flux decrease with an $e$-folding timescale of 10 days.
A set of ten simulated light curves is produced using noise levels between 10~mmag and 100~mmag, the one with $\sigma_\mathrm{noise}=50$~mmag is shown in Fig.~\ref{Fig:simuTransientTs}, sixth panel from top.

The light curves of the Be+neutron star system occupy the upper-left part of the $\diagram$ diagram (green circles in Fig.~\ref{Fig:simuBurstsDiagram}) for noise levels less than 50~mmag, and moves to higher $\Ab$ values for larger noise levels.
Light curves with bursts weaker than those simulated for dwarf novae thus populate regions in the diagram covering a wide range of $\Ab$ values, depending on the noise level (larger $\Ab$ values for noisier time series), but at relatively high $\excessAb$ values depending on the burst recurrence timescale relative to $\Delta T_\mathrm{sub}$ (larger $\excessAb$ values for larger recurrence times).

In summary, light curves with bursting events are predicted to populate the left part of the $\diagram$ diagram with $\Ab\lesssim 0.5$ and $\excessAb\gtrsim 0.2$, unless the bursts have a recurrence timescale shorter than $\Delta T_\mathrm{sub}$ (sequence "c1" in Fig.~\ref{Fig:simuBurstsDiagram}) or are drowned in photometric noise (simulations with the largest noise levels in sequence "d").

\subsubsection{R Coronae Borealis light curves}
\label{Sect:simuTransientsRCrB}

R Coronae Borealis stars have light curves characterized by extinctions due to dust, occurring at unpredictable times and resulting in fast luminosity declines followed by slow recoveries;
low-amplitude periodic variations may also be present because of intrinsic pulsations \citep{TisserandMarquetteWood_etal08}.

The expected location of RCrB stars in the $\diagram$ diagram is investigated with two sets of simulated light curves, one without intrinsic pulsations and one with pulsations.
In the first set of light curves four dip events are simulated, each dip characterized by three phases:
a fast linear fading phase starting at time $t_\mathrm{start}$ and lasting $\Delta t_\mathrm{fading}$;
a plateau of duration $\Delta t_\mathrm{dip}$ and depth $\Delta m_\mathrm{dip}$ relative to the mean base magnitude;
and an exponential recovery phase characterized by a timescale $\tau_\mathrm{recovery}$.
The values adopted for the four dip events in the simulations are
$t_\mathrm{start} = -300$, 0, 75, and 200~d;
$\Delta t_\mathrm{fading} = 20$, 5, 15, and 5~d;
$\Delta t_\mathrm{dip} = 100$, 20, 50, and 5~d;
$\Delta m_\mathrm{dip} = 3$, 1, 2, and 1~mag;
and  $\tau_\mathrm{recovery} = 100$, 10, 70, and 10~d.
A series of ten simulated light curves are produced with various noise levels between 10~mmag and 100~mmag.

The locations in the $\diagram$ diagram of this first set of simulations are shown in Fig.~\ref{Fig:simuMicrolensingDiagram}.
The $\excessAb$ is moderately high ($0.46<\excessAb<0.59$), and the Abbe values small ($\Ab<0.01$) in all cases.
This positions the simulated light curves in the region of the diagram occupied by the other transient simulations studied in the previous sections.

The second set of simulations is constructed by adding to the first set an intrinsic pulsation composed of two frequencies, reminiscent of the variability of semi-regular variable stars.
The pulsation is simulated with two periods, of 30 and 85~days, with amplitudes of 0.06 and 0.04~mag, respectively.
The light curve with 50~mmag noise level is displayed in Fig.~\ref{Fig:simuTransientTs}, seventh panel from top.

The locations in the $\diagram$ diagram of the second set of simulations are shown in Fig.~\ref{Fig:simuMicrolensingDiagram}, sequence "e2" (open green circles) .
They have $\Ab$ values similar to the ones of the first set without pulsation, but $\excessAb$ now reaches values as low as 0.08 for $\sigma_\mathrm{err}=0.01$~mag.
Such a small value for $\excessAb$ occurs only with the lowest simulated noise level, $\excessAb$ increasing above 0.2 for $\sigma_\mathrm{err}\geqslant 0.02$~mag, and up to 0.55 for $\sigma_\mathrm{err}=0.1$~mag.

In summary, RCrB-type light curves will be located in the left part of the $\diagram$ diagram.
The $\excessAb$ value will be small if a (multi)periodic signal is present with period(s) smaller than $\Delta T_\mathrm{sub}$.
The same conclusions will hold true for Be-type light curves that display outbursts characterized by fast luminosity increases followed by slow declines.
In reality, however, light curves of RCrB and Be stars are much more complex than the simple simulations presented here, and wider regions of the $\diagram$ diagram are expected to be populated (see Sects~\ref{Sect:EROS} and following).

\subsection{Summary}
\label{Sect:diagramSummary}

\begin{figure}
  \centering
  \includegraphics[width=\columnwidth]{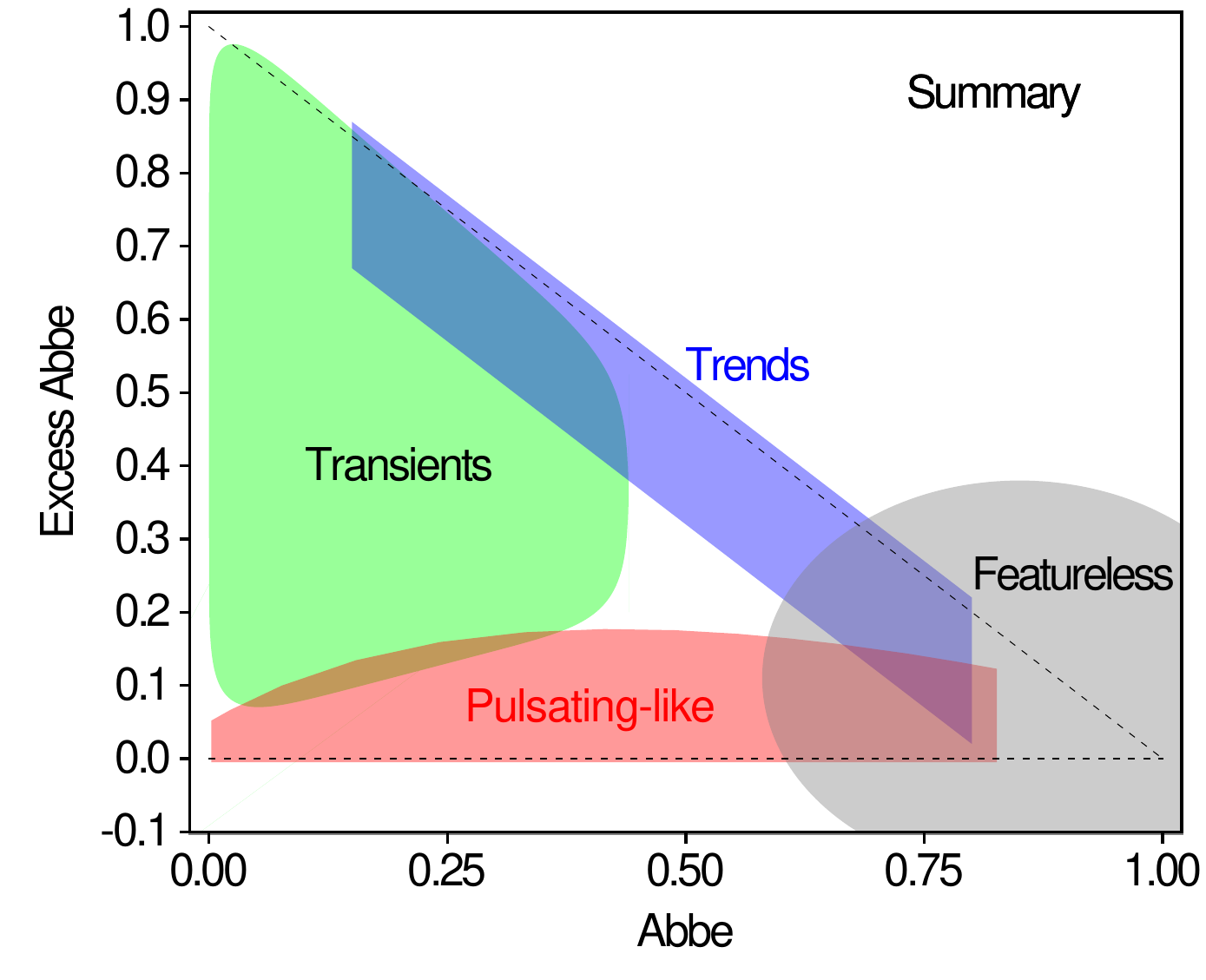}
  \caption{Schematic location of different types of time series expected in the $\diagram$ diagram.
  The lower red region locates time series of pulsating-like light curves, the upper blue diagonal region those with trends, the left green region those of transient candidates, and the right gray region those without any specific features, which includes constant stars.
  }
\label{Fig:summaryDiagram}
\end{figure}

The regions in the $\diagram$ diagram expected to be populated by light curves of pulsating stars, trends, and transients, according to the analysis of Sects~\ref{Sect:simuPeriodic} to \ref{Sect:simuTransients}, are schematized in Fig.~\ref{Fig:summaryDiagram}.
Transients lie in the left region of the diagram.
They are relatively well identified and separated from pulsating-like light curves (lower region of the diagram) and from light curves with trends (upper diagonal limit of the diagram), unless their transient feature has a recurrence timescale shorter than  $\deltaTSub$ (in which case they would have a low $\excessAb$ value; examples are some dwarf novae) or is reminiscent of a trend with a very high slope (in which case they would have $\Ab\sim 0$ and low $\excessAb$ values; examples are novae).

In the rest of this paper, I consider a source to be a transient candidate if it falls in the region $\Ab \leqslant 0.5$ and $\excessAb \geqslant 0.2$.

Time series with trends can contaminate the transients region, if their Abbe values are smaller than 0.5.
Such a contamination is, however, not problematic because these time series would need to be characterized by a rather large trend amplitude (see Sect.~\ref{Sect:simuTrends}).
If such trends are of astrophysical origin, they are likely to characterize transient phenomena.

Pulsating-like light curves can also contaminate the transients region, if the sources show pulsating-like variations on timescales larger than $\Delta T_\mathrm{sub}$.
Simulated examples of sine functions presented in Sect.~\ref{Sect:simuPeriodic} show that $\excessAb \geqslant 0.2$ can be reached for periods larger than $2\,\Delta T_\mathrm{sub}$.
Those time series may therefore be (incorrectly) considered as transient candidates.
This behavior is expected, since the cyclic nature of the variability cannot be established on a $\Delta T_\mathrm{sub}$ time interval smaller than the actual periodicity timescale, as explained in Sect.~\ref{Sect:simuPeriodic}, although it is unfortunate for the identification of transient candidates.
This feature must thus be kept in mind when choosing the value of $\deltaTSub$ and identifying transient candidates in the $\diagram$ diagram.
Real case examples are discussed in the next sections.

The importance of the choice of $\deltaTSub$ for the computation of the excess Abbe, stressed in Sect.~\ref{Sect:method}, is thus confirmed by the simulations presented in this section.
The larger $\deltaTSub$, the smaller the pollution of the transient region by long-timescale pulsating-like sources;
however the excess Abbe values of transients also decrease for increasing $\deltaTSub$.
There is thus a compromise to make between small $\deltaTSub$ to identify transients with shorter variability timescales and large $\deltaTSub$ to avoid pollution by long timescale pulsating-like stars.

\paragraph{The $\diagram$ diagram as a data quality diagnostic tool.}
The above summary is based on the assumption that the variabilities present in the light curves are of astrophysical origin.
This may not be true.
As summarized in Sect.~\ref{Sect:simuTrendDiagnosticTool}, trends resulting from data reduction problems or due to instrument deficiencies (drifts with time due to instrument degradation, for example, that would not have been properly corrected in the data) will lead to an abnormally high density of points on the upper diagonal limit of the $\diagram$ diagram, if all sources of a given survey (or at least a large sample thereof) are plotted at the same time in the diagram.
This makes the $\diagram$ diagram a potentially powerful diagnostic tool to check the data quality of a given survey, as is illustrated in the next section on the EROS-2 database.

\section{EROS sample of the LMC}
\label{Sect:EROS}

In this section and the next two, I check the efficiency of the excess Abbe method to extract transients on a small test region of the LMC.
I apply the method to a subset of the EROS-2 survey (which I will simply call EROS in the rest of this document).
The subset is (arbitrarily) chosen to be the EROS field \field{lm0013m}, spanning right ascensions between 5.41570~hours and 5.44902~hours, and declinations between -69.58063~degrees and -69.41011~degrees (J2000 coordinates).
There are 1309 sources catalogued as variables by the EROS team in this field.
Each source has two light curves, simultaneously recorded in the red ($\EROSR$) and blue ($\EROSB$) wide pass-bands of the EROS instruments \citep{Palanque-DelabrouilleAfonsoAlfert_etal98}.
In the following, I restrict the study to stars brighter than 19.5~mag in each band, the number of which amounts to 1282 and 1244 in the $\EROSR$ and $\EROSB$ bands, respectively.

The light curves in each band are cleaned in several steps to remove bad points and outliers.
All points with uncertainties either larger than 0.5 mag or larger than 3~$\sigma$ above the mean uncertainty are discarded.
Outliers in magnitude are then removed with a sigma-clipping procedure, iteratively applied three times.
A point is considered to be an outlier if its magnitude lies at more than 5~$\sigma$ from the mean magnitude, either at the faint or at the bright end, and if there are no more than three consecutive points satisfying this condition.
The light curve cleaning procedure leads to the removal of a mean of 24.1 (15.5) measurements per star in the $\EROSR$ ($\EROSB$) band, the number of points removed by the magnitude sigma-clipping procedure being 0.4 (0.6), on average.
The resulting mean number of good points per light curve is 524 (568) in the $\EROSR$ ($\EROSB$) band.
I further restrict the study to light curves that contain at least 100~good measurements.
The number of such light curves amounts to 1279 in $\EROSR$ and 1244 in $\EROSB$.

\subsection
{The $\diagram$ diagrams of the EROS sample}
\label{Sect:ErosDiagram}

I adopt $\deltaTSub=100$~d in all computations of $\excessAb$, i.e., the same value as the one used in the simulations presented in Sect.~\ref{Sect:simulations}.
We therefore have to keep in mind that the region of transient candidates in the $\diagram$ diagram can be polluted by pulsating-like light curves with timescales larger than 100~d, while it can miss transients with timescales shorter than 100~d.

\subsubsection{Full light curves}
\label{Sect:ErosDiagramFullLcs}

\begin{figure}
  \centering
  \includegraphics[width=\columnwidth]{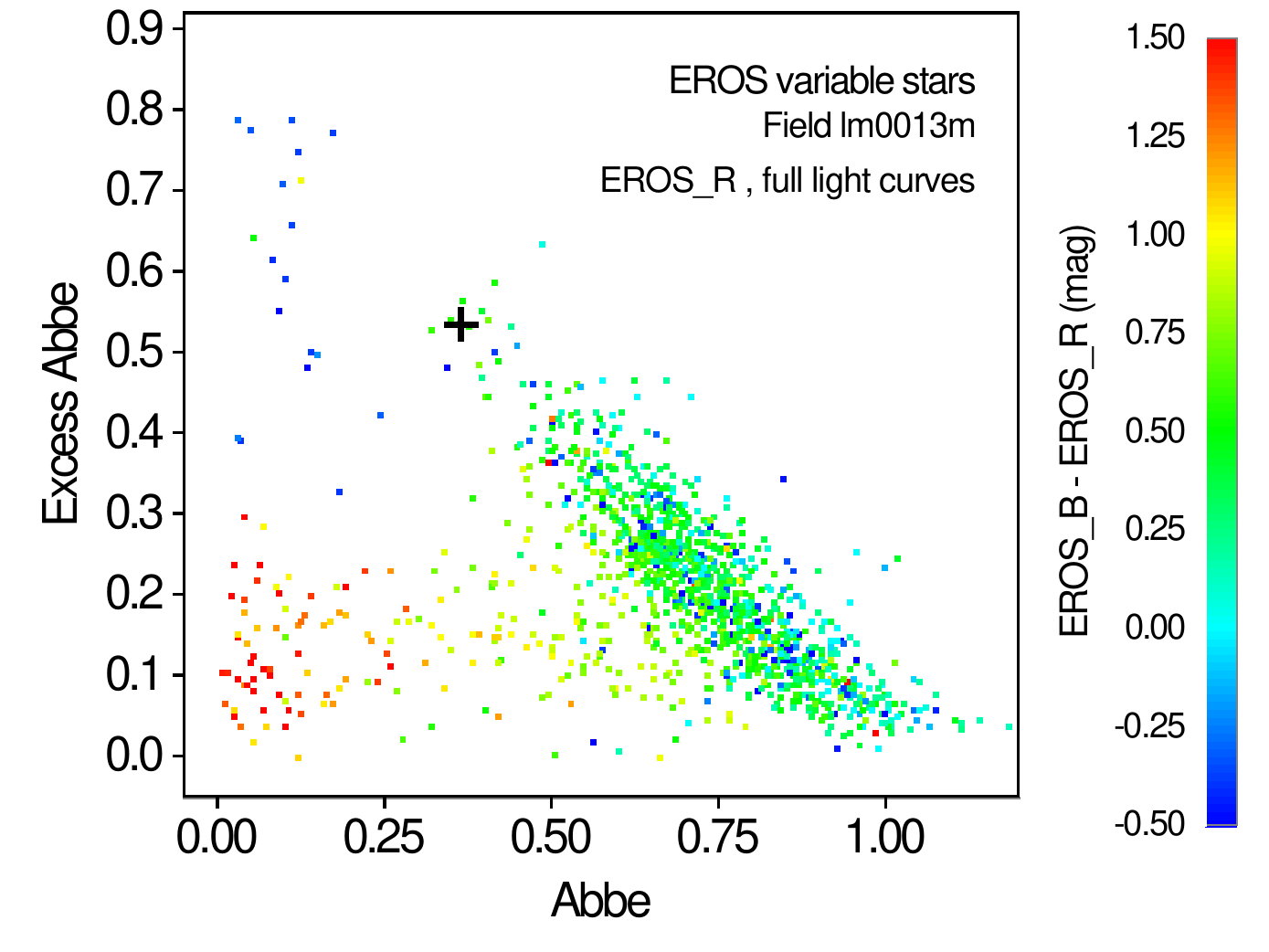}
  \caption{$\diagram$ diagram for EROS variable stars in the EROS field \field{lm0013m}.
           The color scale of each point is proportional to the $\EROSB$ - $\EROSR$ color as shown on the color bar on the right, with all color values above 1.5~mag set to red and all those below -0.5~mag set to blue.
   The plus sign marks the position of star 10578, the light curves of which are displayed in Fig.~\ref{Fig:ErosLcExampleStep}.
  }
\label{Fig:ErosDiagrams}
\end{figure}

\begin{figure}
  \centering
  \includegraphics[width=\columnwidth]{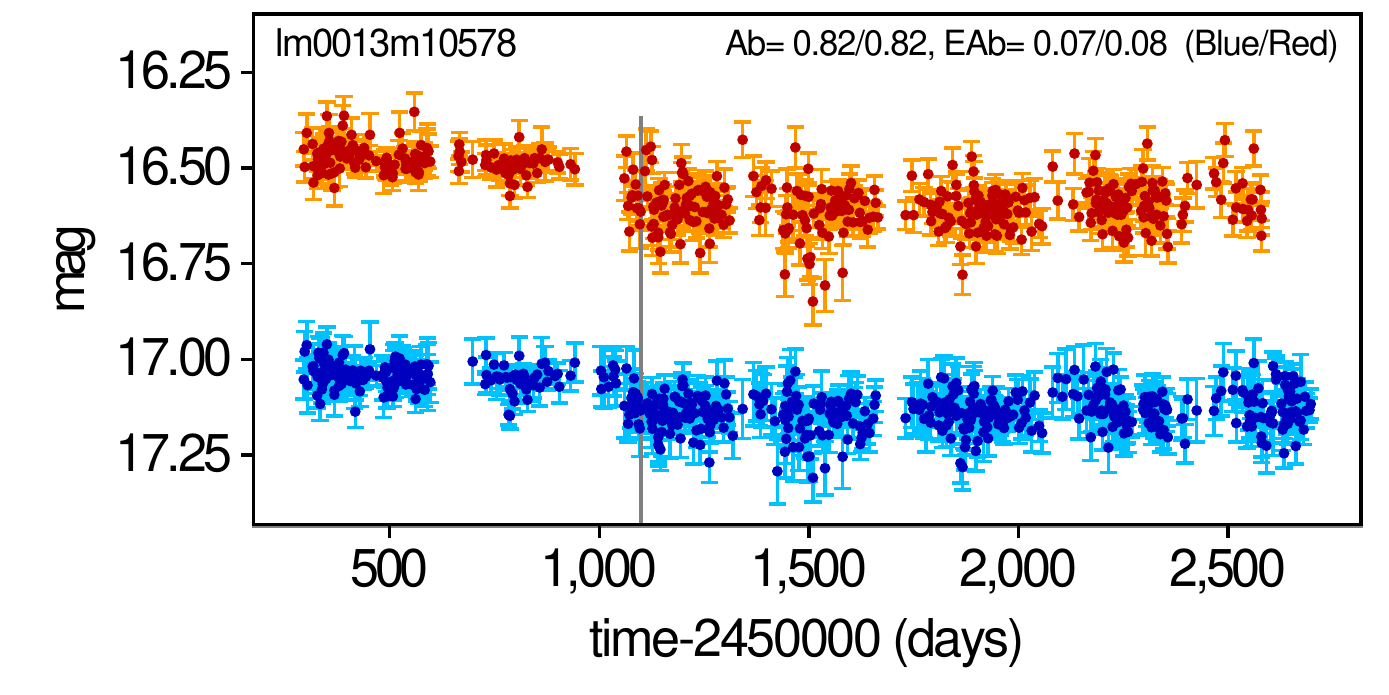}
  \caption{$\EROSR$ (upper red curve) and $\EROSB$ (lower blue curve) light curves of star 10578 highlighted in Figs.~\ref{Fig:ErosDiagrams} and \ref{Fig:ErosDiagramsPartialLcs}.
  Uncertainty bars are drawn in orange for $\EROSR$ measurements and in cyan for $\EROSB$.
  A vertical line is set at time 1100~d that defines the start of the partial light curves mentioned in Sect.~\ref{Sect:partialLcsEROS}.
  The values of $\Ab$ and $\excessAb$ for the $\EROSR$ and $\EROSR$ partial light curves are given in the figure (Ab and EAb, respectively).
   }
\label{Fig:ErosLcExampleStep}
\end{figure}

The $\diagram$ diagram of the EROS sample of variable stars is shown in Fig.~\ref{Fig:ErosDiagrams} for the $\EROSR$ band.
The four regions highlighted in Fig.~\ref{Fig:summaryDiagram} are visible in the diagram.
In particular, the region predicted to be populated by transient candidates is not very crowded, as expected from the known relative rarity of those phenomena in standard stellar populations.
The most populated region is the one of stars with featureless light curves (that I assimilate hereafter to constant stars) at Abbe values above 0.70, followed by the diagonal region at $0.35 \lesssim \Ab \lesssim 0.75$.
The third densest region is the one of pulsating-like stars at $\excessAb \lesssim 0.2$, where pulsating-like variable stars are to be found, among others.

While the high density of points in the region of constant stars and the relatively low density of points in the region of pulsating-like stars meet expectations, the high density of points in the diagonal region is suprising.
Indeed, we do not expect to find so many stars presenting a trend in their light curves.
A closer inspection of the time series reveals a step-like feature in many of them at times\footnote{All times are expressed in Julian days relative to the reference date JD$_0=2450000$~d.
}
around 1100~days.
An example is shown in Fig.~\ref{Fig:ErosLcExampleStep} for source 10578, the position of which in the $\diagram$ diagram is highlighted in Fig.~\ref{Fig:ErosDiagrams}.
The feature is not present in the light curves of all stars, but (if present) it seems to affect both $\EROSR$ and $\EROSB$ time series.
A search in the literature clarified the origin of this feature, attributed to a technical intervention on the cameras of the EROS telescope in May 1998 \citep[][p.~106]{Tisserand04}.
This step-like feature introduces an overall trend in the light curve, leading to the observed density of points in the region of trends in the $\diagram$ diagram.

The diagnostic potential of the $\diagram$ diagram is thus quite interesting; a quick look at the distribution of points in this diagram is able to highlight possible problems in the data.

\subsubsection{Partial light curves}
\label{Sect:partialLcsEROS}

\begin{figure}
  \centering
  \includegraphics[width=\columnwidth]{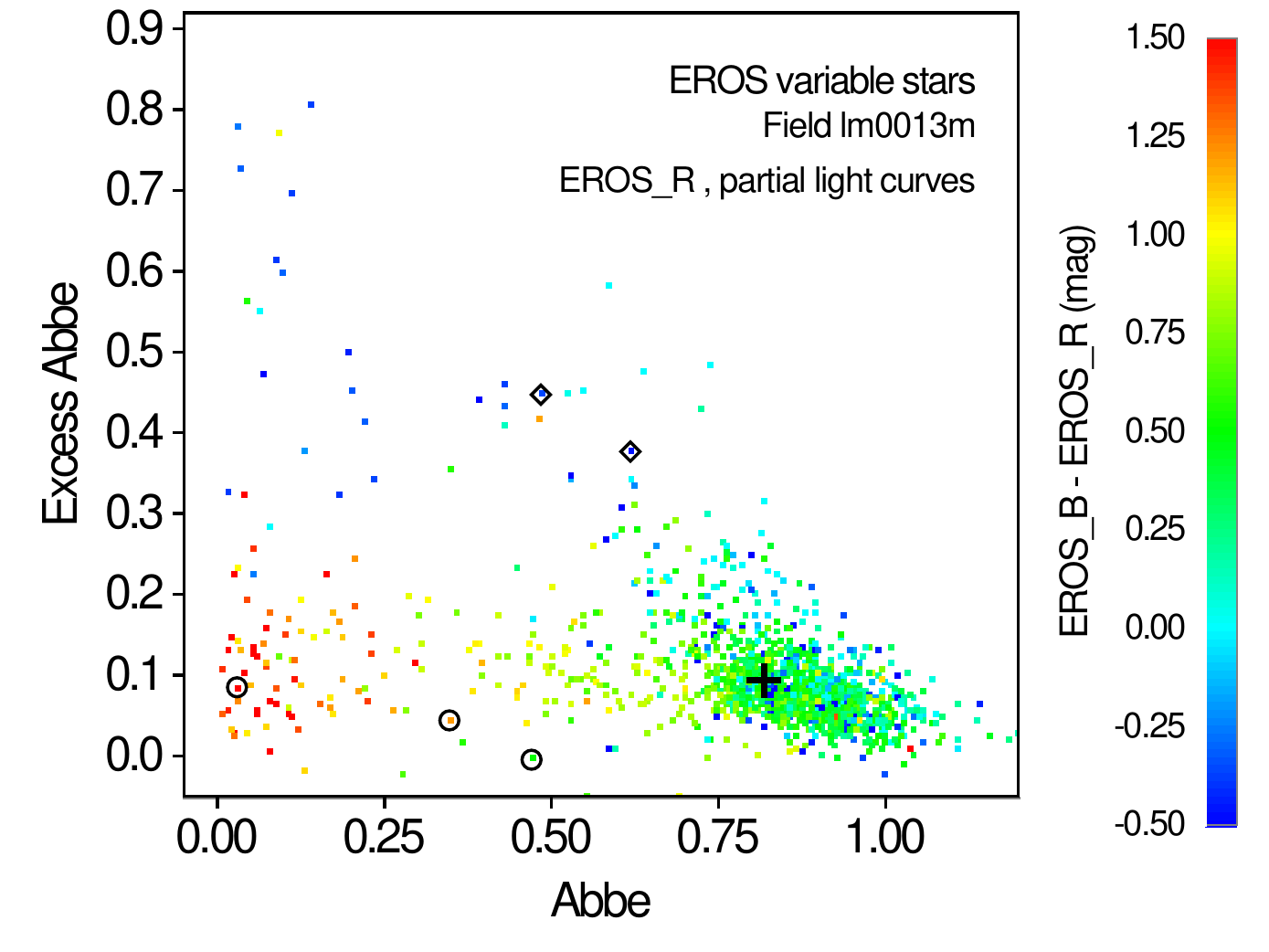}
  \includegraphics[width=\columnwidth]{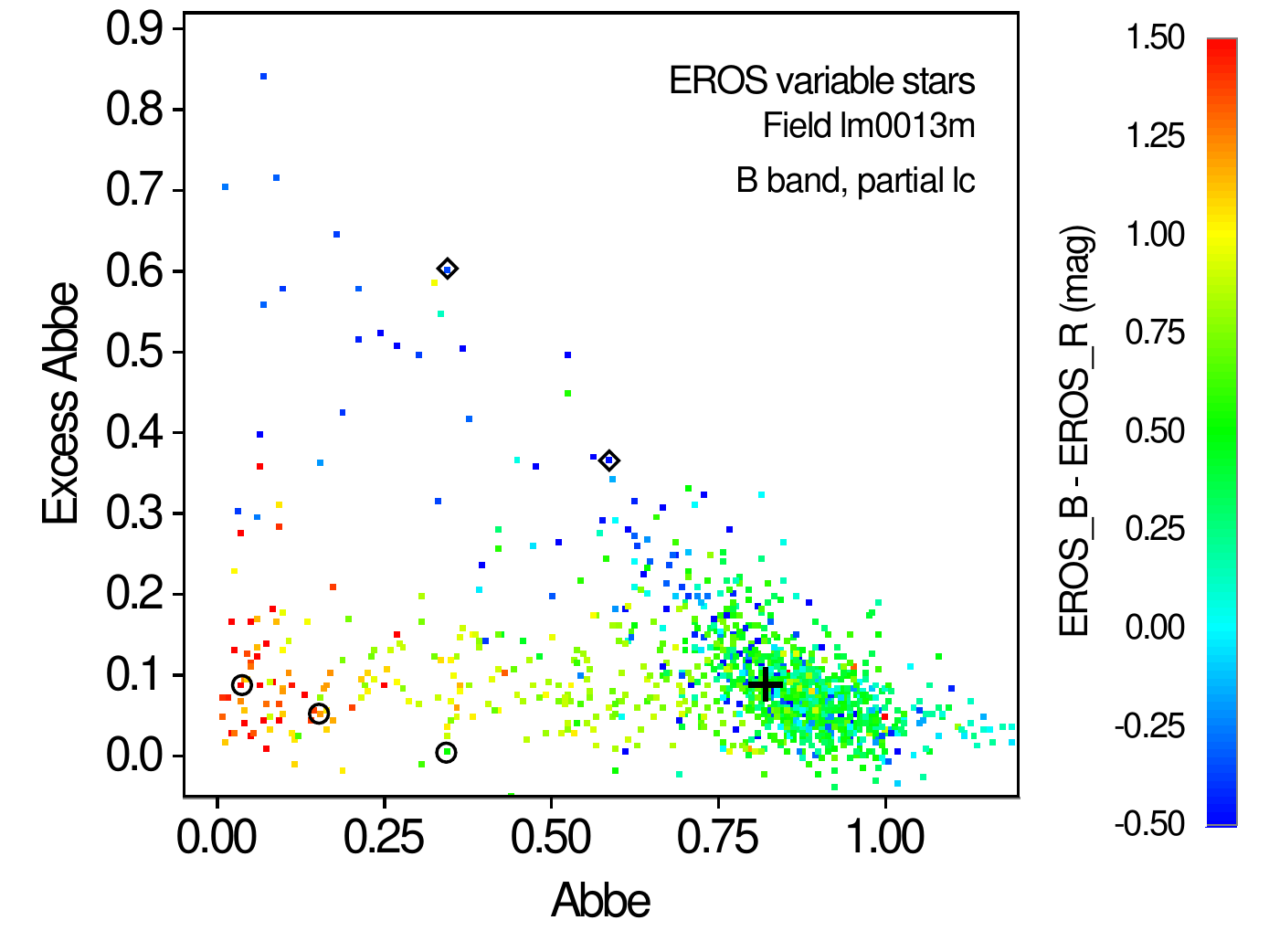}
  \caption{Same as Fig.~\ref{Fig:ErosDiagrams}, but with the Abbe values computed on the partial light curves (i.e., starting at 1100~d).
  The top panel is computed from $\EROSR$ time series, and the bottom panel from $\EROSB$ time series.
   Diamonds and open circles locate example stars with trends and pulsating-like light curves, respectively, the light curves of which are displayed in Figs.~\ref{Fig:ErosLcExamplesTrend} and \ref{Fig:ErosLcExamplesPulsatingLike}, respectively.
   The plus sign marks the position of star 10578, as in Fig.~\ref{Fig:ErosDiagrams}.
  }
\label{Fig:ErosDiagramsPartialLcs}
\end{figure}

\begin{figure}
  \centering
  \includegraphics[width=\columnwidth]{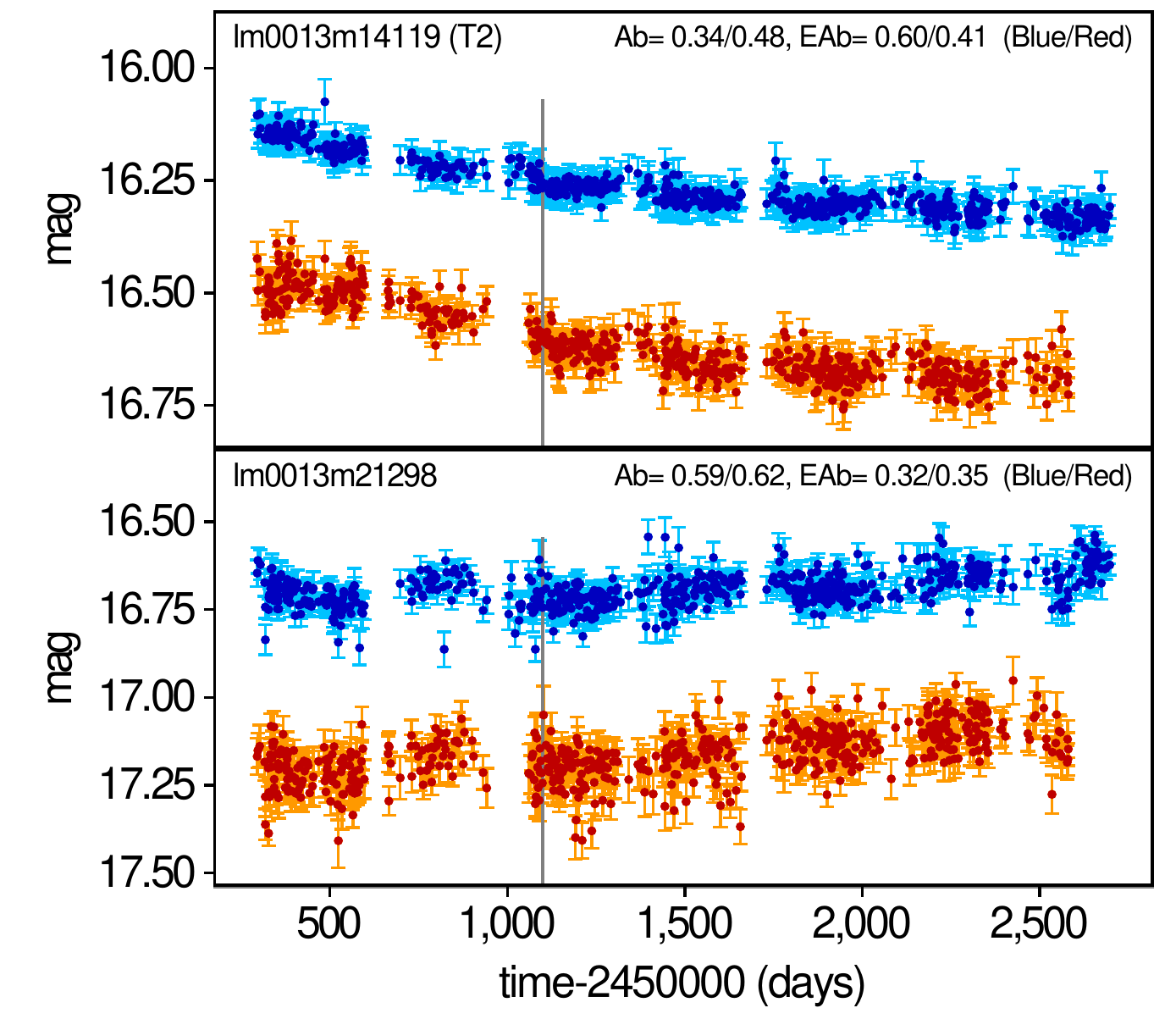}
  \caption{Same as Fig.~\ref{Fig:ErosLcExampleStep}, but for stars 14119 (upper panel) and 21298 (bottom panel) with trends, highlighted in Fig.~\ref{Fig:ErosDiagramsPartialLcs}.
   }
\label{Fig:ErosLcExamplesTrend}
\end{figure}

When data prior to day 1100 are ignored in the time series, the resulting $\EROSR$ and $\EROSB$ $\diagram$ diagrams, shown in Fig.~\ref{Fig:ErosDiagramsPartialLcs}, become much cleaner.
In particular, source 10578, which was in the trend region when using the full light curve (Fig.~\ref{Fig:ErosDiagrams}), is now moved to the region of constant stars where it should be.
The density of points in the region of trends is now much more realistic, and shows the power of the $\diagram$ diagram in assessing the quality of the data.
Two examples of light curves falling in the region of trends are shown in Fig.~\ref{Fig:ErosLcExamplesTrend}, with their representative points in the $\diagram$ diagram indicated in Fig.~\ref{Fig:ErosDiagramsPartialLcs}.
The first one (star 14119, top panel) displays a decreasing trend, and the second (star 21298, bottom panel) an increasing trend.

\begin{figure}
  \centering
  \includegraphics[width=\columnwidth]{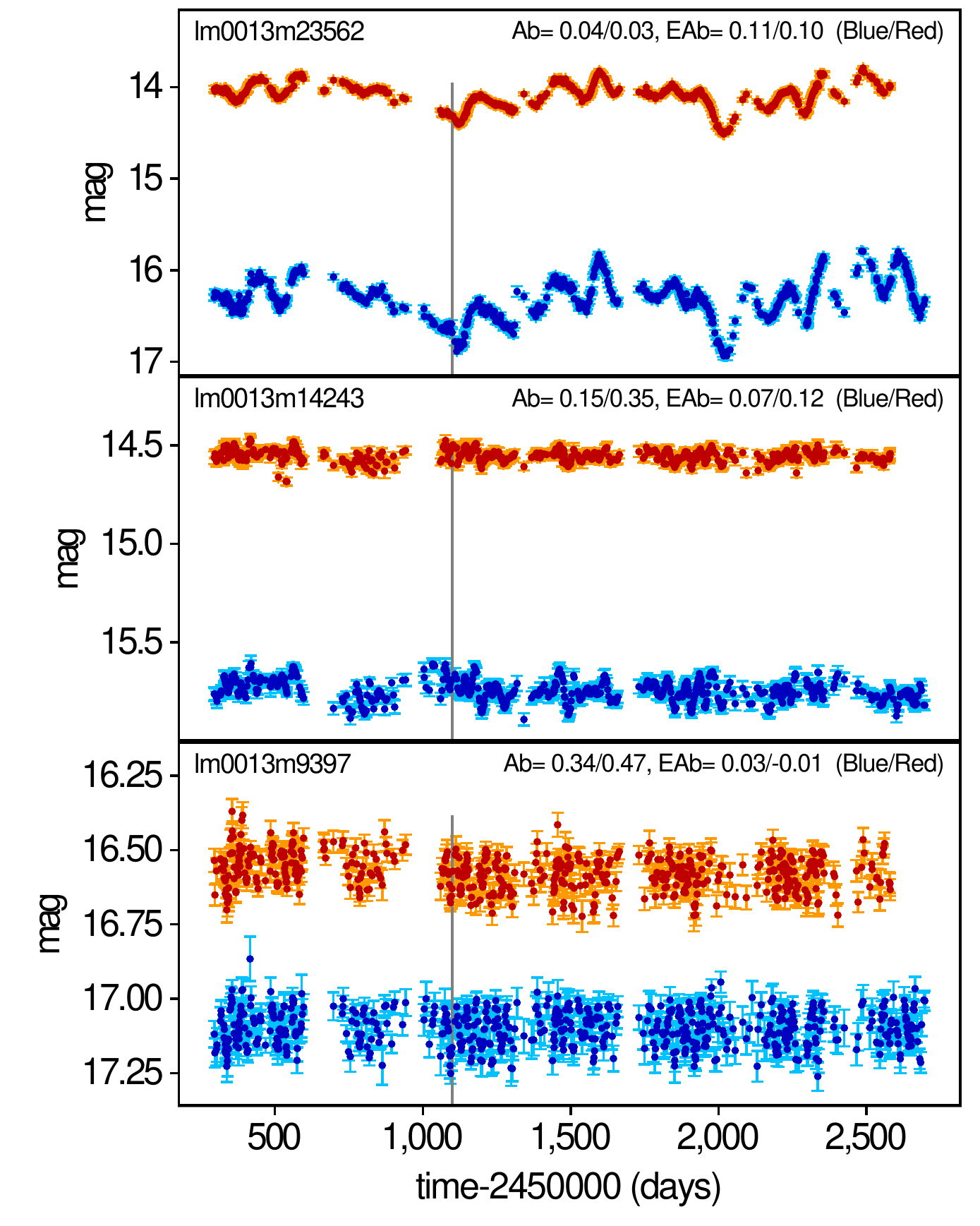}
  \caption{Same as Fig.~\ref{Fig:ErosLcExampleStep}, but for the three pulsating-like stars highlighted in Fig.~\ref{Fig:ErosDiagramsPartialLcs} (top panel: star 23562; middle panel: star 14243; bottom panel: star 9397).
   }
\label{Fig:ErosLcExamplesPulsatingLike}
\end{figure}

The second most populated region in the diagram using partial light curves, after that of constant stars, is the region of pulsating-like light curves at $\excessAb \lesssim 0.15$ and $\Ab \lesssim 0.70$.
Examples of light curves from that region are displayed in Fig.~\ref{Fig:ErosLcExamplesPulsatingLike} for three different Abbe values.
Their variability smoothness increases with decreasing Abbe values from bottom to top panels, with $\Ab=0.34, 0.15$, and 0.04 in the $\EROSB$ band.
Their positions in the $\diagram$ diagram are also indicated in Fig.~\ref{Fig:ErosDiagramsPartialLcs}.

It is interesting to note that most of the stars in the pulsating-like region of the diagram have $\EROSB - \EROSR$ colors above $\sim$1~mag, while transient candidates in the transient and in the trend regions have colors below $\sim$0.1~mag.
This is seen in Fig.~\ref{Fig:ErosDiagramsPartialLcs}, where each point is color-coded according to its $\EROSB - \EROSR$ value.
Transients are thus more frequent among blue stars than among red stars, at least in this field of the LMC, while pulsating-like stars are mainly detected in red stars.
 
Given the improved $\diagram$ diagram obtained after removal of the measurements before time 1100~d in the time series, I restrict the EROS data analysis in the rest of this paper to those partial light curves.
There are 1278 (1234) stars whose $\EROSR$ ($\EROSB$) partial light curves have at least 100 good measurements after day 1100.

\subsection{EROS transient candidates}
\label{Sect:ErosTransientCandidates}

\begin{figure}
  \centering
  \includegraphics[width=\columnwidth]{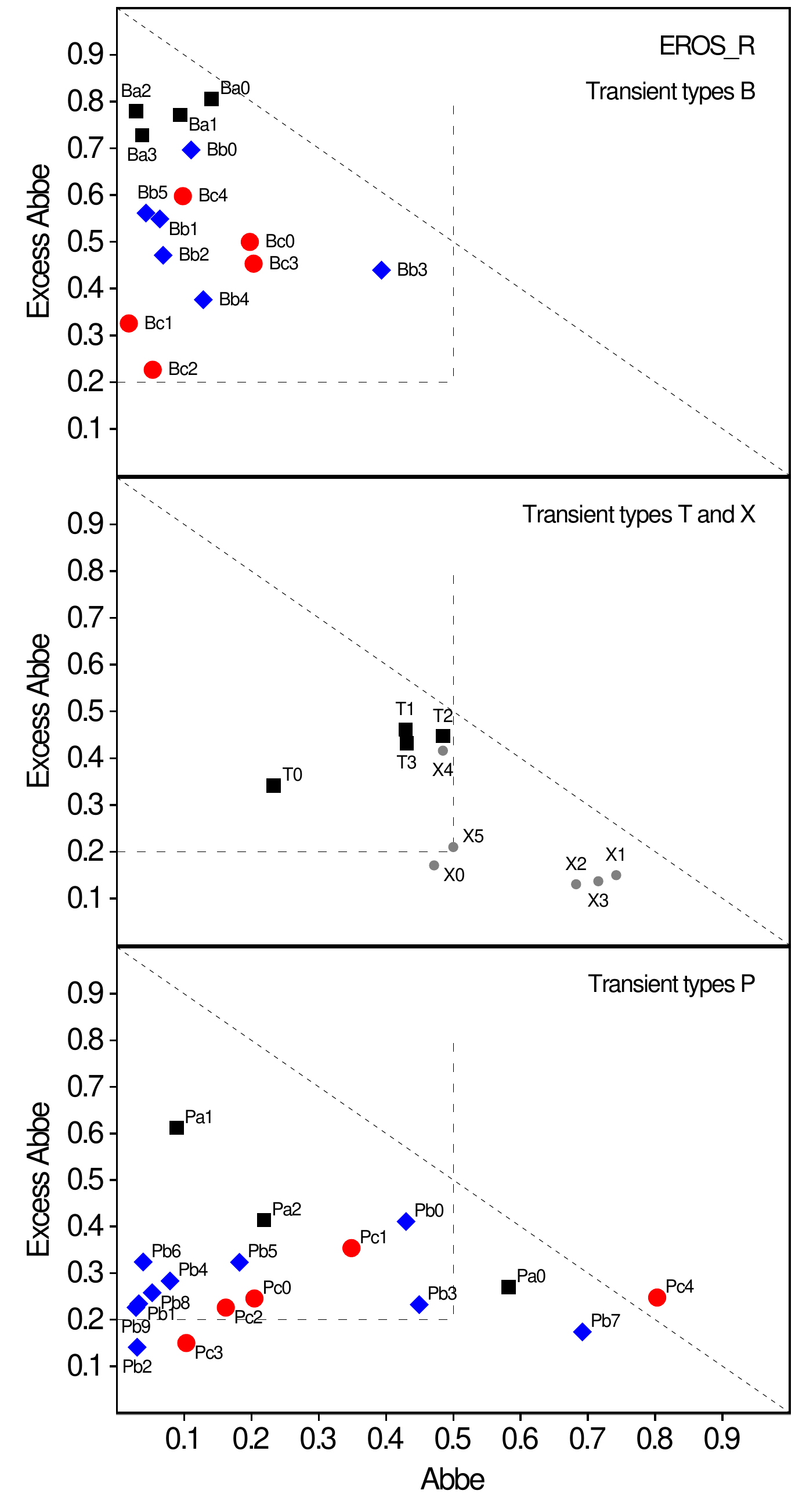}
  \caption{$\diagram$ diagrams for each of the four categories of transient candidates (see text) that fall in the transient region $\Ab \leqslant 0.5$ and $\excessAb \geqslant 0.2$  (delimited by the long-dashed lines) from their $\EROSR$ light curves.
  The short-dashed diagonal line locates the diagonal $\excessAb = 1-\Ab$.
           \textbf{Top panel:} Transient candidates with bursts or outbursts (black squares for type Ba, blue diamonds for type Bb, and red circles for type Bc candidates).
           \textbf{Second panel from top:} Transient candidates with trends (type T, black squares) and unclassified candidates (type X, gray dots).
           \textbf{Bottom panel:} Transient candidates with pulsating-like features in their light curves (black squares for type Pa, blue diamonds for type Pb, and red circles for type Pc candidates).
           The correspondence between the labels in the figure and the EROS source ids is given in Table~\ref{Tab:erosTransients}.
   }
\label{Fig:ErosDiagramTypesR}
\end{figure}

\begin{figure}
  \centering
  \includegraphics[width=\columnwidth]{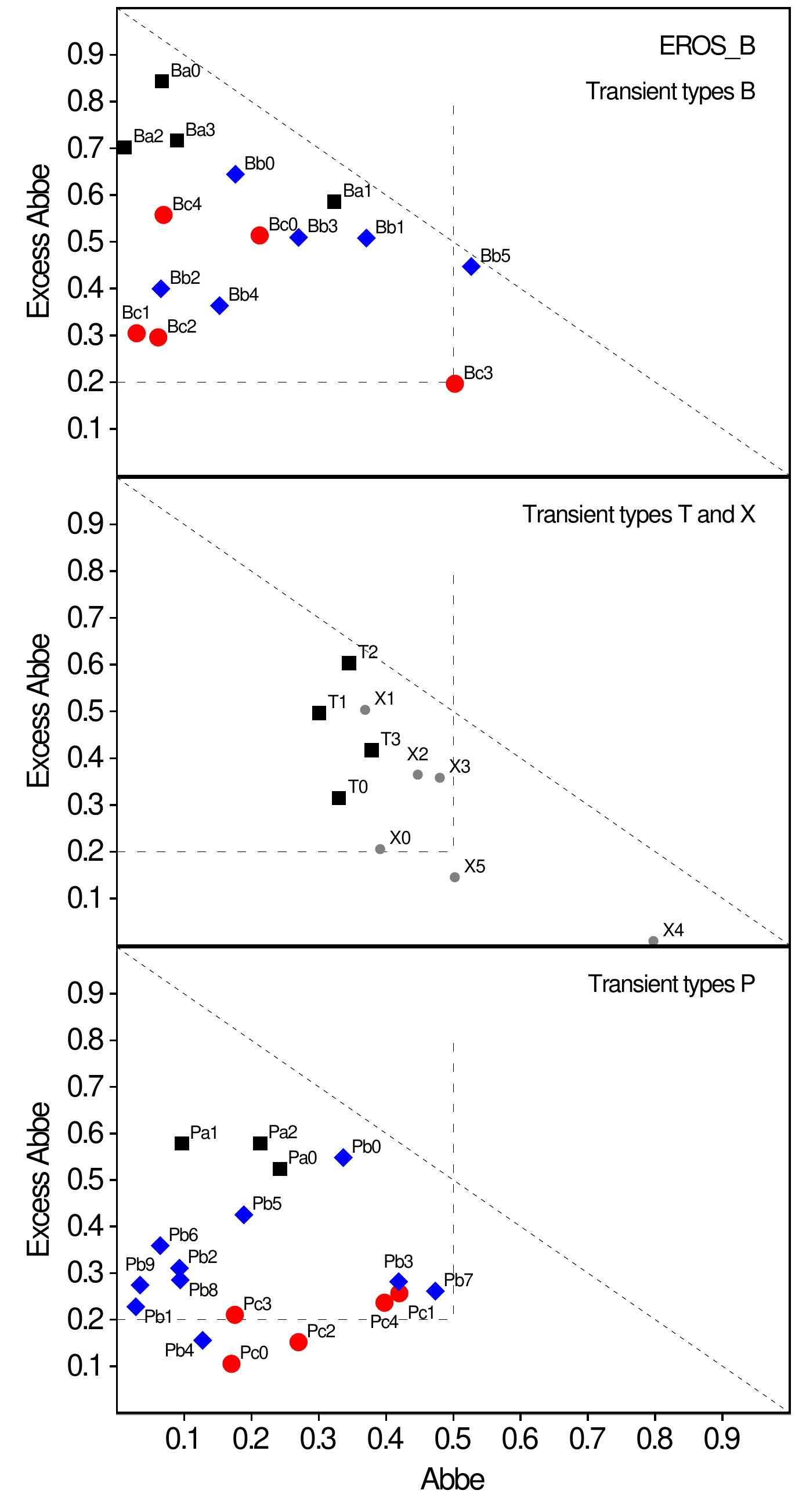}
  \caption{Same as Fig.~\ref{Fig:ErosDiagramTypesR}, but from $\EROSB$ light curves.
   }
\label{Fig:ErosDiagramTypesB}
\end{figure}

\begin{table*}
\centering
\caption{EROS transient candidates (first group of columns, Cols.~2-7; Col~7 gives the ID in Figs.~\ref{Fig:ErosDiagramTypesR} and \ref{Fig:ErosDiagramTypesB}), their OGLE-II matches (second group of columns, Cols.~8-12; OGLE ID annotated with an asterisk if manually matched to the EROS source after visual inspection of the light curves), and their Be star match from \cite{SabogalMennickentPietrzynski_etal05} (third group of columns, Cols.~13-15; Col.~13 gives the Ogle ID listed by those authors, and Col.~15 the distance to the EROS counterpart).
}
 \begin{tabular}{| r | r c@{~}c c@{~}c@{~~}c | r c@{~}c@{~~}c@{~}c | c c@{~}c |}
\hline
  i & ErosId & \Ab(R) & \excessAb(R) & \Ab(B) & \excessAb(B) & TransId & OgleId~~ & \Ab(I) & \excessAb(I) &   Trans? & dist   & Be Star & Be & dist \\
    &        &         &            &          &           &         &          &         &            &      &  (arcsec) &          &  Type  & (arcsec) \\
\hline
  1 &   2642 &  0.11 &  0.70 &  0.18 &  0.64 & Bb0 &    113153~~   &  0.04 &  0.77 & yes & 0.80 & 05253235-6925026 & 1 & 0.76\\
  2 &   5852 &  0.23 &  0.34 &  0.33 &  0.31 & T0  &    352823~~   &  0.10 &  0.24 & yes & 0.86 & 05265146-6925586 & 4 & 0.89\\
  3 &   6072 &  0.14 &  0.81 &  0.07 &  0.84 & Ba0 &    234375~~   &  0.97 &  0.09 &  no & 0.84 & \multicolumn{3}{l|}{\textit{Outburst out of OGLE-II time span}}\\
  4 &   8257 &  0.20 &  0.25 &  0.17 &  0.11 & Pc0 &    227010~~   &  0.17 &  0.08 &  no & 0.80 &   &  &  \\
  5 &  11088 &  0.43 &  0.46 &  0.30 &  0.50 & T1  &    105856~~   &  0.08 &  0.59 & yes & 0.70 & 05250642-6927563 & 1 & 0.56\\
  6 &  13128 &  0.43 &  0.41 &  0.34 &  0.55 & Pb0 &    106521~~   &  0.98 & -0.01 &  no & 0.76 &   &  &  \\
  7 &  14119 &  0.48 &  0.45 &  0.34 &  0.60 & T2  &    346381$^*$ &  0.03 &  0.88 & yes & 1.08 & 05263216-6928470 & 4 & 1.12\\
  8 &  14301 &  0.06 &  0.55 &  0.37 &  0.51 & Bb1 &    346186~~   &  0.03 &  0.50 & yes & 0.98 & 05265491-6928475 & 1 & 0.93\\
  9 &  14409 &  0.20 &  0.50 &  0.21 &  0.51 & Bc0 &    227108~~   &  0.03 &  0.37 & yes & 0.93 & 05255637-6928570 & 1 & 0.89\\
 10 &  14470 &  0.07 &  0.47 &  0.07 &  0.40 & Bb2 &    346184~~   &  0.02 &  0.54 & yes & 1.01 & 05264542-6928521 & 1 & 1.02\\
 11 &  14903 &  0.47 &  0.17 &  0.39 &  0.21 & X0  &    227608~~   &  0.22 &  0.18 &  no & 0.91 & 05255816-6929070 & 4 & 0.93\\
 12 &  15416 &  0.03 &  0.23 &  0.03 &  0.23 & Pb1 &    105747~~   &  0.01 &  0.05 &  no & 0.85 &   &  &  \\
 13 &  15437 &  0.02 &  0.33 &  0.03 &  0.30 & Bc1 &    105768~~   &  0.03 &  0.13 &  no & 0.73 & 05245794-6929243 & 1 & 0.83\\
 14 &  16766 &  0.74 &  0.15 &  0.37 &  0.50 & X1  &    346174~~   &  0.56 &  0.23 &  no & 1.06 &   &  &  \\
 15 &  16784 &  0.03 &  0.14 &  0.09 &  0.31 & Pb2 &    227088~~   &  0.09 &  0.23 & yes & 0.80 &   &  &  \\
 16 &  17299 &  0.68 &  0.13 &  0.45 &  0.36 & X2  &    105744~~   &  0.82 &  0.04 &  no & 0.73 &   &  &  \\
 17 &  17323 &  0.05 &  0.23 &  0.06 &  0.30 & Bc2 &    227083~~   &  0.03 &  0.47 & yes & 0.95 & 05254269-6929570 & 1 & 0.86\\
 18 &  17516 &  0.39 &  0.44 &  0.27 &  0.51 & Bb3 &    346220~~   &  0.22 &  0.62 & yes & 1.04 & 05262110-6929562 & 1 & 0.81\\
 19 &  17551 &  0.35 &  0.35 &  0.42 &  0.26 & Pc1 &    227011~~   &  0.07 &  0.16 &  no & 0.95 &   &  &  \\
 20 &  17790 &  0.58 &  0.27 &  0.24 &  0.52 & Pa0 &    227081~~   &  0.15 &  0.30 & yes & 1.00 & 05260477-6930036 & 4 & 1.06\\
 21 &  18285 &  0.72 &  0.14 &  0.48 &  0.36 & X3  &    346214~~   &  0.76 &  0.08 &  no & 1.08 &   &  &  \\
 22 &  18409 &  0.20 &  0.45 &  0.50 &  0.20 & Bc3 &    227070~~   &  0.07 &  0.57 & yes & 0.98 & 05254452-6930185 & 2 & 0.95\\
 23 &  18510 &  0.48 &  0.42 &  0.80 &  0.01 & X4  &    567315~~   &  0.68 &  0.15 &  no & 0.94 &   &  &  \\
 24 &  18624 &  0.45 &  0.23 &  0.42 &  0.28 & Pb3 &    106890~~   &  0.63 &  0.19 &  no & 0.81 &   &  &  \\
 25 &  19321 &  0.09 &  0.77 &  0.32 &  0.59 & Ba1 &    219805$^*$ &  0.18 &  0.48 & yes & 0.91 & \multicolumn{3}{l|}{\textit{Redder than typical Be stars}}\\
 26 &  19508 &  0.13 &  0.38 &  0.15 &  0.36 & Bb4 &     98092~~   &  0.09 &  0.36 & yes & 0.84 & 05251547-6930430 & 2 & 0.71\\
 27 &  20724 &  0.16 &  0.23 &  0.27 &  0.15 & Pc2 &    219813~~   &  0.08 &  0.01 &  no & 0.95 &   &  &  \\
 28 &  22897 &  0.04 &  0.56 &  0.53 &  0.45 & Bb5 &    339895~~   &  0.07 &  0.44 & yes & 1.00 & \multicolumn{3}{l|}{\textit{Redder than typical Be stars}}\\
 29 &  23052 &  0.03 &  0.78 &  0.01 &  0.70 & Ba2 &     98158~~   &  0.00 &  0.69 & yes & 0.86 & 05251361-6931526 & 2 & 0.84\\
 30 &  23946 &  0.09 &  0.61 &  0.10 &  0.58 & Pa1 &    339157~~   &  0.01 &  0.39 & yes & 1.11 & 05263636-6932003 & 2 & 1.05\\
 31 &  23948 &  0.08 &  0.28 &  0.13 &  0.16 & Pb4 &    339117~~   &  0.06 &  0.28 & yes & 0.97 &   &  &  \\
 32 &  24193 &  0.18 &  0.32 &  0.19 &  0.43 & Pb5 &    339225~~   &  0.08 &  0.14 &  no & 1.11 & 05263046-6932064 & 4 & 1.09\\
 33 &  24220 &  0.50 &  0.21 &  0.50 &  0.15 & X5  &    339156~~   &  0.32 &  0.05 &  no & 1.00 &   &  &  \\
 34 &  26513 &  0.04 &  0.32 &  0.06 &  0.36 & Pb6 &     98034~~   &  0.00 &  0.03 &  no & 0.93 &   &  &  \\
 35 &  26937 &  0.10 &  0.15 &  0.18 &  0.21 & Pc3 &    339123~~   &  0.07 &  0.02 &  no & 1.00 &   &  &  \\
 36 &  27892 &  0.22 &  0.41 &  0.21 &  0.58 & Pa2 &    339139~~   &  0.03 &  0.24 & yes & 1.12 & 05265249-6933172 & 3 & 1.16\\
 37 &  27954 &  0.10 &  0.60 &  0.07 &  0.56 & Bc4 &     98106~~   &  0.49 &  0.35 & yes & 0.89 & \multicolumn{3}{l|}{\textit{Missed by \cite{SabogalMennickentPietrzynski_etal05}?}}\\
 38 &  28029 &  0.69 &  0.17 &  0.47 &  0.26 & Pb7 &    339188~~   &  0.29 &  0.34 & yes & 1.12 &   &  &  \\
 39 &  28855 &  0.05 &  0.26 &  0.09 &  0.29 & Pb8 &    339120~~   &  0.01 &  0.03 &  no & 1.01 &   &  &  \\
 40 &  28894 &  0.43 &  0.43 &  0.38 &  0.42 & T3  &    339184~~   &  0.16 &  0.65 & yes & 1.16 & 05264577-6933377 & 4 & 0.91\\
 41 &  29443 &  0.80 &  0.25 &  0.40 &  0.24 & Pc4 &    220778~~   &  0.95 &  0.03 &  no & 1.00 &   &  &  \\
 42 &  30229 &  0.03 &  0.23 &  0.03 &  0.27 & Pb9 &     89514~~   &  0.02 &  0.05 &  no & 0.89 &   &  &  \\
 43 &  31947 &  0.04 &  0.73 &  0.09 &  0.72 & Ba3 &     89878~~   &  0.02 &  0.72 & yes & 1.05 & 05251910-6934477 & 1 & 1.05\\
\hline
\end{tabular}
\label{Tab:erosTransients}
\end{table*}

There are 43 EROS sources lying in the transient region (defined in Sect.~\ref{Sect:diagramSummary} as $\Ab \le 0.50$ and $\excessAb \ge 0.20$) of the $\diagram$ diagram in either the $\EROSB$ or $\EROSR$ band, and only a few of them are located on the diagonal region reminiscent of trends.
They are all listed in Table~\ref{Tab:erosTransients}.

The transient region may be populated by non-transient sources, as shown in Sect.~\ref{Sect:method}.
In order to assess the importance and nature of this contamination, the light curves of all transient candidates have been visually classified into one of the following types:
\begin{itemize}
\item \textbf{type B:} light curves containing bursts or outbursts;
\vskip 1mm
\item \textbf{type T:} light curves displaying a trend;
\vskip 1mm
\item \textbf{type P:} pulsating-like light curves;
\vskip 1mm
\item \textbf{type X:} other light curves (i.e., not of type B, T, or P).
\end{itemize}
The result of this visual classification is given in Table~\ref{Tab:erosTransients} (column "TransId").
Their distributions in the $\diagram$ diagram are shown in Figs.~\ref{Fig:ErosDiagramTypesR} and \ref{Fig:ErosDiagramTypesB} for the $\EROSR$ and $\EROSB$ bands, respectively.
They are analyzed in Sects~\ref{Sect:EROS_typeB} to \ref{Sect:EROS_typeX}, and a summary is provided in Sect.~\ref{Sect:EROS_typesSummary}.

\subsubsection{Transient candidates with bursts or outbursts (type B)}
\label{Sect:EROS_typeB}

\begin{figure}
  \centering
  \includegraphics[width=0.95\columnwidth]{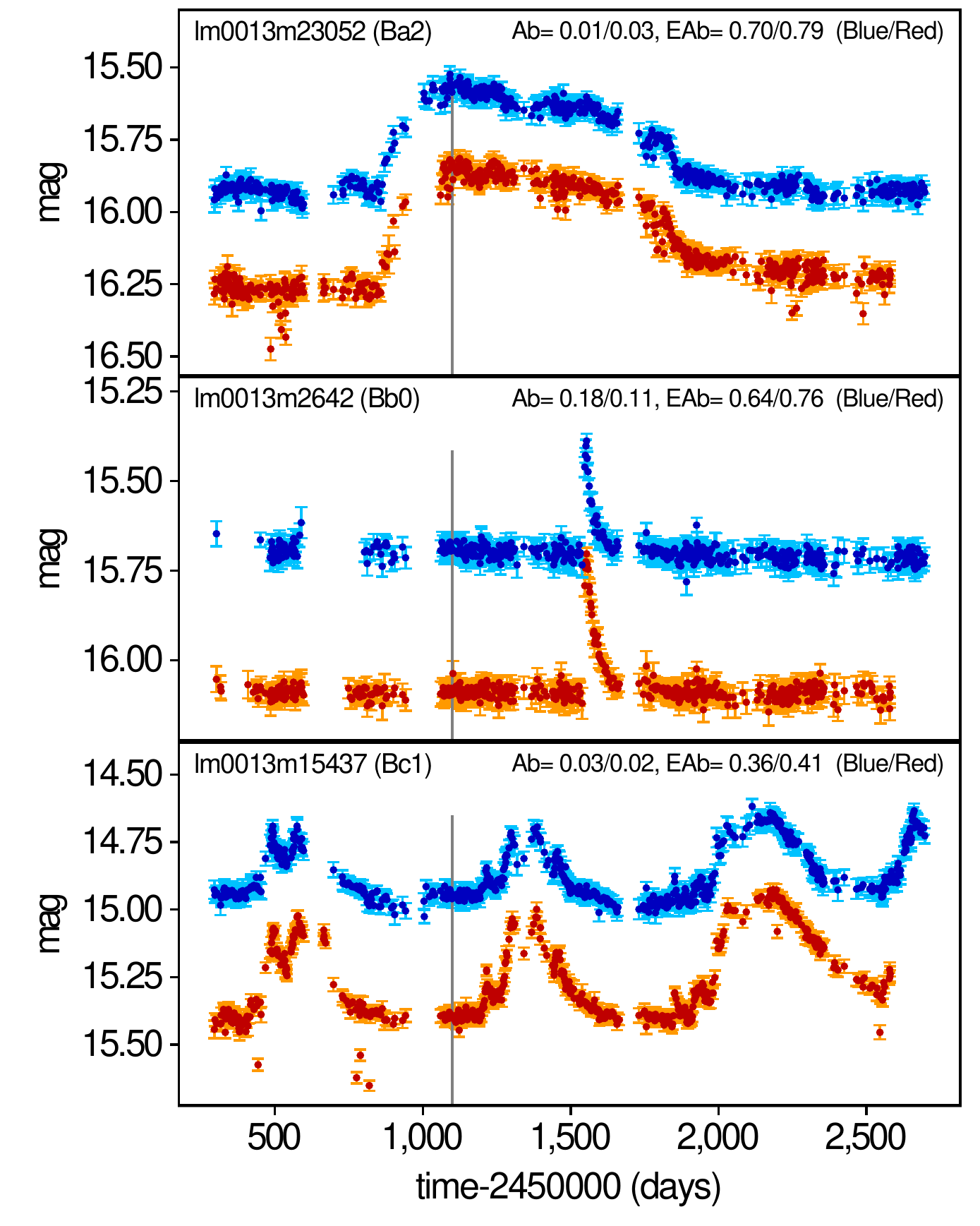}
  \caption{Same as Fig.~\ref{Fig:ErosLcExampleStep}, but for source examples having an outburst (source 23052, top panel), a burst (source 2642, second panel from top), and a mixture of outbursts and bursts (source 15437, third panel from top) in their light curves.
  The label used to identify the transient candidates in Figs.~\ref{Fig:ErosDiagramTypesR} and \ref{Fig:ErosDiagramTypesB} are given in parentheses next to the source name in each panel.
   }
\label{Fig:ErosLcExamplesOutburstAndBursts}
\end{figure}

The distribution of transient candidates that present burst or outburst features in their light curves is shown in the top panels of Figs.~\ref{Fig:ErosDiagramTypesR} and \ref{Fig:ErosDiagramTypesB}.
They have further been visually classified into one of the three following subtypes:
\begin{itemize}
\item \textbf{type Ba:} transient candidates displaying outburst feature(s) in their light curves.
Outburst events are characterized by fluxes that keep a high value for some time.
An example is shown in Fig.~\ref{Fig:ErosLcExamplesOutburstAndBursts} (top panel) for source 23052;
\vskip 1mm
\item \textbf{type Bb:} transient candidates displaying burst feature(s) in their light curves.
Contrary to outbursts, burst events display a rising phase followed almost immediately by a decreasing phase.
An example is shown in Fig.~\ref{Fig:ErosLcExamplesOutburstAndBursts} (middle panel) for source 2642;
\vskip 1mm
\item \textbf{type Bc:} transient candidates displaying both outburst and burst features in their light curves.
An example is shown in Fig.~\ref{Fig:ErosLcExamplesOutburstAndBursts} (bottom panel).
\end{itemize}

Four transient candidates are classified as type Ba, six as type Bb, and five as type Bc in the EROS LMC field studied here.
They are labeled accordingly in Figs.~\ref{Fig:ErosDiagramTypesR} and \ref{Fig:ErosDiagramTypesB}, and are shown in Table~\ref{Tab:erosTransients}.
Their light curves are all shown in Figs.~\ref{Fig:lcsErosBa}, \ref{Fig:lcsErosBb}, and \ref{Fig:lcsErosBc} in Appendix~\ref{SectAppendix:ErosTransients}.

All four Ba-type candidates lie in the uppermost part of the transient region in the $\diagram$ diagram.
Three of them (stars 6072, 23052, and 31947) show a clear outburst feature, the first one showing only the increasing part of the outburst.
The fourth Ba-type star, source 19321, shows a bell-shaped light curve over the whole duration of the survey.
Its Abbe value is much higher in the blue ($\Ab_\mathrm{B}=0.32$) than in the red ($\Ab_\mathrm{R}=0.09$) band because of the larger uncertainties of the measurements in $\EROSB$ than in $\EROSR$;
but its excess Abbe value remains close to the maximum value it can have, putting the star near the diagonal in the $\diagram$ diagram.

The distributions of types Bb and Bc transient candidates in the $\diagram$ diagram are quite similar to that of Ba transient candidates, with the Bb types located, on average, at slightly higher $\excessAb$ values than Bc types.
Stars 15437 (Bc1) and 17323 (Bc2), on the other hand, lie at low $\excessAb$ values close to the region of pulsating-like light curves.
Inspection of their light curves (shown in Fig.~\ref{Fig:lcsErosBc}) indeed reveals a pulsating-like pattern for the first star (also shown in Fig.~\ref{Fig:ErosLcExamplesOutburstAndBursts}, bottom panel), with recurrent outbursts that last about 300~days occurring every 700-800~days.
Star~17323, on the other hand, shows recurrent short burst events superposed on a declining outburst event.
The locations of these two stars in the $\diagram$ diagram are thus compatible with their light curve patterns.

Star 18409 (Bc3) shows an interesting behavior (see Fig.~\ref{Fig:lcsErosBc} in Appendix~\ref{SectAppendix:ErosTransients}), displaying a clear starting outburst event in the red band (with burst events on top of it), while the amplitude of the outburst is much smaller in the blue band.
Consequently, the star is at the limit of being detected as a transient candidate in the $\EROSB$ $\diagram$ diagram, while it is clearly detected in $\EROSR$ (see their respective positions in Figs.~\ref{Fig:ErosDiagramTypesR} and \ref{Fig:ErosDiagramTypesB}).

\subsubsection{Transient candidates with trends (type T)}
\label{Sect:EROS_typeT}

Four of the transient candidates show a trend in their light curves.
They are shown in Fig.~\ref{Fig:lcsErosT} in Appendix~\ref{SectAppendix:ErosTransients}.
They appear clearly in the $\EROSB$ $\diagram$ diagram at $\Ab$ between 0.3 and 0.4 and $\excessAb$ above 0.3 (see Fig.~\ref{Fig:ErosDiagramTypesB}, for example).
The situation is not as good in the diagram based on the $\EROSR$ light curves, mainly because of the larger point-to-point scatter of the measurements in that band compared to the scatter in $\EROSB$, which leads to $\Ab$ values above 0.4.
Only one star, star 5852 (T0), stands out clearly in the diagram, at $\Ab=0.23$, because of a short-term variability pattern in its light curve (see Fig.~\ref{Fig:lcsErosT}).

The situation would have been better had I used the full rather than the partial light curves, as expected from the analysis of light curves with trends presented in Sect.~\ref{Sect:simuTrends}.
This is particularly true for source 14119 (T2), which has, in the $\EROSR$ band, $(\Ab,\excessAb)$ values of $(0.11, 0.79)$ using the full light curve, instead of $(0.48, 0.45)$ with the partial light curve used here.
This clearly shows the obvious advantage of long surveys in identifying sources displaying trends in their light curves.

\subsubsection{Pulsating-like transient candidates (type P)}
\label{Sect:EROS_typeP}

Among the transient candidates, 18 are visually classified as having pulsating-like light curves in either $\EROSR$, $\EROSB$, or in both.
The typical timescale of pulsating-like variables was shown in Sect.~\ref{Sect:simuPeriodic} to be an important parameter in their location in the $\diagram$ diagram.
Time scales larger than $\deltaTSub$ were shown to lead to large $\excessAb$ values.
It is therefore useful to further classify type P transient candidates according to the typical variability timescale observed in their light curves.
I divide them into three categories with the following criteria, knowing that $\deltaTSub=100$~d:
\begin{itemize}
\item \textbf{type Pa:} pulsating-like transient candidates with very long variability timescales, of more than two years (light curves shown in Fig.~\ref{Fig:lcsErosPa} in Appendix~\ref{SectAppendix:ErosTransients});
\vskip 1mm
\item \textbf{type Pb:} pulsating-like transient candidates with variability timescales shorter than the ones of type Pa stars, but above 100~d (Fig.~\ref{Fig:lcsErosPb});
\vskip 1mm
\item \textbf{type Pc:} pulsating-like transient candidates with variability timescales shorter than 100~d (Fig.~\ref{Fig:lcsErosPc}).
\end{itemize}

Inspection of Figs.~\ref{Fig:ErosDiagramTypesR} and \ref{Fig:ErosDiagramTypesB} (bottom panels) shows that the middle part of the region of transient candidates is only populated by pulsating-like transients with long variability timescales (mainly type Pa and few type Pb stars).
All other type P stars populate the lower region of the transient region, close to the region of pulsating-like stars.
These results basically confirm the predictions made in Sect.~\ref{Sect:simuPeriodic}.

\subsubsection{Transient candidates with unclear features (type X)}
\label{Sect:EROS_typeX}

Six transient candidates could not be visually classified into one of the B, P, or T types described above.
Their light curves are shown in Fig.~\ref{Fig:lcsErosX} in Appendix~\ref{SectAppendix:ErosTransients}.
In the $\diagram$ diagram, they all lie in the lower-right part of the transient region (small gray dots in the middle panel of Figs.~\ref{Fig:ErosDiagramTypesR} and \ref{Fig:ErosDiagramTypesB}).
They would thus not be confused with transient candidates that present burst or outburst features in their light curves.

\subsubsection{Conclusions}
\label{Sect:EROS_typesSummary}

The detailed analyses of the EROS time series presented in this section confirm the potential of the $\diagram$ diagram to identify transients in a survey.
Transients that display burst or outburst features in their light curves (type B) and those showing a long-term trend (type T) populate the region of the diagram at high $\excessAb$ values.
Pulsating-like stars (type P) that have variability timescales greater than the duration $\deltaTSub$  of the subtime intervals used in the computation of the $\excessAb$ values, also populate this region of the diagram, as expected from predictions made in Sect.~\ref{Sect:simuPeriodic}.

A classification procedure is thus necessary to disentangle transient targets from identified and rejected type P transient candidates, and to further subclassify type B transients into subtypes Ba, Bb, and Bc.
Such a classification has been performed visually in this study, but an automated procedure would be desirable to process large-scale surveys.
An automated period search, for example, can greatly contribute to the identification (and rejection) of type P candidates.
Further considerations that may help in the establishment of such an automated procedure are given in Appendix~\ref{SectAppendix:CommentsSubClassification}.

\section{OGLE-II sample of the LMC}
\label{Sect:OGLE-II}

In this section, I compare the EROS $\diagram$ diagram with that obtained from the OGLE-II survey \citep{UdalskiKubiakSzymanski97}, which covers the EROS field \field{lm0013m} of the LMC sky and at similar epochs.
The OGLE-II data is first presented in Sect.~\ref{Sect:OgleData}, and the $\diagram$ diagram in Sect.~\ref{Sect:OgleDiagram}.
Transient candidates are then analyzed in Sect.~\ref{Sect:OgleTransients}, where they are also compared to those of EROS.
Conclusions are summarized in Sect.~\ref{Sect:OgleTransientsSummary}.

\subsection{OGLE-II data}
\label{Sect:OgleData}

The EROS field \field{lm0013m} is entirely covered by the OGLE-II field \texttt{LMC\_SC4}.
Part of OGLE-II field \texttt{LMC\_SC5} also overlaps the EROS field (and duplicates sources in \texttt{LMC\_SC4}), but I restrict the analysis to the time series recorded in \texttt{LMC\_SC4}.

All OGLE-II data have been downloaded from the OGLE-II web site\footnote{
\texttt{http://ogledb.astrouw.edu.pl/$\sim$ogle/photdb}
}
\citep{Szymanski05}.
From the \textsl{Photometry Database} query page, sources are selected from I-band Differential Image Analysis (DIA) photometry that fall in the sky region RA=[5.41570,5.44902] h and Decl=[-69.58063,-69.41011] deg, all from field=LMC\_SC4.
The list is restricted to sources with a mean I-band magnitude $I \le 19$~mag.
Furthermore, the option `\textsl{No catalog flag objects only}' was set in the query page, as recommended by \cite{Szymanski05} to exclude multiple identifications.
The query returned 17671 sources.
The DIA photometry time series of those sources were downloaded requesting good points only \citep{Szymanski05}.

The OGLE-II time series were cleaned with the same procedure as the one used to process the EROS time series (see Sect.~\ref{Sect:EROS}), but without removal of data prior to day 1100. 
The number of OGLE sources with at least 100 good points in their light curves amounts to 17663.
The mean uncertainty on their measurements is 22~mmag and the mean of their standard deviations 36~mmag.
These numbers are to be compared with 85.6 and 100~mmag, respectively, for the $\EROSR$ partial light curves (cf. Sect.~\ref{Sect:ErosDiagram}).
The OGLE-II light curves are thus globally more precise than the EROS ones, based on the mean uncertaintie, but, as is described in the next sections, there are a larger number of spurious sources in OGLE-II (at least from DIA photometry) than in EROS.
This most probably results from the fact that OGLE-II data base contains all sources detected by the DIA technique and has not been cleaned from potentially spurious sources.

\subsection{The $\Ab - \excessAb$ diagram of the OGLE-II sample}
\label{Sect:OgleDiagram}

\begin{figure}
  \centering
  \includegraphics[width=\columnwidth]{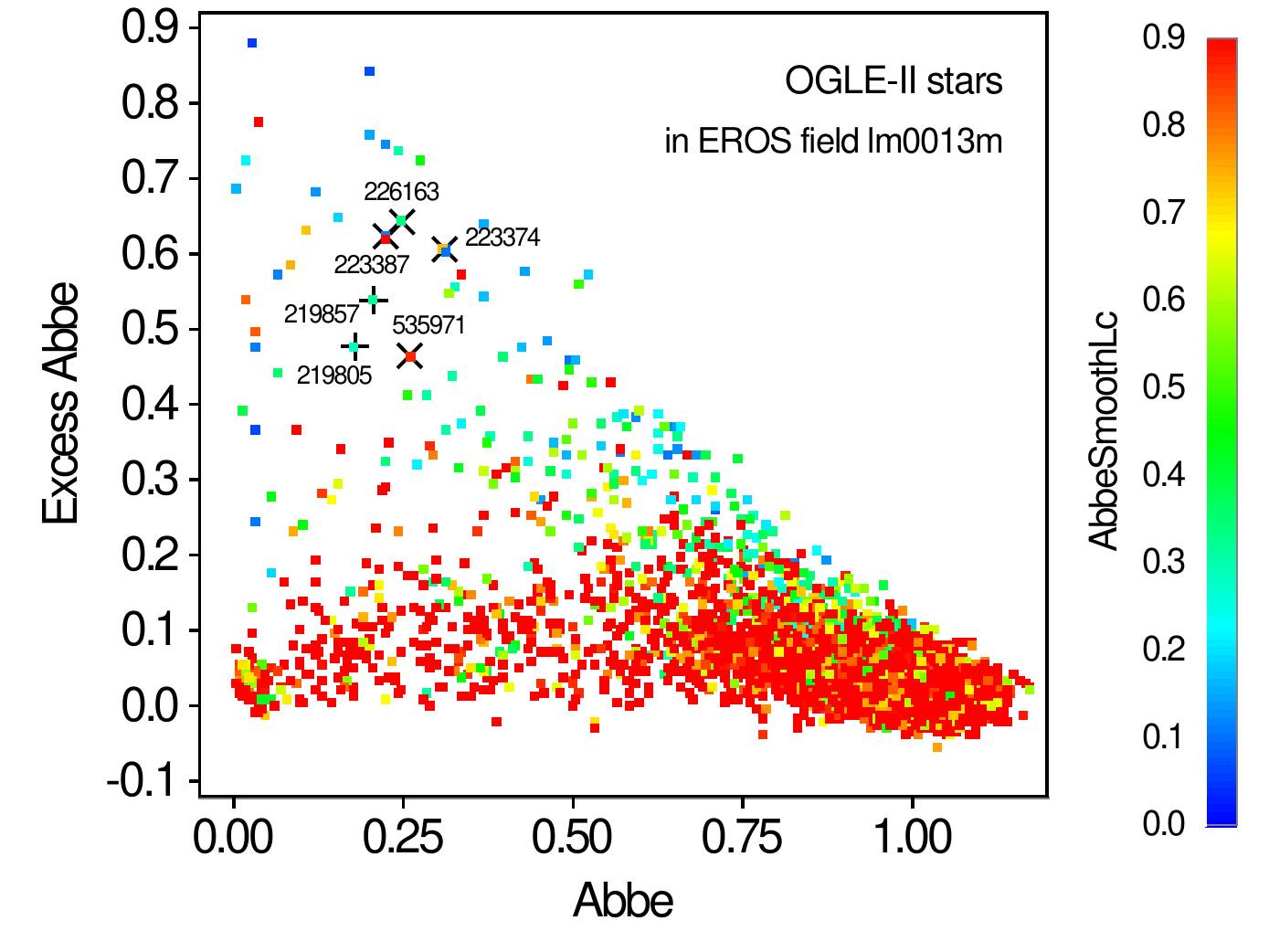}
  \caption{$\Ab$ versus $\excessAb$ diagram for the OGLE-II variable stars lying in the EROS field \field{lm0013m}.
           The color scale of each point is proportional to the Abbe value of the 100~d smoothed light curve as shown on the color scale on the right, with all Abbe values greater than 0.9 set to red.
           For illustration purposes, the locations of the components of an OGLE-II contaminating/contaminated pair (see text) are highlighted with `$+$' markers, with the source IDs labeled next to the markers.
           Similarly, the locations of one of the groups of suspicious pulsating-like time series with trends, displayed in Fig.~\ref{Fig:lcsClumpOgle} in Appendix~\ref{SectAppendix:suspiciousOgleTransients}, are highlighted with `$\times$' markers.
   }
\label{Fig:OgleDiagram}
\end{figure}

The $\diagram$ diagram for the OGLE-II stars is shown in Fig.~\ref{Fig:OgleDiagram}.
We globally observe the same distribution of points in this diagram as found with EROS time series (cf. Fig.~\ref{Fig:OgleDiagram} and Fig.~\ref{Fig:ErosDiagramsPartialLcs}), but with a higher density of points in all regions of the diagram.
The higher quality (smaller uncertainties) of OGLE-II data compared to EROS data is certainly at the origin of the higher density of points in the pulsating-like region, since low-amplitude variables will be more easily detectable in the OGLE-II survey than in EROS, but the higher number of points on the diagonal in Fig.~\ref{Fig:OgleDiagram} look suspicious, and may point to the existence of trends of instrumental and/or data reduction origin, a problem that may also affect the density of points in the region of transient candidates.
This question is addressed in the next section.

Figure~\ref{Fig:OgleDiagram} presents a new quantity computed on the time series, $\AbSmooth$, color coded according to the color scale displayed on the right of the figure.
This quantity is the Abbe value computed on smoothed light curves that are constructed by averaging measurements of the initial time series over time intervals of 100~days.
While I do not use this quantity in the body of this paper, I introduce it in Appendix~\ref{SectAppendix:CommentsSubClassification} as a potential attribute to be used in automated classification of transient candidates.
It is interesting to note in Fig.~\ref{Fig:OgleDiagram} that light curves with small $\AbSmooth$ values are predominantly found in the regions of trends and transient candidates, in agreement with the conclusions in Appendix~\ref{SectAppendix:CommentsSubClassification};
however, I will not analyze further this option here, as it would lead beyond the scope of this article.

\subsection{OGLE-II transient candidates}
\label{Sect:OgleTransients}

\begin{table*}
\centering
\caption{OGLE-II transient candidates (first group of columns, Cols.~2 and 3) that are not listed in Table~\ref{Tab:erosTransients}, but which have a (non-transient) match in the database of EROS variables.
The characteristics of the EROS matches are given in the second group of columns (Cols.~4-7; EROS ID annotated with an asterisk if manually matched to the OGLE-II source after visual inspection of the light curves), and their distance to the OGLE-II transient candidate is given in Col.~7.
}
 \begin{tabular}{| r | r c@{~}c | r c@{~}c c@{~}c c |}
\hline
  i & OgleId & \Ab(I) & \excessAb(I)& ErosId & \Ab(R) & \excessAb(R) & \Ab(B) & \excessAb(B) & dist \\
    &   &    &  &   &   &   &   &   & (arcsec) \\
\hline
 44 &  107165 &  0.26 &  0.41 &     35706~~   &  0.62 &  0.31 &  0.70 &  0.18 & 1.07\\
 45 &  113110 &  0.33 &  0.55 &      6475~~   &  0.83 &  0.05 &  0.70 &  0.21 & 0.64\\
 46 &  113128 &  0.14 &  0.27 &      5330$^*$ &  0.76 &  0.16 &  0.67 &  0.21 & 0.68\\
 47 &  212949 &  0.47 &  0.23 &     31721~~   &  0.84 &  0.09 &  0.87 &  0.12 & 1.01\\
 48 &  220111 &  0.32 &  0.44 &     21298~~   &  0.62 &  0.38 &  0.59 &  0.37 & 0.99\\
 49 &  227200 &  0.47 &  0.34 &     16631~~   &  0.73 &  0.20 &  0.76 &  0.25 & 0.89\\
 50 &  339356 &  0.12 &  0.68 &     25348~~   &  0.80 &  0.20 &  0.68 &  0.24 & 1.14\\
 51 &  346243 &  0.10 &  0.63 &     14453~~   &  0.61 &  0.31 &  0.52 &  0.49 & 0.97\\
 52 &  346331 &  0.20 &  0.84 &     17230~~   &  0.83 &  0.11 &  0.63 &  0.26 & 1.04\\
\hline
\end{tabular}
\label{Tab:ogleTransientsWithErosMatch}
\end{table*}

There are 98 OGLE-II transient candidates in the transient region defined by $\Ab \le 0.5$ and $\excessAb \ge 0.2$.
This is about a factor of two larger than the number of transient candidates extracted from the EROS database.
Given that OGLE-II uncertainties are smaller than EROS uncertainties, we may expect transients to be detected in the OGLE-II data base that may have remained hidden in the EROS database, but this does not explain the factor of two difference.
I thus start the analysis by identifying in Sect.~\ref{Sect:spuriousOgleTransients}, and removing from further study, spurious light curves that populate the region of transient candidates in the $\diagram$ diagram.

Section~\ref{Sect:OgleMatchesToErosTransients} then presents transient candidates common to both OGLE-II and EROS surveys.
OGLE-II transient candidates that have a non-transient EROS counterpart are then analyzed in Sect.~\ref{Sect:OgleTransientsWithErosMatches}, and the remaining OGLE-II transient candidates, for which no EROS counterpart is found, are discussed in Sect.~\ref{Sect:OgleTransientsWithoutErosMatches}.
A summary of the results is given in Sect.~\ref{Sect:OgleTransientsSummary}.

\subsubsection{OGLE-II spurious transient candidates}
\label{Sect:spuriousOgleTransients}

\begin{figure*}
  \centering
  \includegraphics[width=2\columnwidth]{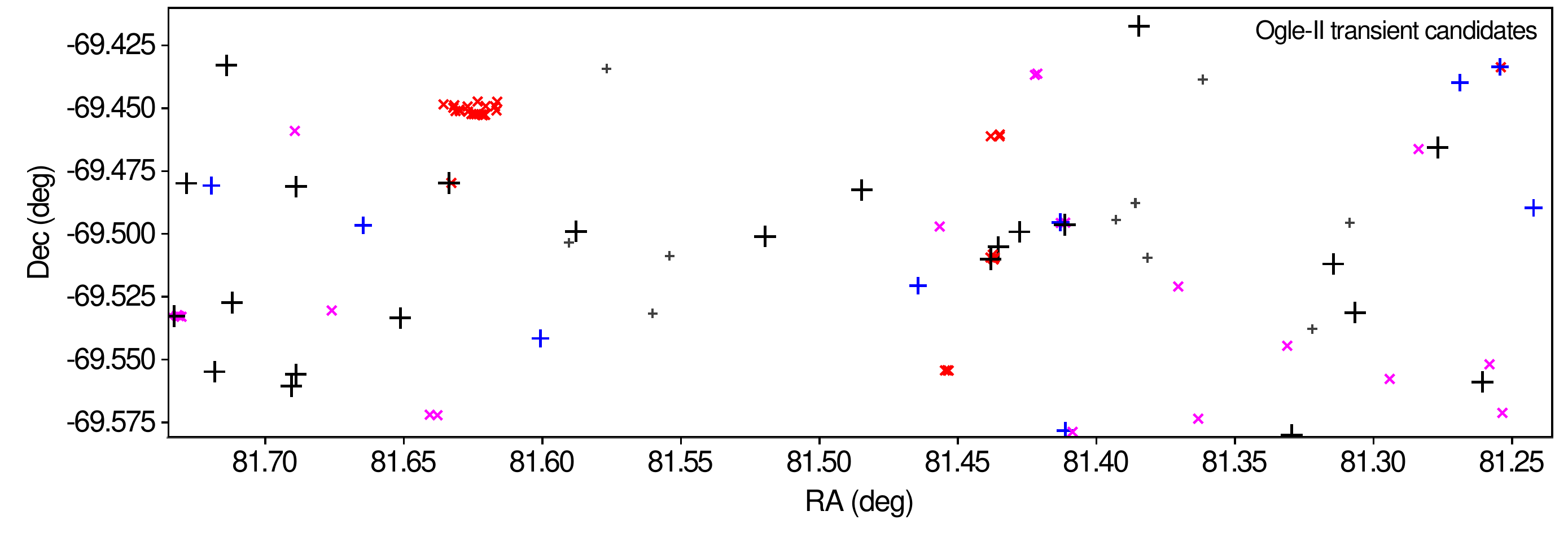}
  \caption{Positions on the sky of OGLE-II transient candidates.
  Big dark $'+'$ markers identify the ones that match an EROS transient candidate (Table~\ref{Tab:erosTransients}),
  medium-sized blue $'+'$ markers the ones with an EROS non-transient star (Table~\ref{Tab:ogleTransientsWithErosMatch}), and small gray $'+'$ markers the ones with no match in the EROS database (Table~\ref{Tab:ogleTransientsWithNoErosMatch}).
  Red and magenta $'\times'$ markers locate spurious OGLE-II transient candidates, in red for those listed in Tables~\ref{Tab:ogleSpuriousTransientsWithErosTransients} and \ref{Tab:ogleSpuriousTransientsFromLcsWithoutEros}, and in magenta for those listed in Table~\ref{Tab:ogleSpuriousTransientsWithErosNonTransients}.
  }
\label{Fig:skyOgleTransients}
\end{figure*}

A close inspection of the light curves of OGLE-II transient candidates reveals the presence of 56 potentially spurious cases, identified from either suspicious light curve similarities or the detection of suspiciously close objects on the sky.
They are listed in Appendix~\ref{SectAppendix:suspiciousOgleTransients} where their light curves are also shown.
Three categories are distinguished.
\begin{enumerate}[a)]
\item 25 sources are contaminated by the light of a bright, close-by OGLE-II star.
The contaminated sources are recognized by the similarity of their light curves with that of the contaminating, brighter star.
Their spurious nature is further supported by the absence of any EROS counterpart at their positions on the sky, in contrast to contaminating stars.
For illustration, the positions in the $\diagram$ diagram of the OGLE-II (contaminating/contaminated) pair 219805/219857 are shown in Fig.~\ref{Fig:OgleDiagram} (light curves shown in Fig.~\ref{Fig:lcsContaminatedOgle}).
The contaminating stars are transient candidates for 3 of them (Table~\ref{Tab:ogleSpuriousTransientsWithErosTransients}), and non-transients for the remaining 22 (Table~\ref{Tab:ogleSpuriousTransientsWithErosNonTransients}).
\item 10 sources, divided in three groups, show variability pattern similarities that suggest common contamination, but no clear contaminating can be identified (Table~\ref{Tab:ogleSpuriousTransientsFromLcsWithoutEros}).
No EROS counterpart is found either.
The origin of their light curve similarities may point to a data reduction problem.
For illustration, the positions in the $\diagram$ diagram of OGLE-II stars 226163, 223387, 223374, and 535971, forming one of these groups of suspicious stars, are shown in Fig.~\ref{Fig:OgleDiagram}.
\item 21 sources lie within 15'' in a clump on the sky around direction $\alpha$=81.63~deg and $\delta$=-69.45~deg, clearly visible in Fig.~\ref{Fig:skyOgleTransients} showing the distribution on the sky of all OGLE-II transient candidates discussed in this section.
They show light curve similarities by groups, suggesting a non-astrophysical origin.
No EROS match is found for any of these stars.
These 21 sources are also listed in Table~\ref{Tab:ogleSpuriousTransientsFromLcsWithoutEros} in Appendix~\ref{SectAppendix:suspiciousOgleTransients}.

The question was raised whether the variability patterns observed in the OGLE-II light curves of the stars clustered around $\alpha$=81.63~deg and $\delta$=-69.45~deg in Fig.~\ref{Fig:skyOgleTransients} could be due to light echos from SN1987A \citep[e.g.,][]{SugermanCrottsKunkel_etal05} scattered on a dense nebula (bok globule), SN1987A being only at 48.5~arcmin from that direction.
That this is unlikely is suggested by the different times (by groups of stars) at which the light curves show flux enhancements (see Fig.~\ref{Fig:lcsClumpOgle}).
In addition, the clump is too far on the sky from SN1987A considering the $\sim$10 years (3~pc for light) separating the supernova explosion and the OGLE-II survey.
Indeed, if we consider a depth of 4 kpc for the LMC \citep{SubramanianSubramaniam09}, and assuming the most favorable case where SN1987A would be at the farthest distance on the line of sight, the maximum angular separation between SN1987A and a bok globule from where a light echo would originate at the time of the OGLE-II survey would be $\sim$10 arcmin
\citep[e.g., Eq.~III of]
[with $t=3$~pc, $b=4$~kpc, and the distance $\theta$ to the LMC $=51.2$~kpc]
{Couderc39}.
This is a factor of about five smaller than the projected distance between the clump and SN1987A.
%
%
%
%
%
%
\end{enumerate}

These sources with potentially spurious light curves are simply removed from the list of OGLE-II transient candidates in the next sections.

\subsubsection{OGLE-II transients with transient EROS matches}
\label{Sect:OgleMatchesToErosTransients}

Of the 42 remaining, potentially non-suprious, OGLE-II transient candidates, 23 match\footnote{
The matching procedure consists in finding the OGLE-II source closest to the EROS transient, and in checking that the reverse procedure, i.e., finding the EROS source closest to the OGLE-II match candidate, falls back on the initial EROS transient.
If this consistency check fails, no match is retained.
Of course, a consistent match does not guarantee a correct match from an astrophysical point of view, and a visual check is performed to validate the matching candidate.
This led to a correction of the matches in some cases.
I also reject matches that are more than 2~arcsec apart on the sky.
}
one of the 43 EROS transient candidates.
They are indicated as such in Table~\ref{Tab:erosTransients}.

The other 20 EROS transient candidates also have a match in the OGLE-II database\footnote{
In two cases when more than one OGLE-II source are present in the vicinity of an EROS source, a visual inspection of the light curve of the consistent match revealed an actual mismatch by the automatic matching procedure.
They are flagged in Table~\ref{Tab:erosTransients} by an asterisk added to the OgleId.
}
(also indicated in Table~\ref{Tab:erosTransients}), but they are not detected as transient candidates from the OGLE-II light curves.
In other words, 20 EROS transient candidates fail to be detected as such in the OGLE-II database.
Visual inspection of the EROS and OGLE-II light curves (Figs.~\ref{Fig:lcsErosBa} to \ref{Fig:lcsErosX} in Appendix~\ref{SectAppendix:ErosTransients}) leads to the following understanding of these cases, sorted by EROS transient type.

\begin{itemize}
\item \textbf{Types B and T} (Figs.~\ref{Fig:lcsErosBa} to \ref{Fig:lcsErosT})\textbf{:} they are all, except one, detected as transient candidates in OGLE-II.
The only exception is EROS source \field{lm0013m6072}, which has a burst starting at a time after the end of the OGLE-II survey, and which could thus not have been detected in the OGLE-II data.
\vskip 1mm
\item \textbf{Type Pa} (Fig.~\ref{Fig:lcsErosPa})\textbf{:} all three EROS transient candidates of this type are detected in OGLE-II as transient candidates.
\vskip 1mm
\item \textbf{Types Pb and Pc} (Figs.~\ref{Fig:lcsErosPb} and \ref{Fig:lcsErosPc})\textbf{:} the vast majority of them are not detected as transient candidates in the OGLE-II survey.
In most cases, the OGLE-II excess Abbe values fall below 0.20, making them pulsating-like candidates rather than transient candidates, which I understand to be the result of the better quality of OGLE-II measurements.
This is a useful result as it cleans the region of transient candidates from polluting pulsating-like stars.
\vskip 1mm
\item \textbf{Type X} (Fig.~\ref{Fig:lcsErosX})\textbf{:} their OGLE-II counterparts are either moved towards the region of constant stars at $\Ab>0.5$, or fall in the region of pulsating-like stars with $\excessAb<0.2$.
Here too, the OGLE-II survey provides a better classification of these sources.
\end{itemize}

In summary, all EROS transient candidates that are either real transients (types B and T) or reminiscent of real transients (type Pa) are correctly identified as transient candidates in the OGLE-II data base as well.
The other EROS transient candidates that are not detected as such in OGLE-II are either pulsating-like variables or constant stars.
Their non-detection in OGLE-II thus confirms the potential of the method presented in this paper to detect real transients; the pollution of the region of transient candidates by non transients decreases with increasing quality of the data.

\subsubsection{OGLE-II transients with non-transient EROS matches}
\label{Sect:OgleTransientsWithErosMatches}

Among the 19 remaining OGLE-II transient candidates that are not spurious and do not have a match in the list of EROS transient candidates, 9 have EROS matches that are not EROS transient candidates.
They are listed in Table~\ref{Tab:ogleTransientsWithErosMatch}, and their OGLE-II and EROS light curves are shown in Fig.~\ref{Fig:lcsOgleWithErosMatch} in Appendix~\ref{SectAppendix:extraOgleTransients}.

They are not recognized as transient candidates in EROS because of the noisier EROS data, compared to the OGLE-II data, and the consequently higher Abbe values (all above 0.5).
The case is particularly evident for EROS sources 5330 and 14453 (OGLE-II sources 113128 and 346243, respectively), which display clear bursts in the OGLE-II light curves that are only marginally visible in their EROS counterparts (see Fig.~\ref{Fig:lcsOgleWithErosMatch}).
The same is true for EROS source 21298, which is successfully detected in its OGLE-II match 220111.

These examples show that, while the $\diagram$ technique is a very powerful way to detect transients, it also has its limitations, as all methods do.
In the case of EROS, the limitation is imposed by the large uncertainties of the measurements.

\subsubsection{OGLE-II transient candidates with no EROS match}
\label{Sect:OgleTransientsWithoutErosMatches}

\begin{table}
\centering
\caption{OGLE-II transient candidates with no match in the list of EROS variable stars.
}
 \begin{tabular}{r r c@{~}c}
\hline
  i & OgleId & \Ab(I) & \excessAb(I) \\
    &        &         &             \\
\hline
 53 &  100211 &  0.46 &  0.27 \\
 54 &  105930 &  0.46 &  0.27 \\
 55 &  105939 &  0.27 &  0.32 \\
 56 &  107195 &  0.49 &  0.33 \\
 57 &  113113 &  0.41 &  0.26 \\
 58 &  219958 &  0.38 &  0.36 \\
 59 &  221478 &  0.49 &  0.46 \\
 60 &  346536 &  0.50 &  0.44 \\
 61 &  352990 &  0.22 &  0.33 \\
 62 &  98412 &  0.37 &  0.64 \\
\hline
\end{tabular}
\label{Tab:ogleTransientsWithNoErosMatch}
\end{table}

Finally, there are ten OGLE-II transient candidates that have no EROS match at all, and for which I could not reliably find any neighboring contaminating OGLE-II star.
They are listed in Table~\ref{Tab:ogleTransientsWithNoErosMatch}, and their light curves are shown in Fig.~\ref{Fig:lcsOgleWithoutErosMatch} in Appendix~\ref{SectAppendix:extraOgleTransients}.
Some of them, like source 98412 that displays a trend in its light curve and 219958 that displays an outburst feature, should have been detected by EROS.
The fact that no match is present in the EROS catalog of variable stars for any of these OGLE-II sources may suggest either spurious cases in OGLE-II, or a failure to detect them as variables in EROS, for one reason or another.
A deeper investigation into this problem would be interesting, but is beyond the scope of this article.

\subsection{Conclusions}
\label{Sect:OgleTransientsSummary}

\begin{figure}
  \centering
  \includegraphics[width=\columnwidth]{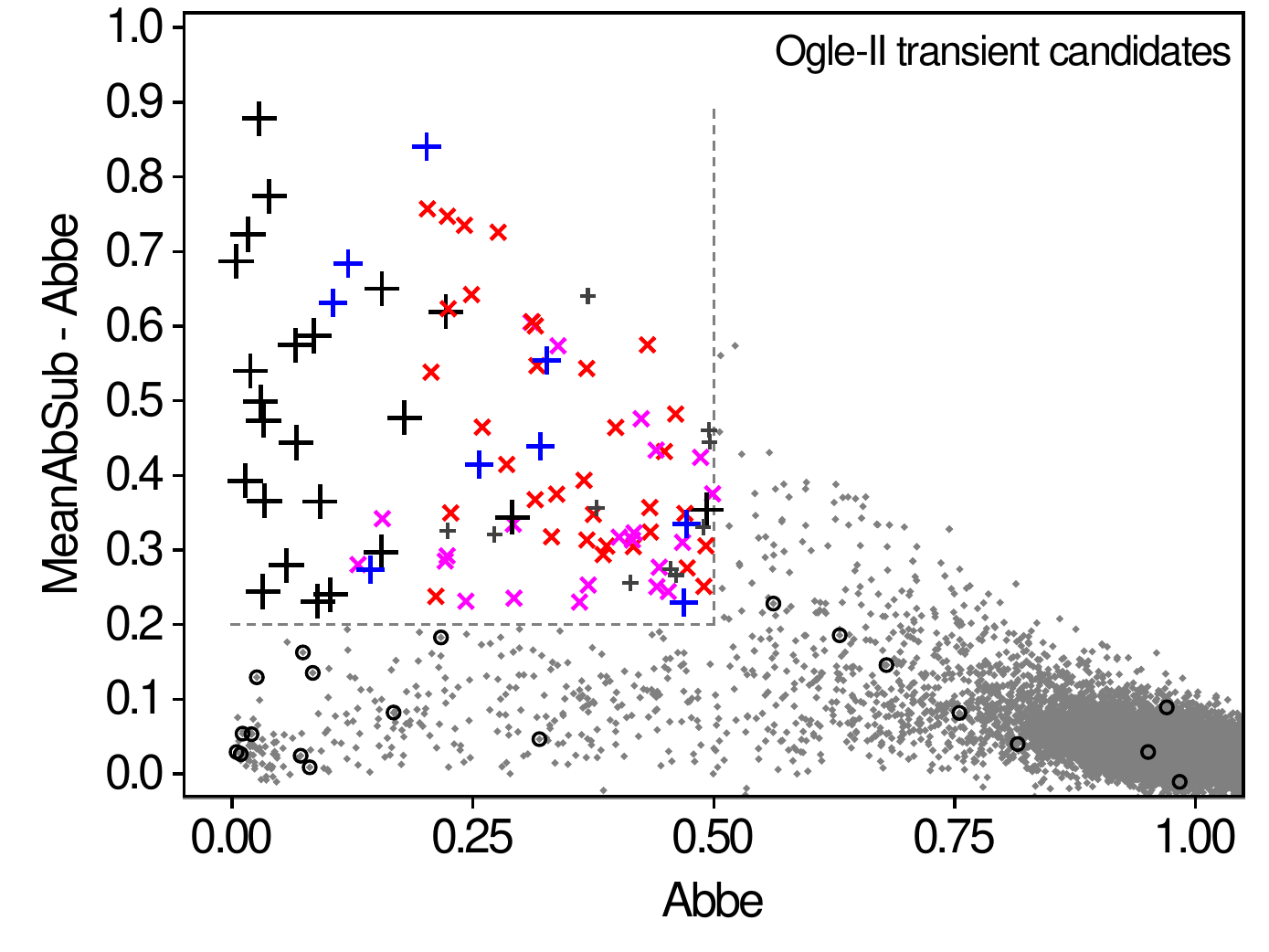}
  \caption{$\Ab$ versus $\excessAb$ diagram of OGLE-II transient candidates.
  Markers have the same meaning as in Fig.~\ref{Fig:skyOgleTransients}.
  The dashed lines delimit the region of transient candidates.
  OGLE-II sources falling out of this region are plotted in gray dots.
  Open circles identify non-transient OGLE-II sources that match EROS transient candidates.
   }
\label{Fig:OgleDiagramTransients}
\end{figure}

The results of the analysis of OGLE-II transient candidates is presented in Fig.~\ref{Fig:OgleDiagramTransients}.
They can be summarized as follows.

\begin{itemize}
\item All EROS transient candidates have an OGLE-II match (Table~\ref{Tab:erosTransients}).
This means that no EROS transient candidate is a spurious source.
\vskip 1mm
\item All EROS transients of type B (with bursts and/or outbursts) and T (with trend) are detected as transient candidates in OGLE-II as well (black plus markers in Fig.~\ref{Fig:OgleDiagramTransients}).
The only exception concerns a source with an outburst appearing outside the observation time frame of the OGLE-II survey.
This means that our transient detection procedure is efficient in detecting transients.
\vskip 1mm
\item All other EROS transient candidates that are not detected as such in OGLE-II turn out not to be transients; they are either of type P (pulsating-like) or X (not classifiable).
They are shown by black open circles in Fig.~\ref{Fig:OgleDiagramTransients}.
OGLE-II is successful in classifying them correctly, thanks to the higher precision of the measurements compared to EROS.
Data of good quality thus leads to less contamination of the region of transient candidates by non-transients sources.
\vskip 1mm
\item OGLE-II finds nine transient candidates that are not detected as such in their EROS counterparts (blue plus markers in Fig.~\ref{Fig:OgleDiagramTransients}).
Their failure to be detected as transient candidates in EROS is due to the larger uncertainties in this survey.
A good quality of the data is thus also essential in order not to miss real transients.
\vskip 1mm
\item Most OGLE-II transient candidates that have no EROS match are spurious sources (red and magenta crosses in Fig.~\ref{Fig:OgleDiagramTransients}), while the spuriousness of the remaining few (small dark gray plus markers in Fig.~\ref{Fig:OgleDiagramTransients}) should be ascertained as they might be real sources.
The distinction between these two groups is, however, irrelevant for the purpose of this study.
\vskip 1mm
\item Most spurious transient candidates in Fig.~\ref{Fig:OgleDiagramTransients} have $\Ab \gtrsim 0.2$.
\end{itemize}

\section{Comparison with known Be stars}
\label{Sect:literature}

Be stars are main-sequence B-type stars that show emission lines in their spectra.
They are known to undergo phases of bursts/outbursts that can last from days to years \citep[see][for a general review]{PorterRivinius03}.
\cite{Harmanec83} identifies four different timescales on which the photometric light of these objects varies, from ultra-rapid variations on the order of minutes to long-term variations on the order of years to tens of years.
The long- and medium-term (several days to months) variations lead to transient signatures in their light curves, which should be detected by our transient search procedure.

\cite{SabogalMennickentPietrzynski_etal05} (hereafter SMPG05) catalogued 2446 Be stars in the LMC based on OGLE-II data, of which 42 are in the \field{lm0013m} field of EROS considered in this paper.
Those authors subclassified Be stars into four types according to their light curves, following \cite{MennickentPietrzynskiGieren_etal02}.
Type~1 stars are characterized by outbursts.
Type~2 stars show sudden luminosity jumps in their light curves (from high to low or from low to high).
Type~3 stars show periodic or near-periodic variations.
Type~4 stars show stochastic variability on timescales running from days to years.

In this section I consider the 42 Be stars of SMPG05 that are in our field of study, and check them against the list of transient candidates found in the EROS light curves.
All type~1 and 2 Be stars from SMPG05 should have been identified as transient candidates by my method, but I do not expect so for their type~3 and type~4 Be stars.

I start in Sect.~\ref{Sect:BeStarsWithErosTransient} by summarizing all Be stars in SMPG05 that match the EROS transient candidates.
I then discuss in Sect.~\ref{Sect:BeStarsWithNoErosTransient} the ones that have not been identified by the transient extractor.
I finally investigate in Sect.~\ref{Sect:ErosTransientsWithNoBeStar} the EROS transient candidates that are not reported by SMPG05, and which could potentially be new Be stars.

\subsection{Detected Be stars}
\label{Sect:BeStarsWithErosTransient}

Among the 42 Be stars cataloged by SMPG05 that fall in the \field{lm0013m} field, 22 match an EROS transient candidate.
They are indicated in the last columns of Table~\ref{Tab:erosTransients}.

Fourteen of them are of type~1 or 2 in the classification schema used by SMPG05, i.e., characterized by outbursts or sudden luminosity jumps.
They are all classified as type B or T in my classification schema, as expected, expect one, star 23946 matching OGLE Be star 05263636-6932003, which is classified as type Pa (i.e., pulsating-like).
Its EROS light curves, shown in Fig.~\ref{Fig:lcsErosPa}, reveal a clear pulsating-like behavior on a timescale of about 1000 days.
The OGLE data, however, covers only 1.5 cycles of the pulsation due to the shorter duration of the OGLE survey compared to the EROS-2 survey.
This would explain why the star has been classified as type~2 by SMPG05 rather than type~3.
It must however be noted that this OGLE Be star has two identical entries in SMPG05, the duplicated entry being given type 4.

One of the Be stars with an EROS transient match is of type~3 in SMPG05, i.e., showing (almost) periodic variability.
The matching EROS transient candidate, star 27893, is consistently classified as type Pa in my classification schema.

The seven remaining Be stars are cataloged as type~4 in SMPG05.
The matching EROS transient candidates are of type T for three of them, of type P for another three (including the duplicated OGLE Be star 05263636-6932003 mentioned above), and of type X for the last one.
Since type~4 Be stars are stochastically variable, it is not surprising to have them classified as type~X or P in our classification schema.
That three of them are of type~T (stars 5852, 14119, and 28894) is a little unexpected, but probably highlights the difficulty of any classification schema, where confusion between different types may exist.

In summary, the 22 stars in common between the SMPG05 Be stars and the EROS transient candidates identified in this work are consistently classified in both studies.

\subsection{Missed Be stars}
\label{Sect:BeStarsWithNoErosTransient}

\begin{table*}
\centering
\caption{Be stars from \cite{SabogalMennickentPietrzynski_etal05} that do not match an EROS transient candidate in Table~\ref{Tab:erosTransients}.
Their EROS matches, if any, are given in the second group of columns (Cols. 3 and 4), and their OGLE-II matches in the third group of columns (Cols. 5 and 6).
EROS and OGLE-II transient candidates have their IDs put in bold.
Columns 4 and 6 give the distances of the Be star to the EROS and OGLE-II matches, respectively.
}
 \begin{tabular}{| l c | r r | r r | l |}
\hline
  Be         &  Be  & Eros &   dist   & Ogle &  dist    & Notes \\
 source name & Type &  ID  & (arcsec) &  ID  & (arcsec) &       \\
\hline
05250104-6926007 & 1 & 5330 &  0.61 & \textbf{113128} &  0.07 & EROS non-transient but OGLE-II transient (see Table~\ref{Tab:ogleTransientsWithErosMatch})\\
05250106-6926007 & 1 & 5330 &  0.72 & \textbf{113128} &  0.04 & Double of previous source?\\
05250647-6927562 & 1 & \textbf{11088} &  0.81 & \textbf{105856} &  0.13 & Double of source 05250642-6927563 in Table~\ref{Tab:erosTransients}?\\
05265275-6928508 & 2 & 14453 &  0.98 & \textbf{346243} &  0.04 & EROS non-transient but OGLE-II transient (see Table~\ref{Tab:ogleTransientsWithErosMatch})\\
05261048-6933284 & 4 & -- &  & 219804 &  0.47 & Would need confirmation\\
05261368-6934266 & 4 & -- &  & 212146 &  0.60 & Would need confirmation\\
05261860-6927448 & 4 & 10925 &  1.26 & 346433 &  0.31 & \\
05262269-6927098 & 4 & -- &  & 346287 &  0.62 & Would need confirmation\\
05263374-6924463 & 4 & 2227 &  0.63 & 352831 &  0.37 & \\
05263636-6932003 & 4 & \textbf{23946} &  1.05 & \textbf{339157} &  0.07 & Exact duplicate of source 05263636-6932003 in Table~\ref{Tab:erosTransients}\\
05263959-6929477 & 4 & 17230 &  1.07 & \textbf{346331} &  0.11 & EROS non-transient but OGLE-II transient (see Table~\ref{Tab:ogleTransientsWithErosMatch})\\
05263965-6924594 & 4 & 2846 &  0.76 & 352909 &  0.48 & \\
05250881-6925541 & 4 & -- &  & 113134 &  0.59 & Would need confirmation\\
05252514-6926243 & 4 & 6593 &  1.29 & 113055 &  0.52 & \\
05253221-6934396 & 4 & 31605 &  0.91 & 89893 &  0.15 & \\
05253248-6933238 & 4 & -- &  & 98113 &  0.64 & Would need confirmation\\
05253330-6925478 & 4 & 4878 &  0.57 & 113045 &  0.38 & \\
05255034-6934404 & 4 & -- &  & 212108 &  0.34 & Would need confirmation\\
05255060-6927327 & 4 & -- &  & 227136 &  0.55 & Would need confirmation\\
05255148-6931144 & 4 & 21298 &  0.97 & \textbf{220111} &  0.08 & EROS non-transient but OGLE-II transient (see Table~\ref{Tab:ogleTransientsWithErosMatch})\\
05255493-6927580 & 4 & -- &  & 227297 &  0.63 & Would need confirmation\\
05261038-6932021 & 4 & -- &  & 219917 &  0.48 & Would need confirmation\\
\hline
\end{tabular}
\label{Tab:beWithNoErosTransientMatch}
\end{table*}

Twenty Be stars reported by SMPG05
are not identified in the EROS light curves of variable stars by the transient search method, and here I investigate why.
They are listed in Table~\ref{Tab:beWithNoErosTransientMatch}, together with their EROS (non-transient) matches, if any, and their OGLE-II DIA matches used in this study.

Among the twenty Be stars from SMPG05 not detected as transient candidates in EROS, five are detected in OGLE-II (see Table~\ref{Tab:ogleTransientsWithErosMatch}).
As explained in Sect.~\ref{Sect:OgleTransientsWithErosMatches}, they are not identified as transient candidates in EROS because of the larger uncertainties in the EROS survey compared to the OGLE-II survey.
It must be noted that only three of them are of type~1 or 2, and that the two type~1 sources, 05250104-6926007 and 05250106-6926007, are most probably the same object in the sky.

Six other Be stars from SMPG05 are not detected as transient candidates in either EROS or OGLE-II, although matches are found in both databases.
They are EROS sources 2227, 2846, 4878, 6593, 10925, and 31605.
Their light curves, displayed in Fig.~\ref{Fig:NonTransientErosAndOgleMatchesToBeLcs} in Appendix~\ref{SectAppendix:notMatchingBeLcs}, do not show any obvious transient feature.
If the matches to the SMPG05 Be stars are correct, it would indeed have been very hard to identify them as transients from their EROS or OGLE-II DIA light curves.
It must be noted that SMPG05 classify those OGLE Be stars as of type~4.

The nine remaining SMPG05 Be stars listed in Table~\ref{Tab:beWithNoErosTransientMatch}, for which no matching star is found in the EROS catalog of variable stars, are all classified as type~4 by SMPG05.
Their OGLE-II light curves, displayed in Fig.~\ref{Fig:NonTransientOgleMatchesToBeLcs} in Appendix~\ref{SectAppendix:notMatchingBeLcs}, reveal no clear transient signatures, explaining why they are not identified by the transient extractor.
Discovering why they are not listed in the EROS catalog of variable stars, if they are indeed light curves of real stars, would require deeper investigation in the EROS data, which goes beyond the purpose of this paper. 

In summary, the automatic transient extractor presented in this paper did not miss any Be stars with detectable transient features in their light curves.

\subsection{New Be stars}
\label{Sect:ErosTransientsWithNoBeStar}

Twenty-two transient candidates listed in Table~\ref{Tab:erosTransients} are not present in the SMPG05 catalog of Be stars, four of which are of type~B or T.
Since these four stars might be new Be stars, I analyze them individually in this section, listing them by their EROS number in field \field{lm0013m}.

\begin{itemize}
\item \textbf{EROS star 6072} (type Ba) is a clear Be candidate (see Fig.~\ref{Fig:lcsErosBa}).
Its outburst was recorded in the EROS survey after the end of the OGLE-II survey and could thus not have been detected by SMPG05;
\vskip 2mm
\item \textbf{EROS star 19321} (type Ba) has an outburst feature in its light curves on a timescale of more than 2000 days (see Fig.~\ref{Fig:lcsErosBa}).
The fact that both EROS and OGLE-II record a similar light curve shape confirms its astrophysical origin.
The source has a negative $R-B$ color, however, and may thus have failed to pass the initial selection criterium by SMPG05, who considered only stars bluer than a given color limit.
It is probably not a Be star, but it is clearly a transient;
\vskip 2mm
\item \textbf{EROS star 22897} (type Bb) has clear outbursts on timescales of few hundred days (see Fig.~\ref{Fig:lcsErosBb}), but, like EROS star 19321, it may be too red to have been considered by SMPG05 as a Be star;
\vskip 2mm
\item \textbf{EROS star 27954} (type Bc) has a clear outburst during the last 1000 days of the EROS survey, with bursts superposed on the outburst (see Fig.~\ref{Fig:lcsErosBc}).
At least two of the bursts, just before and at the beginning of the outburst, are present in the OGLE light curve as well, but the outburst itself occurred after the end of the OGLE-II survey, probably explaining why the star has not been picked by SMPG05 as a Be star.
The subsequent outburst visible in the EROS data and the blue mean color of the source make it a very good Be star candidate.
\end{itemize}

In conclusion, four new transient candidates are found in this study, two Be stars with a very high probability (EROS stars 6072 and 27954), the two other Be star candidates needing confirmation due to their red colors (EROS stars 19321 and 22897).

\section{Conclusions}
\label{Sect:conclusions}

A new method has been presented in the first part of this paper to extract transient candidates in a survey from their light curves (Sects~\ref{Sect:method} and \ref{Sect:simulations}).
The method is based on the Abbe value $\Ab$ that estimates the smoothness of a light curve, and on the newly introduced excess Abbe value $\excessAb$ that estimates the regularity of the light curve variability pattern over the duration of the observation.

The basic tool is the $\diagram$ diagram introduced in Sect.~\ref{Sect:method}.
Based on simulated light curves presented in Sect.~\ref{Sect:simulations}, four regions have been identified in that diagram, depending on the type of variability present in the light curves.
They distinguish light curves with transient features, with pulsating-like features, with trends, and without any particular feature.
The regions are summarized in Fig.~\ref{Fig:summaryDiagram}.
In this paper, transient candidates have been defined with $\Ab<0.5$ and $\excessAb>0.2$.

A crucial parameter in the computation of $\excessAb$ is $\deltaTSub$.
The choice of its value is dictated by the timescales of the transient phenomena that are to be studied.
Typically, all transients with timescales greater than the value of this parameter will be extracted as transient candidates by the method, but the results will also be polluted by pulsating-like variables with periods larger than this value.
In this paper, we have taken $\deltaTSub=100$~d.

In the second part of the paper, the method has been successfully tested on a subset of EROS and OGLE-II surveys (Sects~\ref{Sect:EROS} and \ref{Sect:OGLE-II}, respectively) covering a $0.50^{\circ} \times 0.17^{\circ}$ field of the LMC centered at RA(J2000)=5h25m56.5s and DEC(J2000)=-69d29m43.3s.
Forty-three transient candidates have been extracted by the method in EROS (Table~\ref{Tab:erosTransients}), and nine more in OGLE-II (Table~\ref{Tab:ogleTransientsWithErosMatch}) thanks to the smaller measurement uncertainties in OGLE-II than in EROS.
The smaller uncertainties of OGLE-II measurements also lead to less pollution of the transient region in the $\diagram$ diagram by non-transients.

Sixty-six more transient candidates have been identified in OGLE-II, that have no EROS variable counterpart.
Most of them are spurious sources (Sect.~\ref{Sect:spuriousOgleTransients}), but the case of ten of them is unclear and would require further confirmation to assess their reliability (Sect.~\ref{Sect:OgleTransientsWithoutErosMatches}, Table~\ref{Tab:ogleTransientsWithNoErosMatch}).
As a result, the region of transient candidates in the OGLE $\diagram$ diagram is about twice as populated as in the EROS diagram.

The efficiency of the transient candidate extraction method was also tested in Sect.~\ref{Sect:literature} against known Be stars in the literature for the relevant field of the LMC.
The comparison has shown that all known Be stars with detectable bursts or outbursts in their light curves were successfully extracted by the method.
In addition, two new Be stars and two new transients that could be Be candidates have been identified.
Those four transients have not yet been reported in the literature, to the best of my knowledge.

In a massive use of the method on large-scale surveys, the method should be combined with a period search method to identify (and exclude) periodic variables from the list of transient candidates, and an automated classification method should be used to subclassify the later ones into transient subgroups like outbursting events, (multi-)bursting events, or events with trends.
The detailed study presented in this paper on a small sky area of the LMC can serve as a test case to validate automated subclassification methods.
Some discussion of attributes that can help such a subclassification is provided in Appendix~\ref{SectAppendix:CommentsSubClassification}.

Finally, the $\diagram$ diagram has been shown in Sects~\ref{Sect:simuTrendDiagnosticTool} and \ref{Sect:diagramSummary} to be a potentially powerful tool to check the data quality of a survey.
Data reduction problems or drifts of instrumental origin may lead to a high density of points in the trend region of the $\diagram$ diagram.
Inspection of the EROS $\diagram$ diagram constructed from the full light curves suggested the existence of such a data reduction problem in the EROS survey, which turns out to be confirmed in the literature (Sect.~\ref{Sect:ErosDiagramFullLcs}).

\begin{acknowledgements}

I thank the EROS team for providing the data base of variable stars. I am also very grateful to L. Rimoldini and M. S\"uveges for their careful reading and commenting of the paper.

\end{acknowledgements}

\bibliographystyle{aa}
\bibliography{bibTex}

\newpage
\begin{appendix}

\section{Analytical expression of the Abbe value for time series with trends}
\label{SectAppendix:AbbeWithTrend}

We consider a time series $\{t_i,y_i\}$ regularly sampled at $n$ times from time $t_1=-\frac{1}{2} \Delta T$ to $t_n= \frac{1}{2} \Delta T$.
These times are adopted to make the analytical calculations easier, but do not restrain the generality of the problem.
I assume that the time series has a linear trend of slope $a$,
\begin{equation}
  y_i = f_i + a \; t_i \;,
\label{Eq:tsDefinition}
\end{equation}
where $f_i$ are the detrended components of the time series.
I first compute the variance of the time series, then proceed to the calculation of its Abbe value.

$\bullet$
The variance $\sigma^2$ of $\{y_i\}$ can be decomposed into the variance $\sigma^2_f$ of its detrended component $\{f_i\}$ and the variance $a^2 \sigma^2_t$ of its trend component $\{a\,t_i\}$.
Assuming the detrended component to be uncorrelated with the trend component (a linear correlation is already excluded by construction), we have
\begin{equation}
  \sigma^2 = \sigma_f^2 + a^2 \sigma_t^2 \; .
\label{Eq:uncorrelatedVariances:0}
\end{equation}
The variance of the detrended component of the signal is unknown, but that of the trend can be calculated explicitly given the regularity of the time sampling.
Since
\begin{equation}
  \sigma_t^2 = \frac{1}{n} \sum_{i=1}^{n} \left( t_i - \bar{t} \right)^2
\label{Eq:variance(t_i):0}
\end{equation}
and knowing that, by construction, $t_i = [i-(n+1)/2] \; \delta t$ ($\delta t$ is the time step $t_{i+1}-t_i$) and $\bar{t}=0$, we have
\begin{equation}
  \sigma_t^2 = \frac{1}{n} (\delta t)^2
               \left[\sum_{i=1}^{n} i^2
                     + \sum_{i=1}^{n} \left( \frac{n+1}{2}\right)^2
                     - 2 \sum_{i=1}^{n} i \frac{(n+1)}{2}
               \right] \; .
\label{Eq:variance(t_i):1}
\end{equation}
The terms on the right-hand side of Eq.~\ref{Eq:variance(t_i):1} can respectively be developed as
\begin{equation}
\left\{
\begin{array}{lcl}
  \displaystyle \sum_{i=1}^{n} i^2 & = & \displaystyle \frac{1}{6} \, n \, (n+1) \, (2n+1) \\
  \displaystyle \sum_{i=1}^{n} \left( \frac{n+1}{2}\right)^2 & = & \displaystyle \frac{1}{4} \, n \, (n+1)^2\\
  \displaystyle \sum_{i=1}^{n} i \, (n+1) & = & \displaystyle \frac{1}{2} \, n \, (n+1)^2 \; ,
\end{array}
\right.
\label{Eq:sumsInVariance(t_i)}
\end{equation}
and Eq.~\ref{Eq:variance(t_i):1} becomes
\begin{equation}
  \sigma_t^2 = \frac{1}{12} \, (\delta t)^2 \, n^2 = \frac{1}{12} \, (\Delta T)^2 \; .
\label{Eq:variance(t_i):2}
\end{equation}
We can now write the variance of the time series (Eq.~\ref{Eq:uncorrelatedVariances:0}), using Eq.~\ref{Eq:variance(t_i):2}, as
\begin{eqnarray}
  \sigma^2 & = & \sigma_f^2 + \frac{1}{12}\; (a \,\Delta T)^2 \nonumber\\
           & = & \sigma_f^2 \; (1+\beta) \;\;,
\label{Eq:uncorrelatedVariances:1}
\end{eqnarray}
where I have introduced the parameter
\begin{equation}
  \beta = \frac{1}{12} \frac{(a \, \Delta T)^2}{\sigma_f^2} \;,
\label{Eq:beta}
\end{equation}
which quantifies the amplitude of the trend ($a\,\Delta T$) relative to the variability level of the detrended signal ($\sigma_f$).

$\bullet$
We can now compute the Abbe value of $\{y_i\}$, given by
\begin{equation}
  \Ab = \frac{1}{2\,\sigma^2} \; \frac{1}{n-1} \; 
        \sum_{i=1}^{n-1} \; [(f_{i+1}-f_i) + a \, (t_{i+1}-t_i)]^2 \; .
  \label{Eq:AbbeWithTrend:0}
\end{equation}
The right-hand side of Eq.~\ref{Eq:AbbeWithTrend:0}, after development of the square, has three sums which equal, respectively,
\begin{equation}
\left\{
\begin{array}{lcl}
  \displaystyle \sum_{i=1}^{n-1} (f_{i+1}-f_i)^2 & = & \displaystyle 2 (n-1) \, \sigma_f^2 \, \Ab_f \\
  \displaystyle \sum_{i=1}^{n-1} a^2 \, (t_{i+1}-t_i)^2 & = & \displaystyle a^2 \, (n-1) \, (\delta t)^2 \\
  \displaystyle \sum_{i=1}^{n-1} 2a \, (f_{i+1}-f_i) \, (t_{i+1}-t_i) & = & \displaystyle 2a \, \delta t \, (f_n-f_1) \; .
\end{array}
\right.
\label{Eq:sumsInAbbeWithTrend}
\end{equation}
Using these expressions, Eq.~\ref{Eq:AbbeWithTrend:0} becomes
\begin{equation}
  \Ab = \frac{1}{2\,\sigma^2} \; 
        \left[ 2\, \sigma_f^2 \, \Ab_f + a^2 \, (\delta t)^2 + 2a \, \delta t \, \frac{(f_n-f_1)}{(n-1)} \right] \; .
  \label{Eq:AbbeWithTrend:1}
\end{equation}
We can neglect the third term on the right-hand side of this equation, knowing that $\{f_i\}$ is the detrended component of the signal (and hence $f_n \simeq f_1$) and that the number of points is expected to be large ($n\gg 1$).
Equation~\ref{Eq:AbbeWithTrend:1} then becomes, with $\beta$ given by Eq.~\ref{Eq:beta},
\begin{equation}
  \Ab \simeq \frac{\sigma_f^2}{\sigma^2} \; 
        \left( \Ab_f + 6 \, \frac{\beta}{n^2} \right) \; .
  \label{Eq:AbbeWithTrend:2}
\end{equation}

$\bullet$
We can now combine Eqs.~\ref{Eq:uncorrelatedVariances:1} and \ref{Eq:AbbeWithTrend:2} to obtain the final approximate expression
\begin{equation}
  \Ab \simeq \frac{\Ab_f + 6 \, \beta / n^2}{1 + \beta}
  \label{Eq:AbbeWithTrend:3}
\end{equation}
of the Abbe value of the time series.
A discussion of the equation is given in Sect.~\ref{Sect:simuTrendAnalytical} in the body of the paper.

\clearpage
\section{Light curves of EROS transient candidates}
\label{SectAppendix:ErosTransients}

\begin{figure*}
  \centering
  \includegraphics[width=1.00\columnwidth]{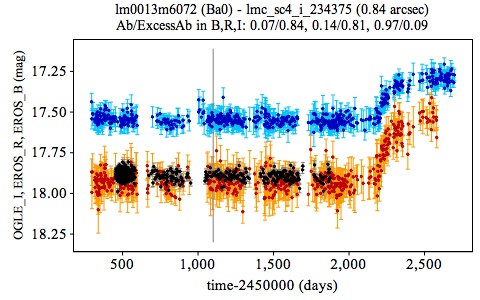}
  \includegraphics[width=1.00\columnwidth]{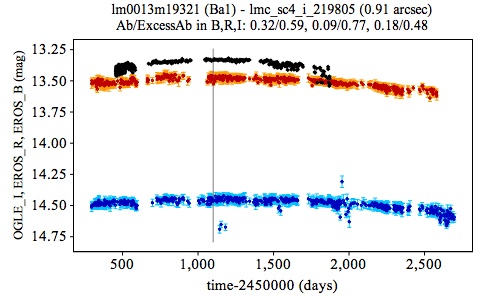}
  \includegraphics[width=1.00\columnwidth]{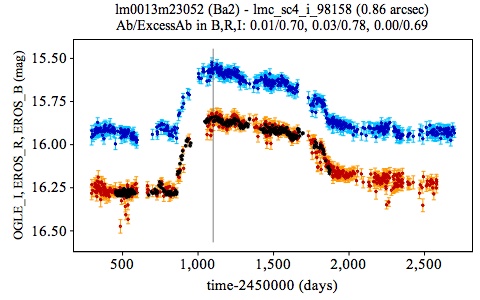}
  \includegraphics[width=1.00\columnwidth]{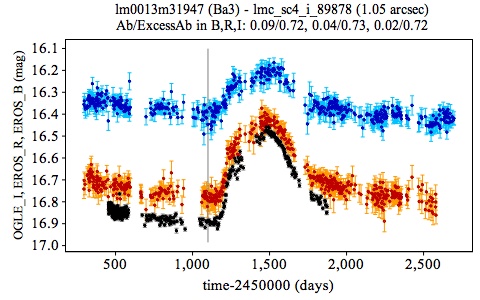}
  \caption{Light curves of EROS transient candidates with outbursts (type Ba) in blue/red for the $\EROSB$/$\EROSR$ photometric band.
  The light curve of the matching OGLE-II source is superposed in black.
  The vertical gray line locates the starting time (1100~d) of partial light curves considered in the paper.
   }
\label{Fig:lcsErosBa}
\end{figure*}

\begin{figure*}
  \centering
  \includegraphics[width=1.00\columnwidth]{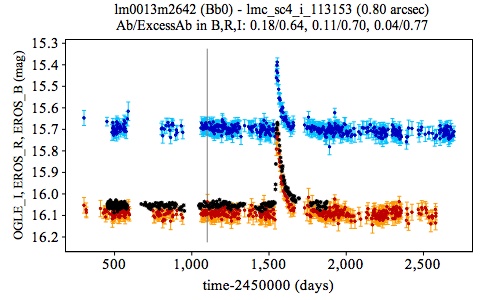}
  \includegraphics[width=1.00\columnwidth]{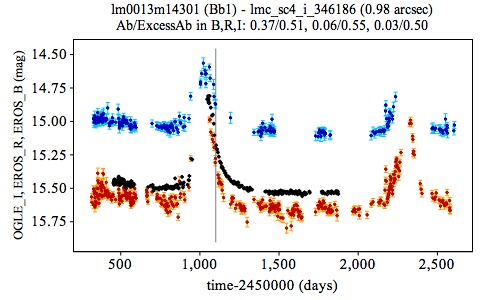}
  \includegraphics[width=1.00\columnwidth]{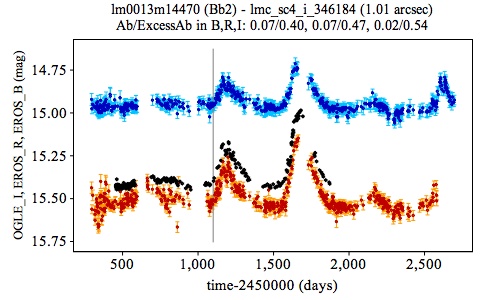}
  \includegraphics[width=1.00\columnwidth]{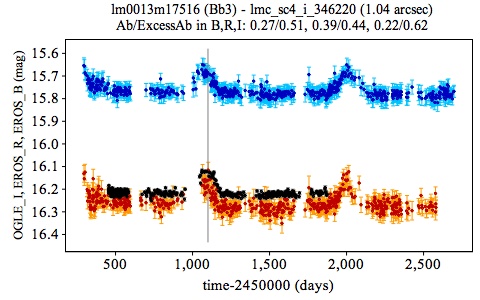}
  \includegraphics[width=1.00\columnwidth]{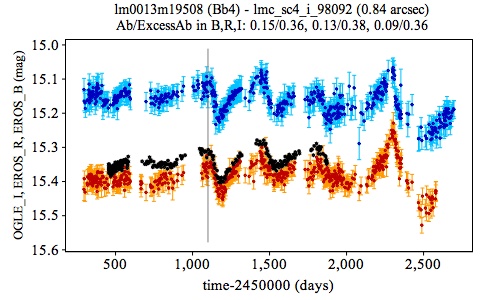}
  \includegraphics[width=1.00\columnwidth]{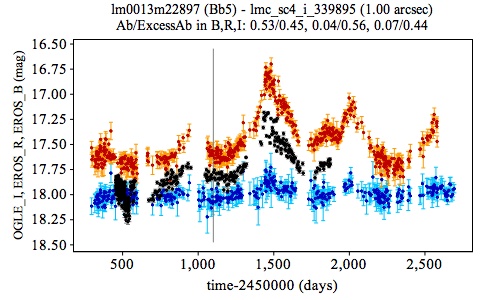}
  \caption{Same as Fig.~\ref{Fig:lcsErosBa}, but for transient candidates with burst-type feature(s) in their light curves (type Bb).}
\label{Fig:lcsErosBb}
\end{figure*}

\begin{figure*}
  \centering
  \includegraphics[width=1.00\columnwidth]{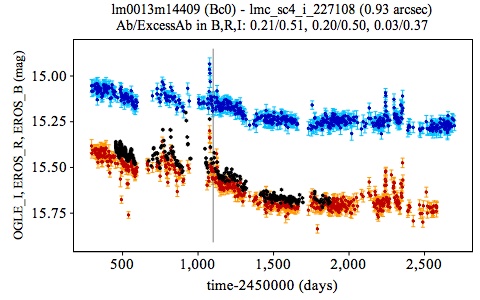}
  \includegraphics[width=1.00\columnwidth]{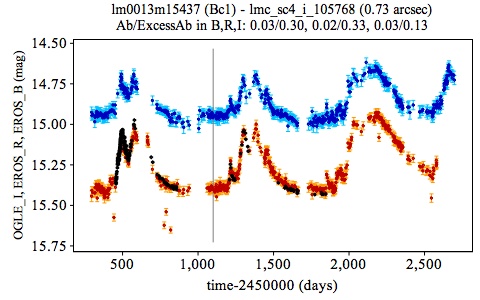}
  \includegraphics[width=1.00\columnwidth]{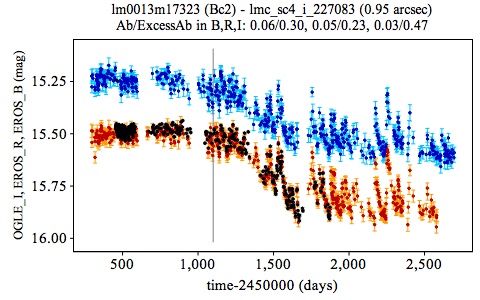}
  \includegraphics[width=1.00\columnwidth]{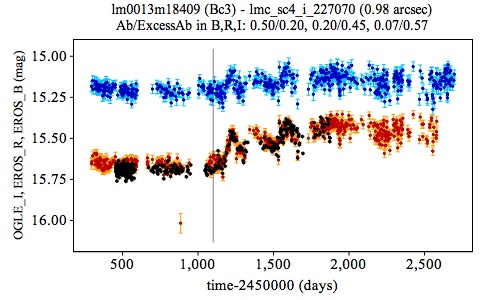}
  \includegraphics[width=1.00\columnwidth]{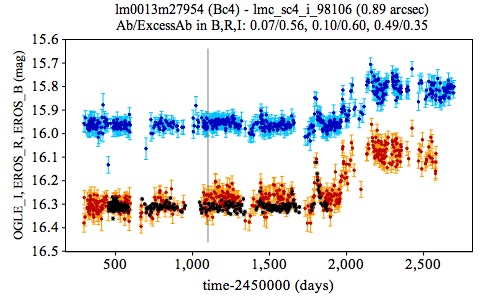}
  \caption{Same as Fig.~\ref{Fig:lcsErosBa}, but for transient candidates displaying both outburst- and burst-type features in their light curves (type Bc).}
\label{Fig:lcsErosBc}
\end{figure*}

\begin{figure*}
  \centering
  \includegraphics[width=1.00\columnwidth]{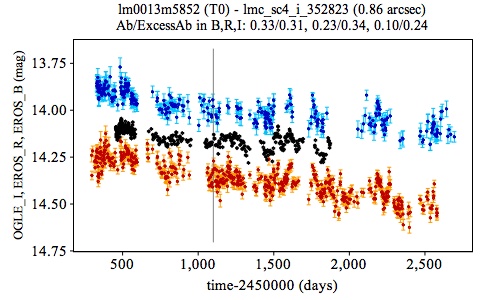}
  \includegraphics[width=1.00\columnwidth]{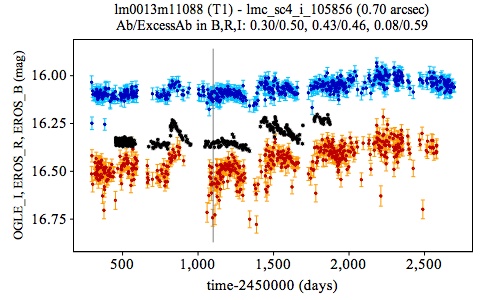}
  \includegraphics[width=1.00\columnwidth]{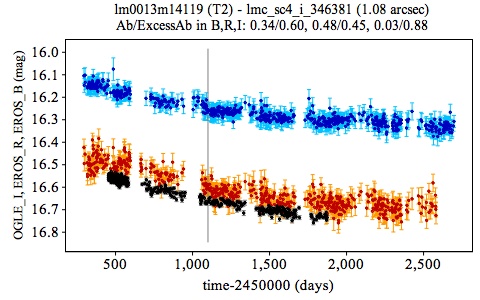}
  \includegraphics[width=1.00\columnwidth]{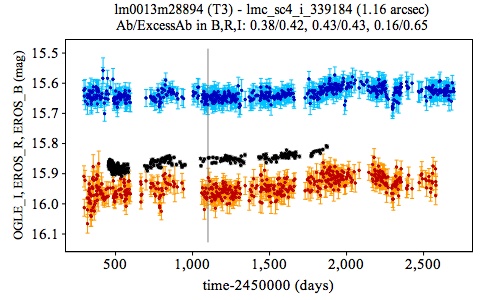}
  \caption{Same as Fig.~\ref{Fig:lcsErosBa}, but for transient candidates displaying trends in their light curves (type T).}
\label{Fig:lcsErosT}
\end{figure*}

\begin{figure*}
  \centering
  \includegraphics[width=1.00\columnwidth]{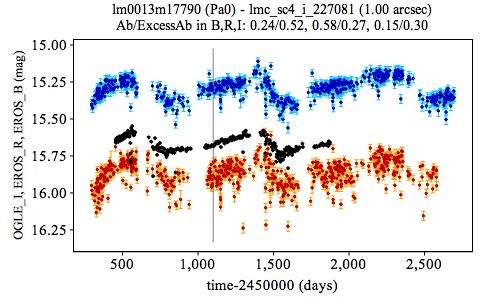}
  \includegraphics[width=1.00\columnwidth]{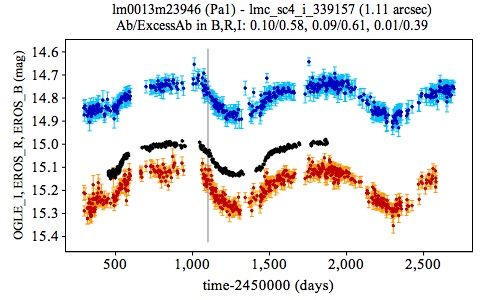}
  \includegraphics[width=1.00\columnwidth]{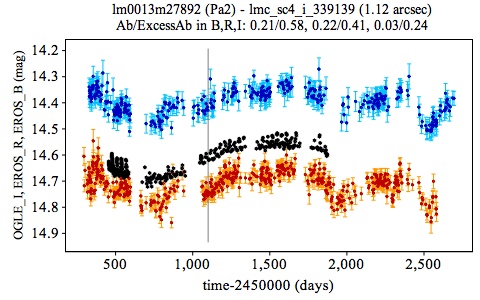}
  \caption{Same as Fig.~\ref{Fig:lcsErosBa}, but for sources with pulsating-like features of type Pa in their light curves.}
\label{Fig:lcsErosPa}
\end{figure*}

\begin{figure*}
  \centering
  \includegraphics[width=1.00\columnwidth]{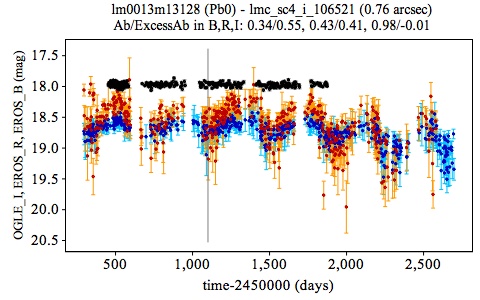}
  \includegraphics[width=1.00\columnwidth]{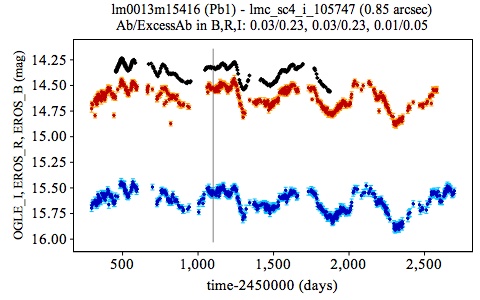}
  \includegraphics[width=1.00\columnwidth]{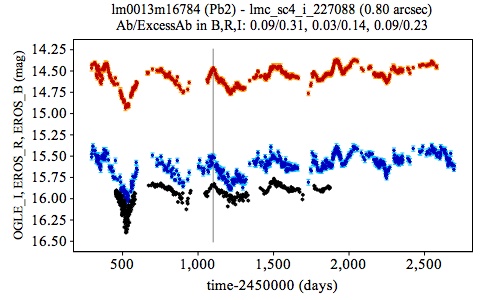}
  \includegraphics[width=1.00\columnwidth]{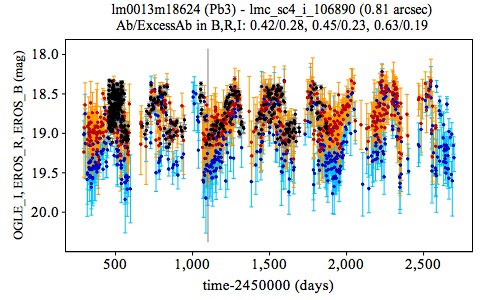}
  \includegraphics[width=1.00\columnwidth]{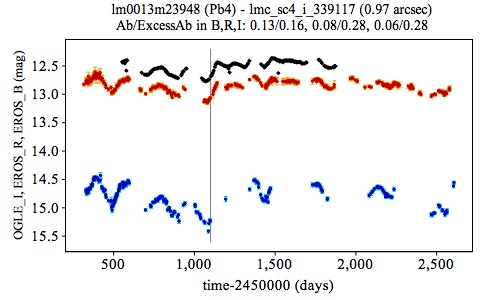}
  \includegraphics[width=1.00\columnwidth]{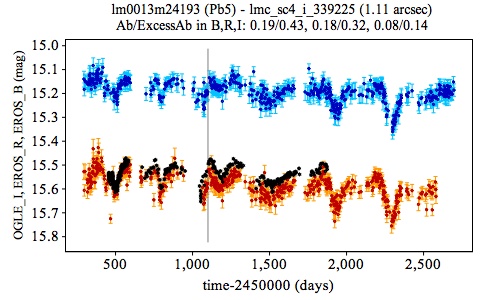}
  \includegraphics[width=1.00\columnwidth]{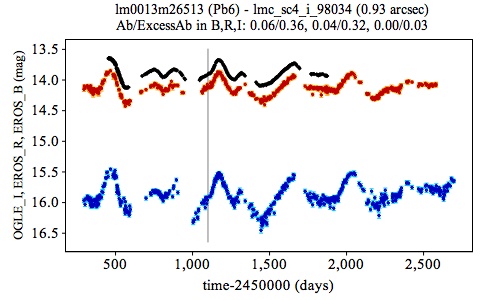}
  \includegraphics[width=1.00\columnwidth]{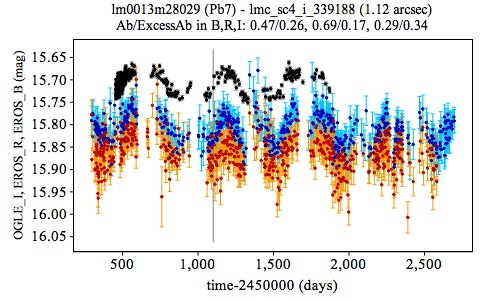}
  \caption{Same as Fig.~\ref{Fig:lcsErosBa}, but for sources with pulsating-like features of type Pb in their light curves.}
\label{Fig:lcsErosPb}
\end{figure*}

\addtocounter{figure}{-1}
\begin{figure*}
  \centering
  \includegraphics[width=1.00\columnwidth]{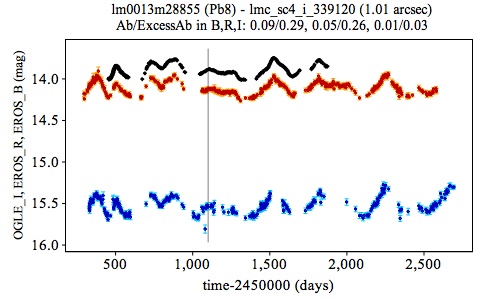}
  \includegraphics[width=1.00\columnwidth]{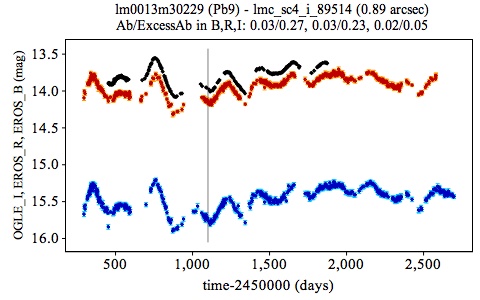}
  \caption{Continued.}
\end{figure*}

\begin{figure*}
  \centering
  \includegraphics[width=1.00\columnwidth]{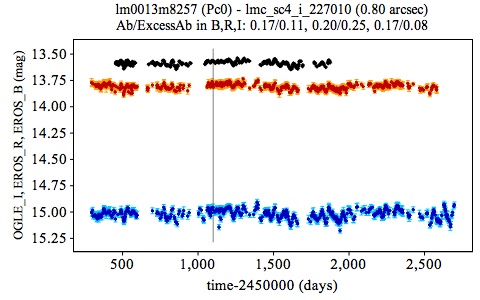}
  \includegraphics[width=1.00\columnwidth]{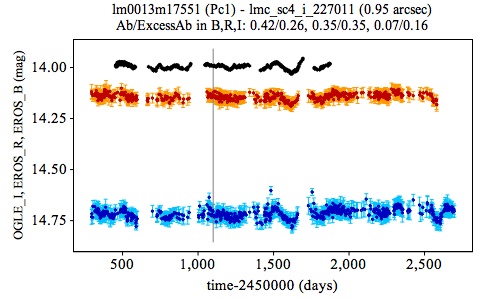}
  \includegraphics[width=1.00\columnwidth]{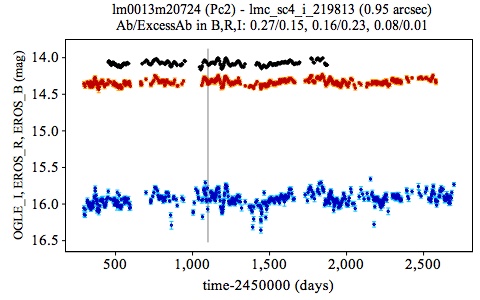}
  \includegraphics[width=1.00\columnwidth]{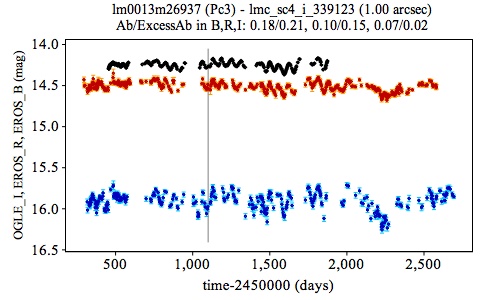}
  \includegraphics[width=1.00\columnwidth]{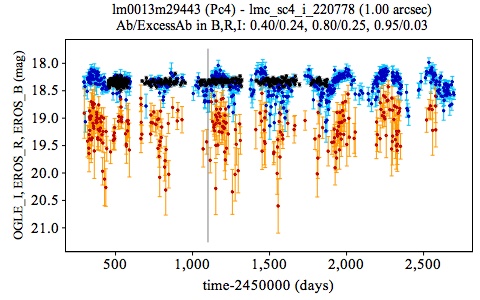}
  \caption{Same as Fig.~\ref{Fig:lcsErosBa}, but for sources with pulsating-like features of type Pc in their light curves.}
\label{Fig:lcsErosPc}
\end{figure*}

\begin{figure*}
  \centering
  \includegraphics[width=1.00\columnwidth]{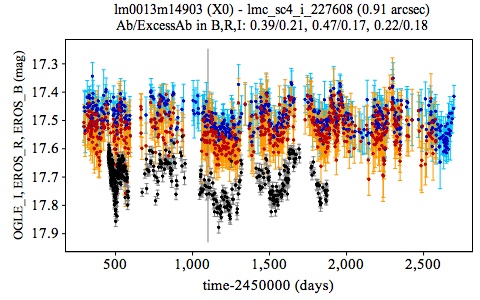}
  \includegraphics[width=1.00\columnwidth]{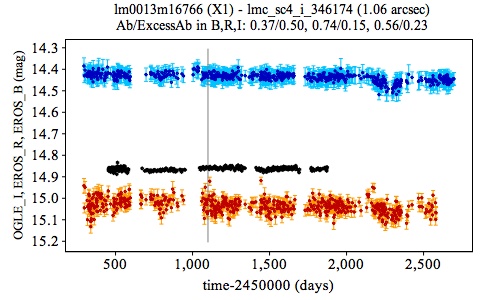}
  \includegraphics[width=1.00\columnwidth]{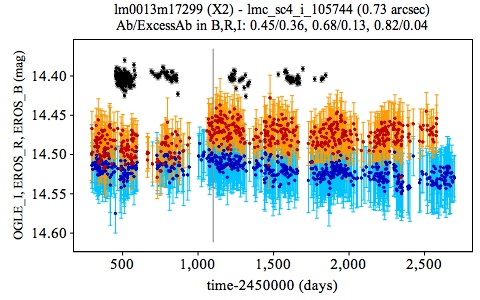}
  \includegraphics[width=1.00\columnwidth]{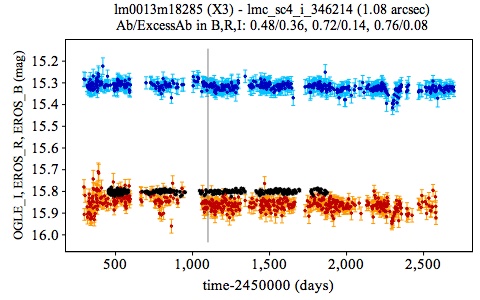}
  \includegraphics[width=1.00\columnwidth]{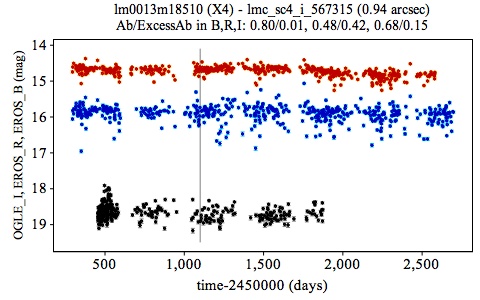}
  \includegraphics[width=1.00\columnwidth]{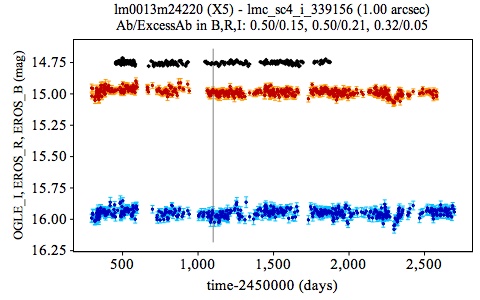}
  
  \caption{Same as Fig.~\ref{Fig:lcsErosBa}, but for transient candidates of type X, i.e., that could not be classified as either type B, P, or T.}
\label{Fig:lcsErosX}
\end{figure*}

The light curves of all EROS transient candidates listed in Table~\ref{Tab:erosTransients} in the body of the paper are shown in Figs.~\ref{Fig:lcsErosBa} to \ref{Fig:lcsErosX}, superposed on the light curves of their OGLE-II matches.
Transient candidates of type~B are shown in Figs.~\ref{Fig:lcsErosBa} (type~Ba), \ref{Fig:lcsErosBb} (type~Bb), and \ref{Fig:lcsErosBc} (type~Bc), while those of type~T are shown in Fig.~\ref{Fig:lcsErosT}.
Transient candidates of type~P are shown in Figs.~\ref{Fig:lcsErosPa} (type~Pa), \ref{Fig:lcsErosPb} (type~Pb), and \ref{Fig:lcsErosPc} (type~Pc).
Finally, those of type~X are shown in Fig. \ref{Fig:lcsErosX}.

\clearpage
\section{Suspicious OGLE-II transient candidates}
\label{SectAppendix:suspiciousOgleTransients}

The suspicious OGLE-II transient candidates identified in Sect.~\ref{Sect:spuriousOgleTransients} are listed in Tables~\ref{Tab:ogleSpuriousTransientsWithErosTransients} to \ref{Tab:ogleSpuriousTransientsFromLcsWithoutEros}.

\begin{figure*}
  \centering
  \includegraphics[width=1.00\columnwidth]{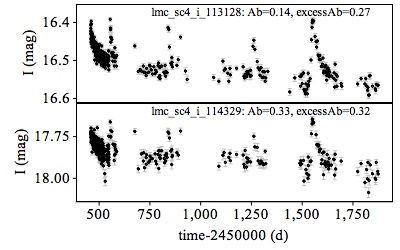}
  \includegraphics[width=1.00\columnwidth]{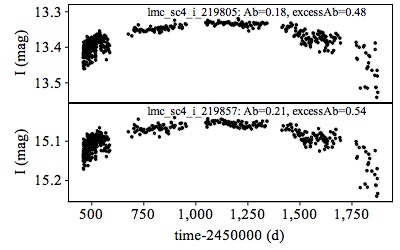}
  \includegraphics[width=1.00\columnwidth]{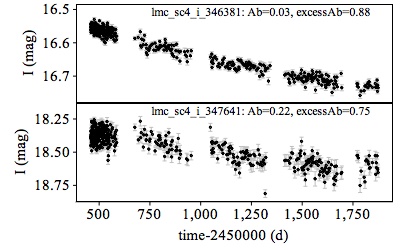}
  \caption{OGLE-II ghost transient candidates (lower panels) whose light curves are contaminated by a confirmed OGLE-II transient candidate (upper panels).
   }
\label{Fig:lcsContaminatedOgle}
\end{figure*}

Table \ref{Tab:ogleSpuriousTransientsWithErosTransients} lists OGLE-II transient candidates that turn out to be ghost light curves contaminated by other OGLE-II transient candidates.
That the former ones are ghosts of the latter ones is attested by the similarity of their light curves and by the fact that the former ones do not have an EROS match while the latter ones do.
The contaminated stars are indicated as such in Table~\ref{Tab:ogleSpuriousTransientsWithErosTransients}, together with the OGLE-II starId in field \texttt{LMC\_SC4} of the contaminating star and the angular distance to it.
The light curves of both contaminating and contaminated stars are shown in Fig.~\ref{Fig:lcsContaminatedOgle}.
It must be noted that the light curves of the two stars in the pair 219857-219805 are distant from one another by 6.55 arcsec (Table~\ref{Tab:ogleSpuriousTransientsWithErosTransients}), which is rather large.
Yet, the light curve variabilities of the two are very much correlated, as can be attested visually from Fig.~\ref{Fig:lcsContaminatedOgle}.

Table \ref{Tab:ogleSpuriousTransientsWithErosNonTransients} also lists ghost OGLE-II transient candidates, but for sources contaminated by the light curve of non-transient OGLE-II sources matching non-transient EROS sources.
The light curves of some of them are shown in Fig.~\ref{Fig:lcsOgleGhostsWithErosMatch}.

Finally, Table~\ref{Tab:ogleSpuriousTransientsFromLcsWithoutEros} lists suspicious OGLE-II transient candidates that do not match any EROS variable source.
Some of their light curves are shown in Figs.~\ref{Fig:lcsSuspiciousOgle} and \ref{Fig:lcsClumpOgle}.
Figure~\ref{Fig:lcsSuspiciousOgle} displays transient candidates that show suspicious similarities in their light curves, pointing to some contamination effects among them.
Figure~\ref{Fig:lcsClumpOgle} displays transient candidates that abnormally fall in a clump on the sky.
The brightest star in each group of such stars is listed first in Table~\ref{Tab:ogleSpuriousTransientsFromLcsWithoutEros}, with ``\textsl{Origin of variability?}'' written in the last column and the related stars listed below it; the angular distance to the brightest star is given in parentheses.

\begin{table}
\centering
\caption{OGLE-II ghost transient candidates contaminated by an OGLE-II transient candidate having an EROS transient candidate match.
}
 \begin{tabular}{l c@{~}c l}
\hline
 OgleId & \Ab(I) & \excessAb(I) & Notes \\
\hline
114329 &  0.33 &  0.32 & Contaminated by 113128 (at 0.83")\\
219857 &  0.21 &  0.54 & Contaminated by 219805 (at 6.55")\\
347641 &  0.22 &  0.75 & Contaminated by 346381 (at 1.06")\\
\hline
\end{tabular}
\label{Tab:ogleSpuriousTransientsWithErosTransients}
\end{table}

\begin{table}
\centering
\caption{OGLE-II ghost transient candidates (Cols. 1 and 2) contaminated by a non-transient OGLE-II source (Cols. 3 and 4) matching a non-transient EROS source (Col. 5).
}
 \begin{tabular}{l c@{~}c l c@{~}c l}
\hline
 OgleId & \Ab(I) & \excessAb(I) &  OgleId & \Ab(I) & \excessAb(I) & ErosId \\
\hline
90672 &  0.42 &  0.48 & 97168 &  0.77 &  0.14 & 30865 \\
98216 &  0.24 &  0.23 & 98036 &  0.01 &  0.03 & 27757 \\
98602 &  0.29 &  0.24 & 98039 &  0.02 &  0.02 & 26609 \\
98696 &  0.37 &  0.25 & 98042 &  0.09 &  0.08 & 25502 \\
98976 &  0.50 &  0.38 & 98046 &  0.01 &  0.06 & 21180 \\
107523 &  0.22 &  0.29 & 105752 &  0.01 &  0.02 & 11185 \\
212943 &  0.45 &  0.24 & 212107 &  0.02 &  0.04 & 31856 \\
227485 &  0.42 &  0.32 & 227026 &  0.01 &  0.04 & 16784 \\
228240 &  0.40 &  0.32 & 227025 &  0.02 &  0.04 & 16944 \\
228268 &  0.22 &  0.29 & 227200 &  0.47 &  0.34 & 16631 \\
234373 &  0.44 &  0.25 & 233964 &  0.02 &  0.04 & 6089 \\
234107 &  0.42 &  0.31 & 233964 &  0.02 &  0.04 & 6089 \\
333668 &  0.44 &  0.28 & 331603 &  0.01 &  0.03 & 31003 \\
335283 &  0.13 &  0.28 & 331603 &  0.01 &  0.03 & 31003 \\
340683 &  0.34 &  0.57 & 339117 &  0.06 &  0.28 & 23948 \\
339832 &  0.47 &  0.31 & 339117 &  0.06 &  0.28 & 23948 \\
339860 &  0.29 &  0.33 & 339117 &  0.06 &  0.28 & 23948 \\
340684 &  0.31 &  0.61 & 339117 &  0.06 &  0.28 & 23948 \\
340691 &  0.49 &  0.42 & 339117 &  0.06 &  0.28 & 23948 \\
566347 &  0.44 &  0.43 & 339117 &  0.06 &  0.28 & 23948 \\
347036 &  0.36 &  0.23 & 346157 &  0.03 &  0.03 & 10530 \\
505630 &  0.16 &  0.34 & 89514 &  0.02 &  0.05 & 30229 \\
\hline
\end{tabular}
\label{Tab:ogleSpuriousTransientsWithErosNonTransients}
\end{table}

\begin{table}
\centering
\caption{Potentially contaminated OGLE-II transient candidates with no EROS match.
}
 \begin{tabular}{l c@{~}c l}
\hline
 OgleId & \Ab(I) & \excessAb(I) & Note \\
\hline
226163 &  0.25 &  0.64 & Related to transient 219805 (at 1.34")?\\
223374 &  0.31 &  0.61 & Related to 226163 (at 2.49")\\
223387 &  0.22 &  0.62 & Related to 226163 (at 3.00")\\
535971 &  0.26 &  0.46 & Related to 226163 (at 1.23")\\
227864 &  0.23 &  0.35 & Origin of variability? \\
228882 &  0.21 &  0.24 & Related to 227864 (at 4.00")\\
230256 &  0.47 &  0.28 & Related to 227864 (at 4.79")\\
347128 &  0.24 &  0.74 & Origin of variability?  -- \textit{in clump}\\
348078 &  0.28 &  0.73 & Related to 347128 (at 1.99") -- \textit{in clump}\\
348089 &  0.32 &  0.55 & Related to 347128 (at 2.58") -- \textit{in clump}\\
348105 &  0.37 &  0.54 & Origin of variability?  -- \textit{in clump}\\
349468 &  0.31 &  0.60 & Related to 348105 (at 3.52") -- \textit{in clump}\\
354463 &  0.43 &  0.57 & Related to 348105 (at 6.49") -- \textit{in clump}\\
354491 &  0.20 &  0.76 & Related to 348105 (at 14.17") -- \textit{in clump}\\
568282 &  0.37 &  0.39 & Related to 348105 (at 14.95") -- \textit{in clump}\\
353696 &  0.46 &  0.48 & Origin of variability?  -- \textit{in clump}\\
568225 &  0.45 &  0.43 & Related to 353696 (at 15.13") -- \textit{in clump}\\
534576 &  0.43 &  0.32 & Origin of variability? \\
534577 &  0.42 &  0.30 & Related to 534576 (at 1.03")\\
534578 &  0.43 &  0.36 & Related to 534576 (at 0.77")\\
568223 &  0.38 &  0.35 & Origin of variability?  -- \textit{in clump}\\
568231 &  0.29 &  0.41 & Related to 568223 (at 1.22") -- \textit{in clump}\\
568233 &  0.37 &  0.31 & Related to 568223 (at 1.12") -- \textit{in clump}\\
568234 &  0.40 &  0.46 & Related to 568223 (at 13.23") -- \textit{in clump}\\
568260 &  0.39 &  0.31 & Related to 568223 (at 1.36") -- \textit{in clump}\\
568262 &  0.39 &  0.29 & Related to 568223 (at 1.84") -- \textit{in clump}\\
568263 &  0.49 &  0.31 & Related to 568223 (at 4.54") -- \textit{in clump}\\
568267 &  0.31 &  0.37 & Related to 568223 (at 2.33") -- \textit{in clump}\\
568268 &  0.34 &  0.37 & Related to 568223 (at 3.45") -- \textit{in clump}\\
568269 &  0.49 &  0.25 & Related to 568223 (at 7.31") -- \textit{in clump}\\
568570 &  0.47 &  0.35 & Origin of variability?  -- \textit{in clump}\\
\hline
\end{tabular}
\label{Tab:ogleSpuriousTransientsFromLcsWithoutEros}
\end{table}

\begin{figure*}
  \centering
  \includegraphics[width=0.90\columnwidth]{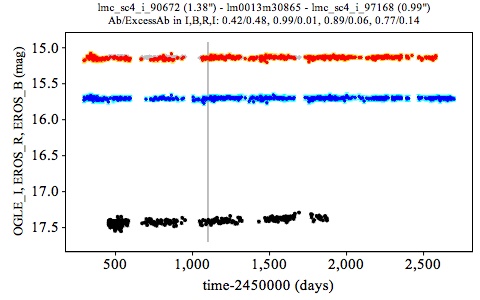}
  \includegraphics[width=0.90\columnwidth]{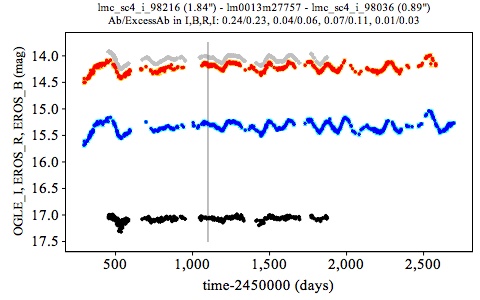}
  \includegraphics[width=0.90\columnwidth]{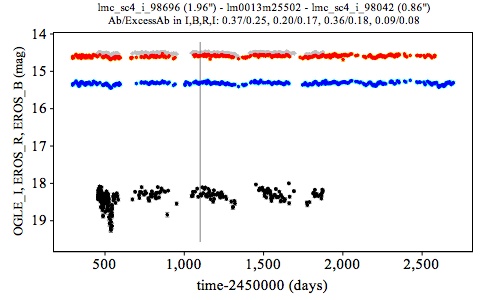}
  \includegraphics[width=0.90\columnwidth]{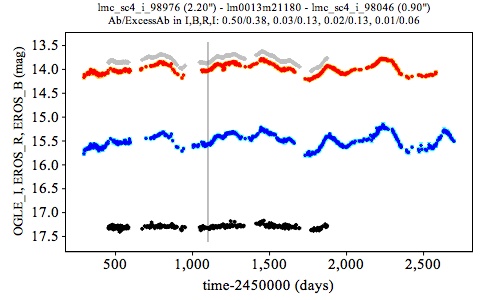}
  \includegraphics[width=0.90\columnwidth]{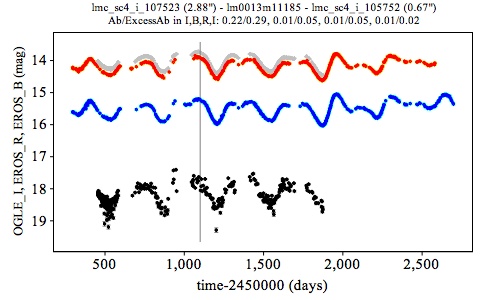}
  \includegraphics[width=0.90\columnwidth]{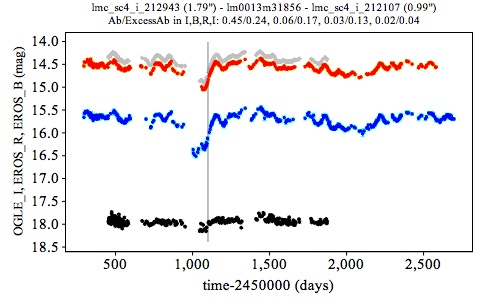}
  \includegraphics[width=0.90\columnwidth]{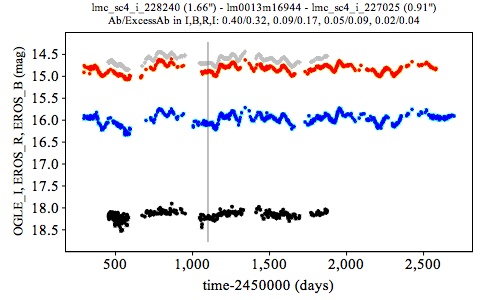}
  \includegraphics[width=0.90\columnwidth]{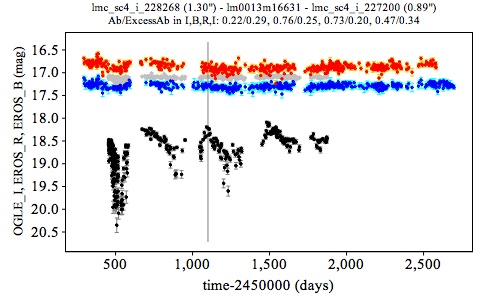}
  \caption{OGLE-II ghost transient candidates (black) whose light curves are contaminated by a non-transient OGLE-II source (gray) that match a non-transient EROS source (blue and red).
   }
\label{Fig:lcsOgleGhostsWithErosMatch}
\end{figure*}

\addtocounter{figure}{-1}
\begin{figure*}
  \centering
  \includegraphics[width=0.90\columnwidth]{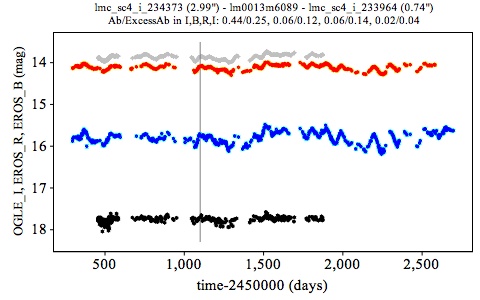}
  \includegraphics[width=0.90\columnwidth]{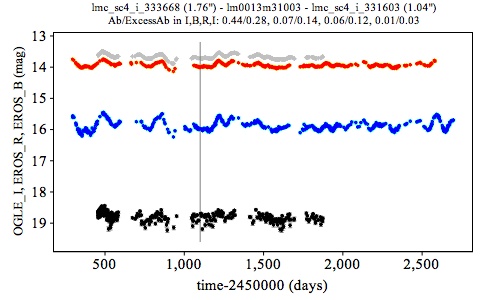}
  \includegraphics[width=0.90\columnwidth]{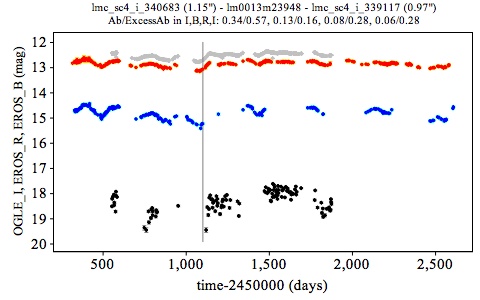}
  \includegraphics[width=0.90\columnwidth]{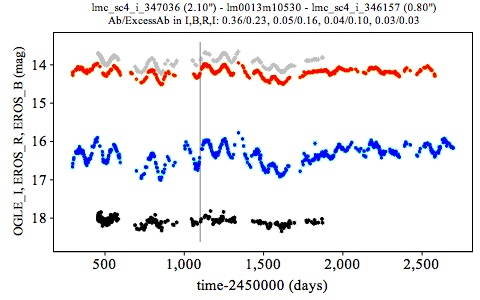}
  \includegraphics[width=0.90\columnwidth]{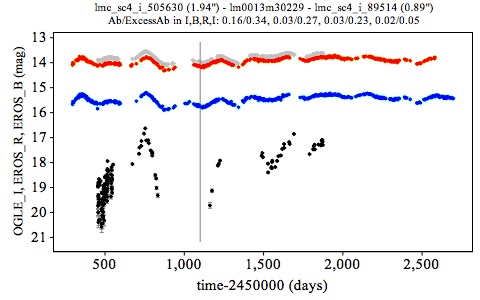}
  \caption{Continued.
   }
\end{figure*}

\begin{figure*}
  \centering
  \begin{tabular}{c}
  \includegraphics[width=0.95\columnwidth]{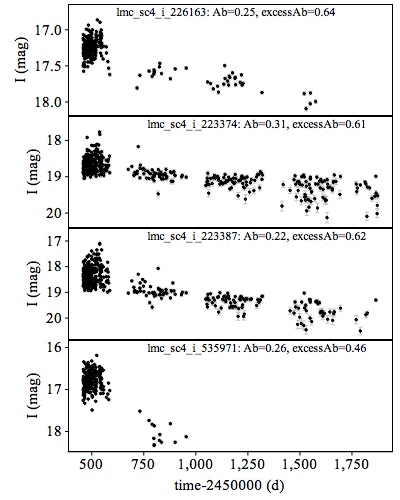}
  \end{tabular}
  \begin{tabular}{c}
    \includegraphics[width=0.95\columnwidth]{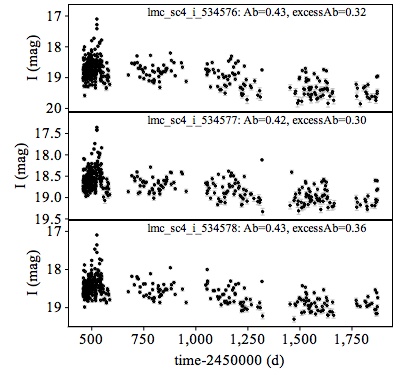} \\
    \includegraphics[width=0.95\columnwidth]{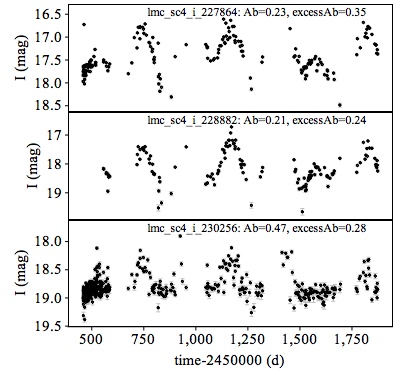}
  \end{tabular}
  \caption{Suspicious OGLE-II transient candidates whose light curves display similar patterns and for which no EROS counterpart is found.
   }
\label{Fig:lcsSuspiciousOgle}
\end{figure*}

\begin{figure*}
  \centering
  \begin{tabular}{c}
    \includegraphics[width=0.80\columnwidth]{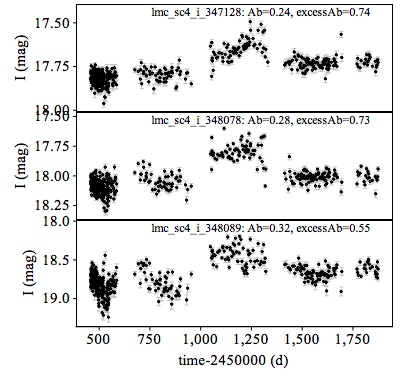} \\
    \includegraphics[width=0.80\columnwidth]{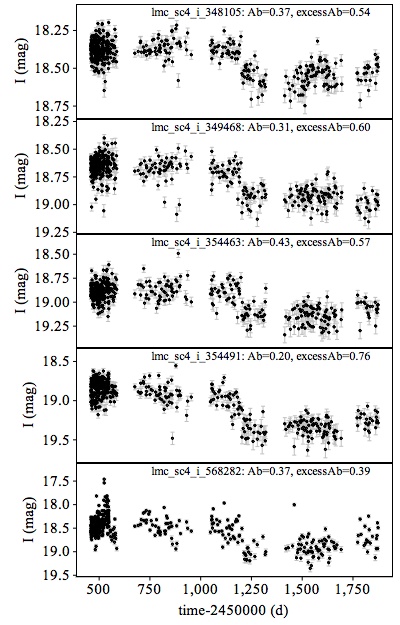} \\
    \includegraphics[width=0.80\columnwidth]{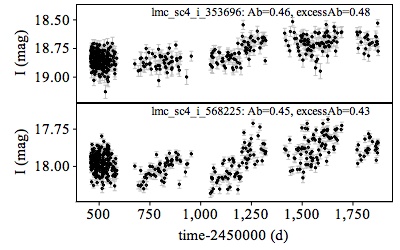} 
  \end{tabular}
  \begin{tabular}{c}
  \includegraphics[width=0.78\columnwidth]{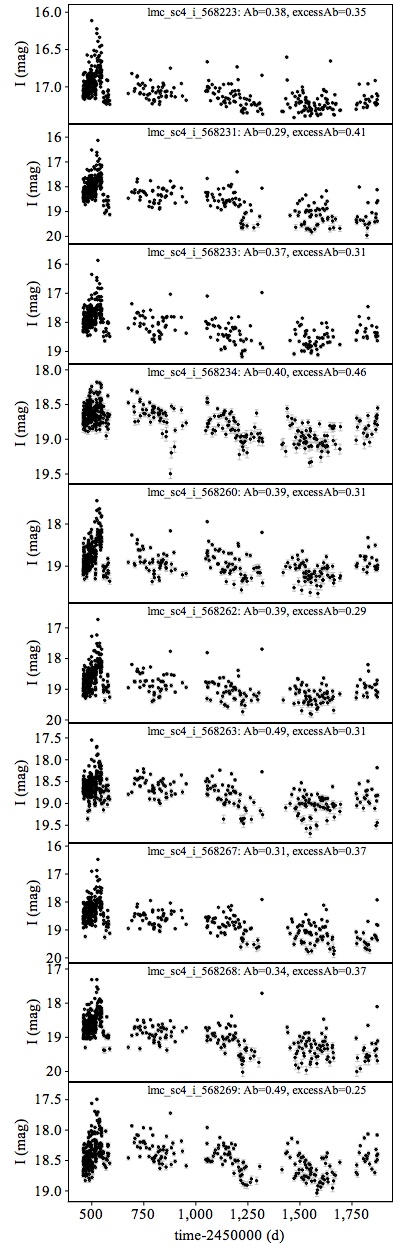} \\
  \includegraphics[width=0.78\columnwidth]{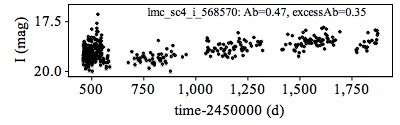}
  \end{tabular}
  \caption{Suspicious OGLE-II transient candidates that fall in a clump on the sky around direction $\alpha=81.63$~deg and $\delta=-69.45$~deg.
   }
\label{Fig:lcsClumpOgle}
\end{figure*}

\clearpage
\section{List of extra OGLE-II transient candidates}
\label{SectAppendix:extraOgleTransients}

This appendix displays the light curves of OGLE-II transient candidates that were not detected as transient candidates in EROS.
Figure~\ref{Fig:lcsOgleWithErosMatch} shows the ones that have a (non-transient) EROS match.
The EROS light curves are also shown in the figures, superposed on the OGLE-II data.
Figure~\ref{Fig:lcsOgleWithoutErosMatch} displays the remaining OGLE-II transient candidates for which no match is found in the EROS data base of variable stars.

\begin{figure*}
  \centering
  \includegraphics[width=1.00\columnwidth]{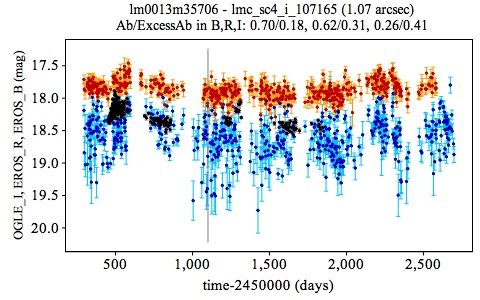}
  \includegraphics[width=1.00\columnwidth]{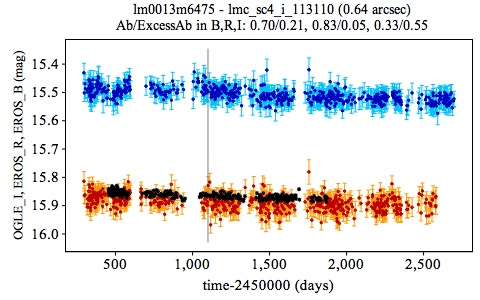}
  \includegraphics[width=1.00\columnwidth]{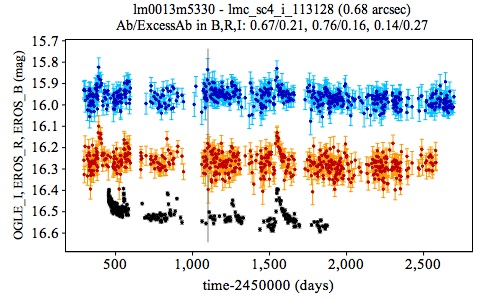}
  \includegraphics[width=1.00\columnwidth]{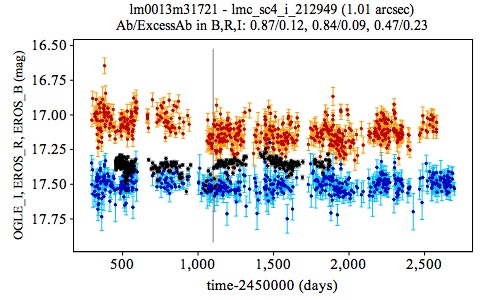}
  \includegraphics[width=1.00\columnwidth]{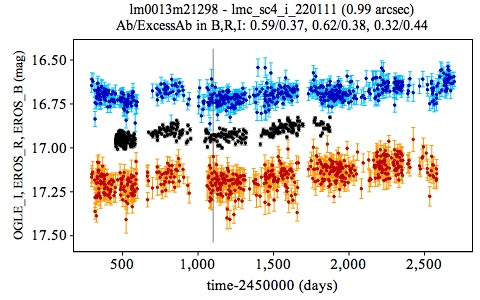}
  \includegraphics[width=1.00\columnwidth]{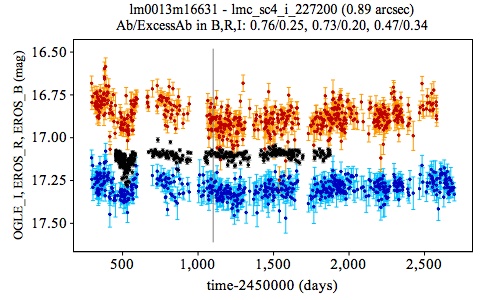}
  \includegraphics[width=1.00\columnwidth]{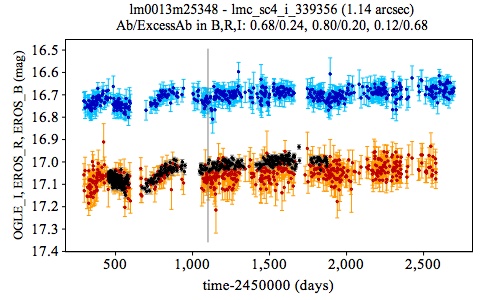}
  \includegraphics[width=1.00\columnwidth]{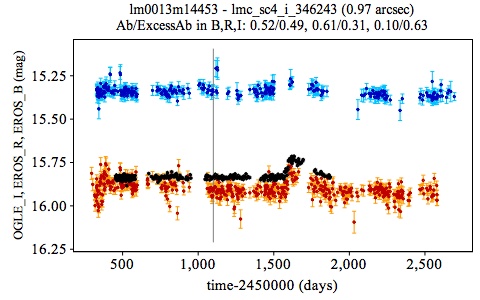}
  \caption{Same as Fig.~\ref{Fig:lcsErosBa}, but for OGLE-II transient candidates that have non-transient EROS matches.
   }
\label{Fig:lcsOgleWithErosMatch}
\end{figure*}

\addtocounter{figure}{-1}

\begin{figure*}
  \centering
  \includegraphics[width=1.00\columnwidth]{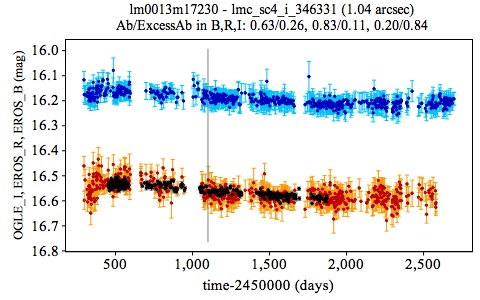}
  \caption{Continued.
   }
\end{figure*}

\newpage
\begin{figure*}
  \centering
  \includegraphics[width=0.65\columnwidth]{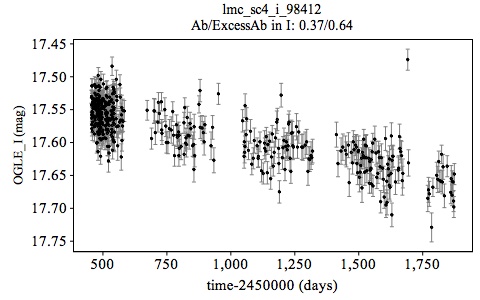}
  \includegraphics[width=0.65\columnwidth]{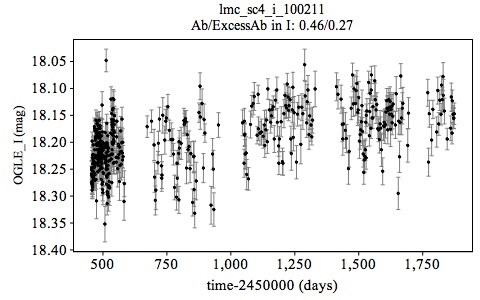}
  \includegraphics[width=0.65\columnwidth]{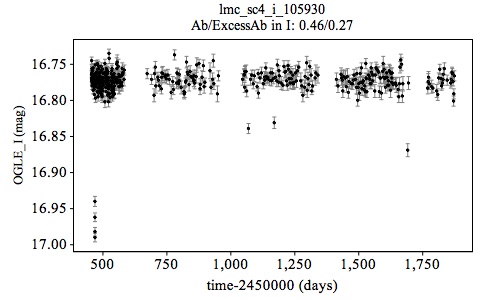}
  \includegraphics[width=0.65\columnwidth]{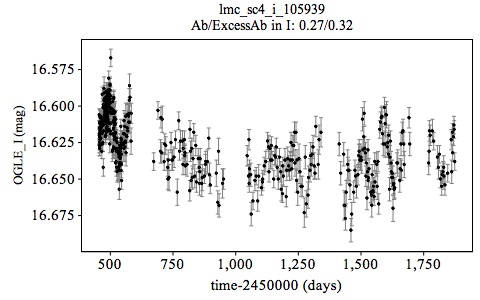}
  \includegraphics[width=0.65\columnwidth]{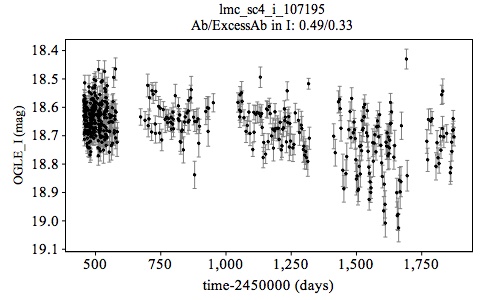}
  \includegraphics[width=0.65\columnwidth]{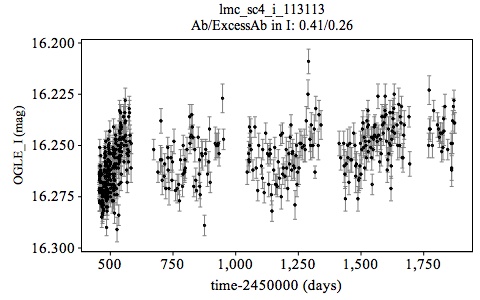}
  \includegraphics[width=0.65\columnwidth]{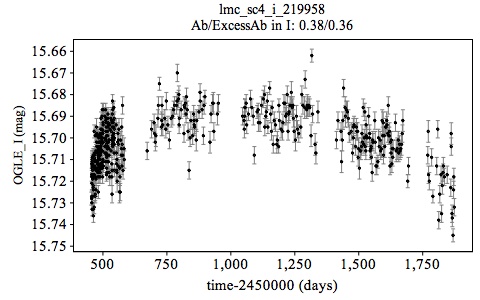}
  \includegraphics[width=0.65\columnwidth]{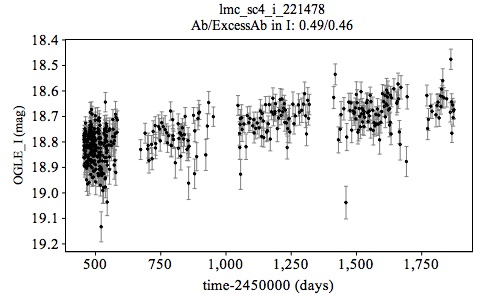}
  \includegraphics[width=0.65\columnwidth]{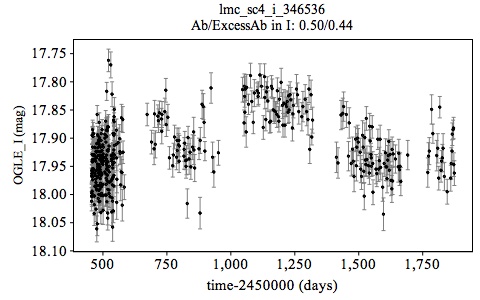}
  \includegraphics[width=0.65\columnwidth]{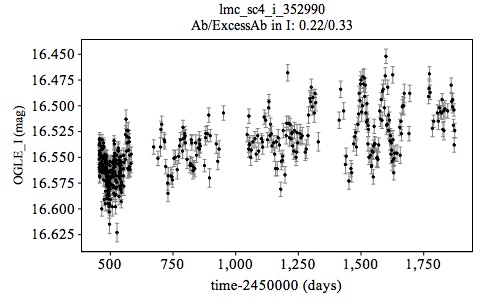}
  \caption{OGLE-II transient candidates that have no EROS match.
   }
\label{Fig:lcsOgleWithoutErosMatch}
\end{figure*}

\clearpage
\section{Be stars candidates with no transient candidate match}
\label{SectAppendix:notMatchingBeLcs}

Figure~\ref{Fig:NonTransientErosAndOgleMatchesToBeLcs} displays OGLE-II and EROS light curves of Be stars of \cite{SabogalMennickentPietrzynski_etal05} that have non-transient EROS matches (see Sect.~\ref{Sect:literature}).
Be stars from \cite{SabogalMennickentPietrzynski_etal05} for which no EROS counterpart is found in the EROS data base of variable stars are shown in Fig.~\ref{Fig:NonTransientOgleMatchesToBeLcs}, where the OGLE-II light curve of their matches are drawn. 

\begin{figure}
  \centering
  \includegraphics[width=1\columnwidth]{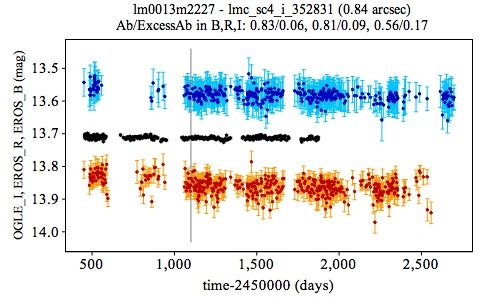}
  \includegraphics[width=1\columnwidth]{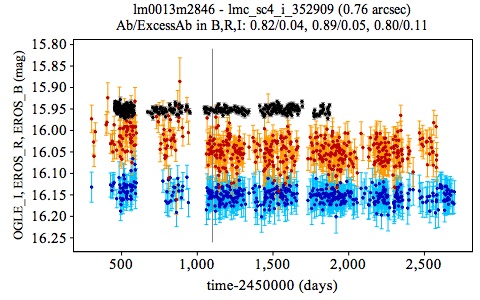}
  \includegraphics[width=1\columnwidth]{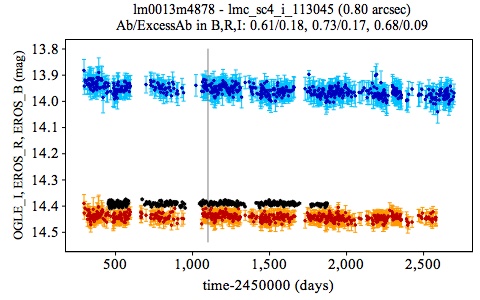}
  \caption{Same as Fig.~\ref{Fig:lcsErosBa}, but for non-transient EROS and OGLE-II matches to Be star candidates of \cite{SabogalMennickentPietrzynski_etal05}.
   }
\label{Fig:NonTransientErosAndOgleMatchesToBeLcs}
\end{figure}  

\addtocounter{figure}{-1}

\begin{figure}
  \centering
  \includegraphics[width=1\columnwidth]{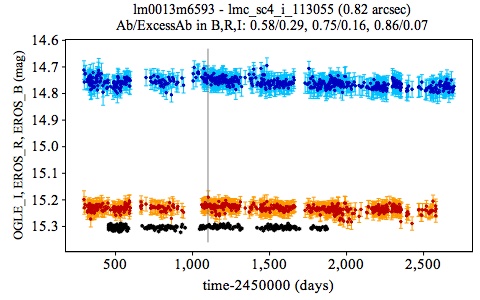}
  \includegraphics[width=1\columnwidth]{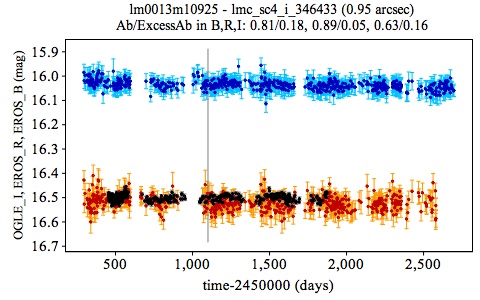}
  \includegraphics[width=1\columnwidth]{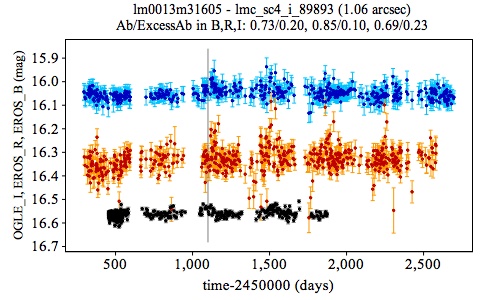}
  \caption{Continued.
   }
\end{figure}

\begin{figure}
  \centering
  \includegraphics[width=0.85\columnwidth]{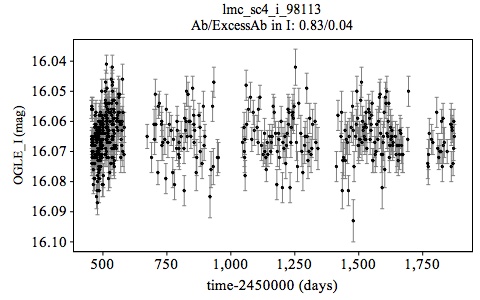}
  \includegraphics[width=0.85\columnwidth]{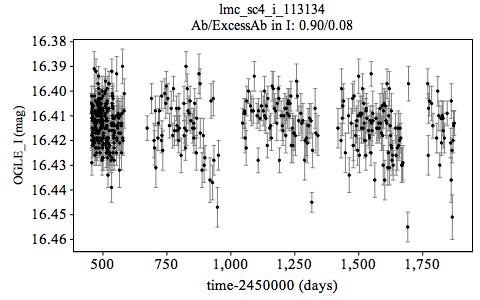}
  \includegraphics[width=0.85\columnwidth]{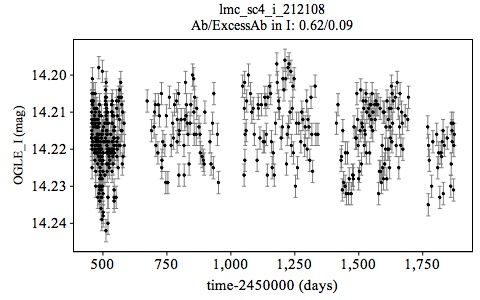}
  \includegraphics[width=0.85\columnwidth]{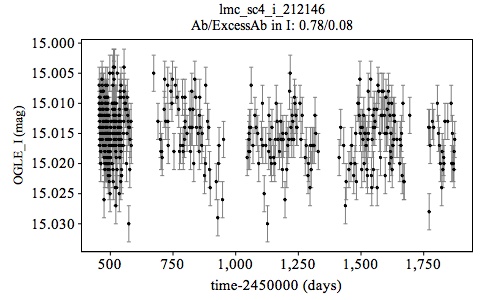}
  \includegraphics[width=0.85\columnwidth]{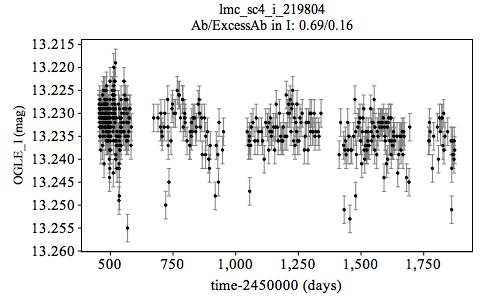}
  \caption{Light curve of OGLE-II matches to Be star candidates of \cite{SabogalMennickentPietrzynski_etal05} that have no detected EROS variable star counterpart.
   }
\label{Fig:NonTransientOgleMatchesToBeLcs}
\end{figure}

\addtocounter{figure}{-1}

\begin{figure}
  \centering
  \includegraphics[width=0.85\columnwidth]{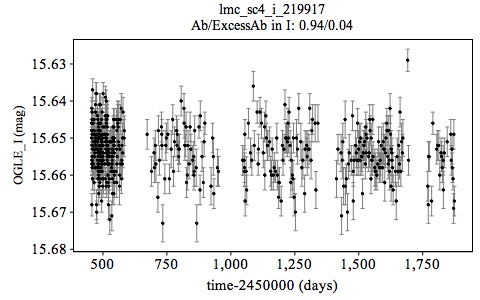}
  \includegraphics[width=0.85\columnwidth]{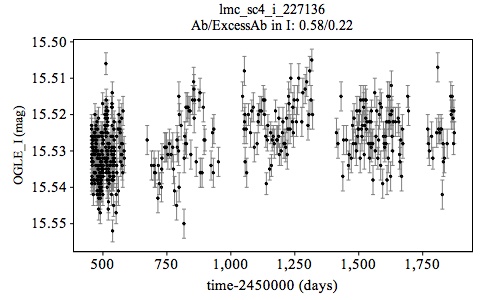}
  \includegraphics[width=0.85\columnwidth]{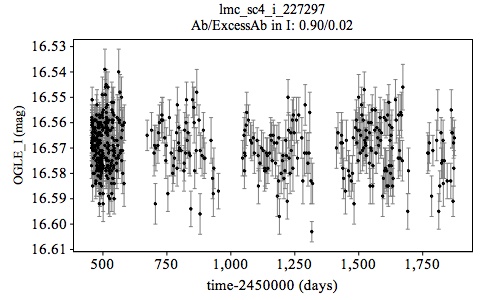}
  \includegraphics[width=0.85\columnwidth]{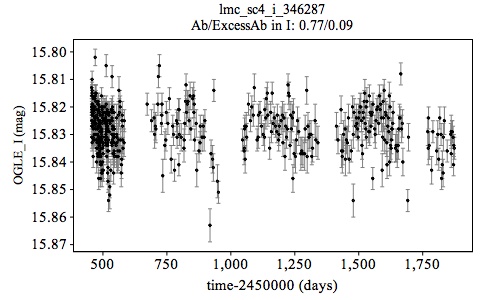}
  \caption{Continued.
   }
\end{figure}

\clearpage
\section{Comments on transient subclassification}
\label{SectAppendix:CommentsSubClassification}

The set up of an automated classification procedure of the transient candidates is beyond the scope of this article, but two points are worth mentioning here that can help setting-up such a procedure.
The first concerns the potential of the Abbe value computed on smoothed light curves to complement $\Ab$ and $\excessAb$ as classification attributes.
The second addresses the potential of combining data from different bands, here $\EROSR$ and $\EROSB$.
They are successively addressed in the next two sections.

\subsection{Abbe value on smoothed light curves}
\label{SectAppendix:AbSmooth}

\begin{figure}
  \centering
  \includegraphics[width=\columnwidth]{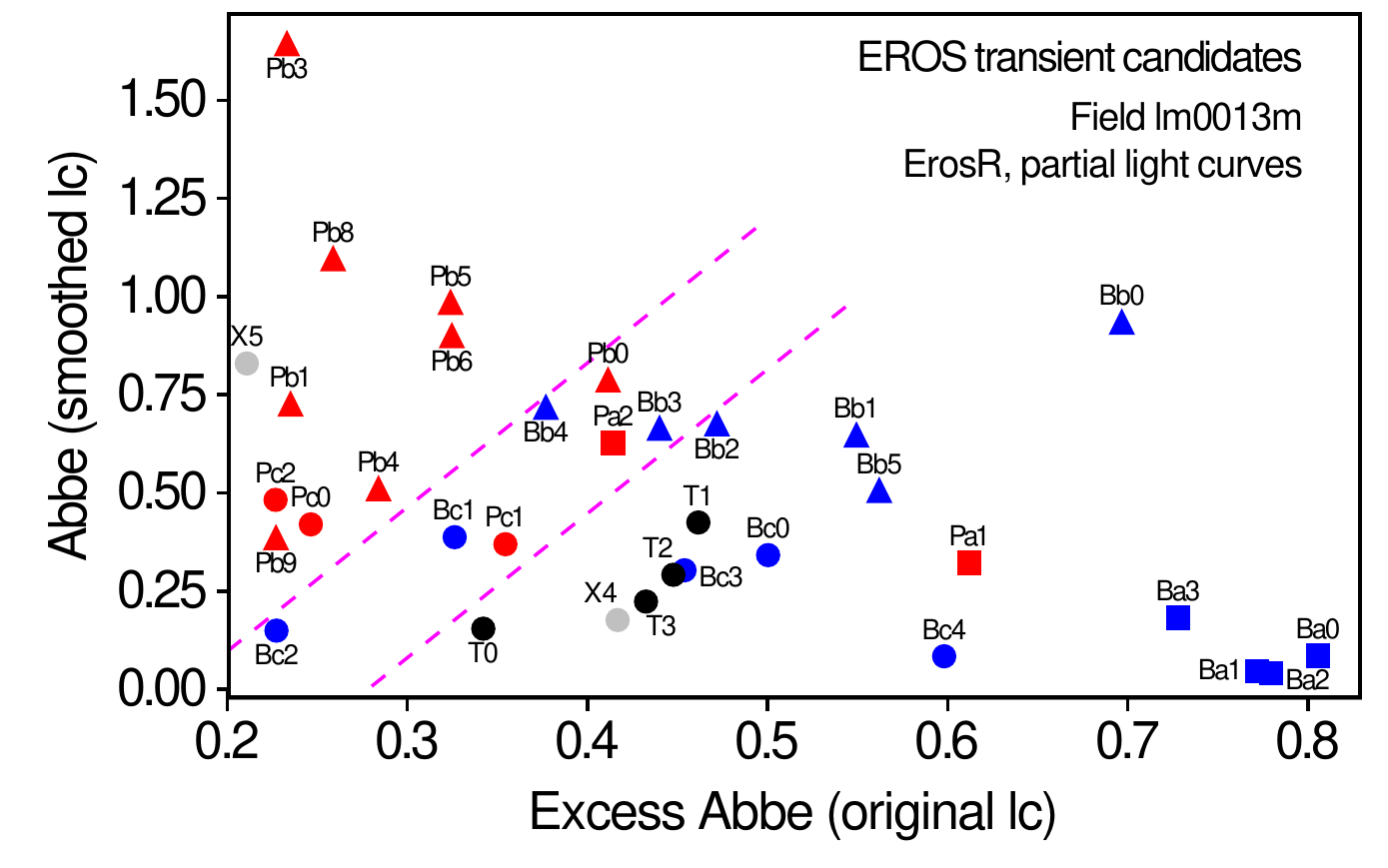}
  \caption{Abbe values of 100~d smoothed $\EROSR$ light curves versus excess Abbe values of their original light curves.
  Only data for transient candidates are shown.
  Type B transient candidates are plotted in blue, using square, triangle and circle markers for Ba, Bb, and Bc types, respectively.
  Type T transient candidates are plotted in black circles.
  Type P transient candidates are plotted in red, using square, triangle, and circle markers for Pa, Pb, and Pc types, respectively.
  Other transient candidates are plotted in gray circles.
  The upper dashed line separates the upper-left part of the diagram, populated by type P transient candidates, while the lower dashed line separates the lower-right part of the diagram, where type B and T transient candidates are found (see text).
  The labels giving the transient IDs are the same as in Figs.~\ref{Fig:ErosDiagramTypesR} and \ref{Fig:ErosDiagramTypesB}.
  }
\label{Fig:ErosAbbeSmooth}
\end{figure}

Smoothed light curves are constructed by averaging measurements in intervals of duration $T_\mathrm{smooth}$ (the mean magnitude in the interval is assigned to the time at the middle of the interval).
I take $T_\mathrm{smooth}=100$~d, and retain the mean magnitude only if the interval contains at least five measurements.
The resulting Abbe values of the smoothed light curves, denoted $\AbSmooth$, of the EROS transient candidates in the $\EROSR$ band are shown in Fig.~\ref{Fig:ErosAbbeSmooth} against the excess Abbe values of their original light curves.
Sources flagged as transient candidates in $\EROSB$ but not in $\EROSR$ are not shown in the figure.

Light curve smoothing removes all variability information on timescales below the time interval $T_\mathrm{smooth}$ used for the smoothing, and enhances the long-term variability pattern.
Therefore, sources displaying outbursts or trends are expected to have, in general, lower $\AbSmooth$ values than sources displaying bursts or other variability patterns.
Figure~\ref{Fig:ErosAbbeSmooth} confirms this expectation:
transient candidates with outbursts (blue squares and circles) or trends (black circles) have $\AbSmooth<0.5$;
pulsating-type stars (red markers) and transients with bursts (blue triangles) have $\AbSmooth>0.3$
and $\gtrsim 0.5$, respectively.
The last two types can further be disentangled to some degree in the $(\excessAb, \AbSmooth)$ diagram, because transients with bursts have larger excess Abbe values than those with pulsating-like variability.
This is shown in Fig.~\ref{Fig:ErosAbbeSmooth}, where types B and T transients are seen to populate the lower-right part of the diagram, while type P transients populate the upper-left part.
A transition region, delimited by the two dashed lines in Fig.~\ref{Fig:ErosAbbeSmooth}, contains both types of stars.
The main exception is star 23946 (Pa1), which lies in the type B region of the figure, with $(\excessAb,\AbSmooth)$ = $(0.61,0.32)$, despite its pulsating-like nature.
Inspection of its light curve, shown in Fig.~\ref{Fig:lcsErosPa}, reveals a very long variability timescale, thus mimicking an outburst pattern of variability.

In conclusion, $\AbSmooth$ may be a good attribute to help classify transient candidates.
An illustration of this potential is given in Sect.~\ref{Sect:OgleDiagram} in the body of the article, where Fig.~\ref{Fig:OgleDiagram} shows that OGLE-II light curves with small $\AbSmooth$ values are predominantly found in the regions of trends and transient candidates, in agreement with the conclusions given here.

\subsection{Combining $\EROSR$ and $\EROSB$ bands}
\label{SectAppendix:ErosTransientCandidatesRB}

\begin{figure}
  \centering
  \includegraphics[width=\columnwidth]{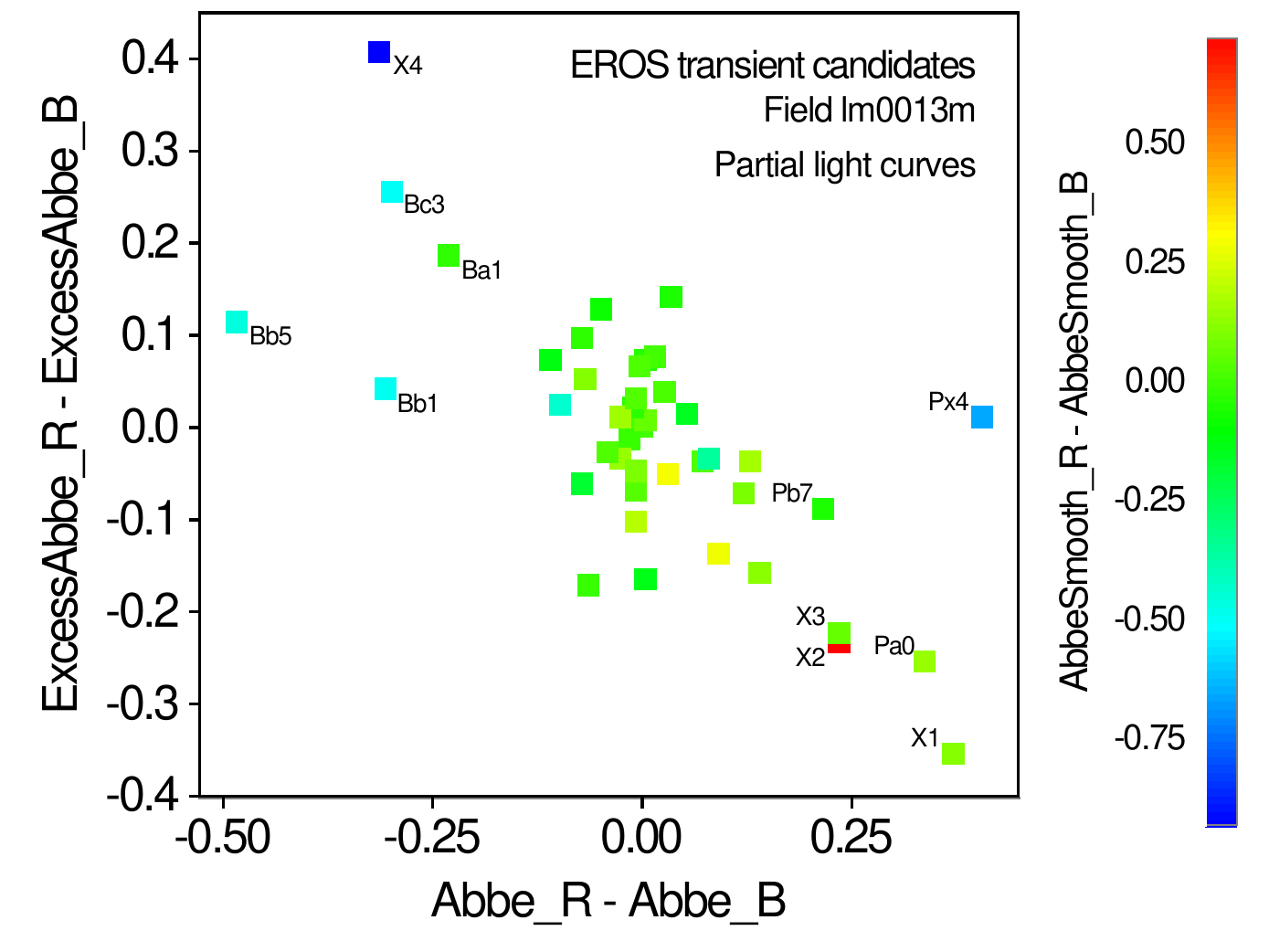}
  \caption{$\Ab(\EROSR) - \Ab(\EROSB)$ versus $\excessAb(\EROSR) - \excessAb(\EROSB)$ of EROS transient candidates.
  The color of the markers scale as $\AbSmooth(\EROSR) - \AbSmooth(\EROSB)$ according to the color scale shown on the right, limited to the range from -0.75 to 0.5. 
  Labels of transient IDs are added to sources which have differences in $\Ab$ either smaller than -0.20 or greater than 0.20.
  Transient IDs are the same as in Figs.~\ref{Fig:ErosDiagramTypesR} and \ref{Fig:ErosDiagramTypesB}.
  }
\label{Fig:ErosBRdiffs}
\end{figure}

If the physics at the origin of a transient phenomenon is achromatic and if the measurement uncertainties are similar in both photometric bands, then we expect similar $\Ab$ and $\excessAb$ values in both bands.
The source would then be found at about the same location in the $\diagram$ diagram, irrespective of the band, but if one of these conditions is not fulfilled, then the Abbe and/or the excess Abbe values may differ significantly in each band.

Most of our transient candidates have similar $\Ab$ and $\excessAb$ values in $\EROSR$ and $\EROSB$, as seen by comparing Figs.~\ref{Fig:ErosDiagramTypesR} and \ref{Fig:ErosDiagramTypesB}.
The differences $\Ab_\mathrm{R}-\Ab_\mathrm{B}$ of the Abbe values and the differences ${\excessAb}_\mathrm{,R}-{\excessAb}_\mathrm{,B}$ of the excess Abbe values in $\EROSR$ and $\EROSB$ are summarized in Fig.~\ref{Fig:ErosBRdiffs}.
For some sources, differences are noteworthy.
In particular, several B-type candidates lie at significantly different positions in the two diagrams.
Two good examples of such cases are sources Bc3 (18409) and Bb5 (22897), highlighted in Fig.~\ref{Fig:ErosBRdiffs}.
Inspection of their light curves, shown in Figs.~\ref{Fig:lcsErosBc} and \ref{Fig:lcsErosBb}, respectively, reveals significant differences in the variability amplitudes in both bands.
In source 18409, an outburst starts around time 1100~d (with some bursts superposed on it), with an amplitude of $\sim$0.3~mag in $\EROSR$, while the amplitude of the burst is a factor of about five less in $\EROSB$.
In source 22897, bursts are seen with amplitudes up to 0.75~mag in $\EROSR$, while no significant burst is visible in $\EROSB$.
Only a small flux enhancement is visible in $\EROSB$ at 1500 and 2000~days, when bursts are recorded in $\EROSR$.
Comparison of Abbe-related data in both bands thus reveals wavelength-dependent amplitudes of transient phenomena.
Both sources are more variable in $\EROSR$ than in $\EROSB$, suggesting that the physical origin of variability, which may differ in both stars, has a greater effect on longer wavelengths than on shorter ones.
We note that the natures of the two stars are different; the first one is a blue star, while the second is a red star.

In total, eleven stars have $|\Ab_R-\Ab_B|>0.20$ (see Fig.~\ref{Fig:ErosBRdiffs}).
In some cases, as warned earlier, the difference in the two bands is simply due to larger point-to-point scatter in one of the light curves compared to the other, probably due to higher noise in that band.
This is the case, for example, for source Pa0 (17790; see light curve in Fig.~\ref{Fig:lcsErosPa}), which has a noisier light curve (higher $\Ab_\mathrm{R}$) in $\EROSR$ than in $\EROSB$.

\end{appendix}

\end{document}